# Supermartingales, Ranking Functions and Probabilistic Lambda Calculus

Andrew Kenyon-Roberts
University of Oxford

C.-H. Luke Ong
University of Oxford

*Abstract*—We introduce a method for proving almost sure termination in the context of lambda calculus with continuous random sampling and explicit recursion, based on ranking supermartingales. This result is extended in three ways. *Antitone ranking functions* have weaker restrictions on how fast they must decrease, and are applicable to a wider range of programs. *Sparse ranking functions* take values only at a subset of the program's reachable states, so they are simpler to define and more flexible. *Ranking functions with respect to alternative reduction strategies* give yet more flexibility, and significantly increase the applicability of the ranking supermartingale approach to proving almost sure termination, thanks to a novel (restricted) confluence result which is of independent interest. The notion of antitone ranking function was inspired by similar work by McIver, Morgan, Kaminski and Katoen in the setting of a first-order imperative language, but adapted to a higher-order functional language. The sparse ranking function and confluent semantics extensions are unique to the higher-order setting. Our methods can be used to prove almost sure termination of programs that are beyond the reach of methods in the literature, including higher-order and non-affine recursion.

## I. INTRODUCTION

Probabilistic (or randomised) programs have long been recognised as essential to the efficient solution of many algorithmic problems [1]. More recently, in probabilistic programming [2–4], probabilistic programs are used to express generative models whose posterior probability can be computed by general purpose inference engines. Sampling from continuous distributions is an essential construct of probabilistic programming languages, such as Church [5], Stan [6], Anglican [7], Gen [8], Pyro [9], Edward [10] and Turing [11]. Another key feature is higher order; in fact some of the most influential probabilistic programming languages are functional.

In this paper we study a central property of probabilistic programs: *termination*. When a probabilistic program implements a solution to an algorithmic problem, we naturally require the computation to terminate with probability 1, in which case the program is called *almost surely terminating* (AST). Indeed, it is standard for designers and implementors of probabilistic programming systems to regard non-AST programs as defining invalid models, and hence inadmissible [3, 5]. (Yet none of these systems provides any support for the development or verification of AST programs.)

Moreover various theorems about probabilistic programs rely on the assumption that the program terminates almost surely (see e.g. [5]). A recent result [12] proves that AST programs have density functions that are differentiable almost everywhere. This is significant for Bayesian machine learning, because almost everywhere differentiability is a precondition for the correctness of some of the most scalable inference algorithms, such as Hamiltonian Monte Carlo [13, 14] and reparameterised gradient variational inference [15].

The AST problem is not just important, but also difficult: Deciding AST of first-order imperative programs with discrete probabilities is $\Pi_2^0$-complete [16]. In recent work [17] the AST problem for higher-order programs with continuous distributions (and suitable primitive functions) is shown to have the same complexity. Though the problem of proving AST of imperative probabilistic programs is much studied [16, 18–23], to our knowledge the problem of proving AST of higher-order programs with continuous distributions is new.

*Contributions*

A possible approach to proving AST is to find some variant on the program state, called *ranking function*, that decreases on average sufficiently quickly that it must at some point reach 0, at which point the program terminates. In other words, the program's behavior is used to define an associated (ranking) *supermartingale*. Proof rules based on relating the program state to supermartingales already exist for first-order imperative programs [21, 24, 25]. This paper's contribution is to extend this method to a higher-order setting.

The language PPCF used in this paper is simply typed, with random continuous sampling, and an explicit recursion primitive, Y. We define a *ranking function* on a term $M$ to be a non-negative measurable function on the reachable terms from $M$ whose expected value decreases or is unchanged as the term is reduced. Because the type system already constrains the terms enough to force termination in the absence of the recursion construct [26, 27], it is only the Y-reduction steps that must be counted, making defining ranking functions somewhat easier. Using supermartingales and Doob's (iterated) Optional Sampling Theorem, we show the soundness of rankability (Thm. III.10): if a term has a ranking function, it is AST.

To define a ranking function, one would typically try to organise the set of reachable terms (of the program being analysed) into a manageable number of syntactic cases. Unfortunately (for any fixed reduction strategy) there are programs whose reachable terms span such a large number of cases that it would be extremely difficult to analyse by hand, if at all possible. How can the construction of ranking functions be made possible or easier?

Our first answer is the notion of *sparse ranking function*. Most of the individual execution steps of a typical program are trivial and easy to mentally skip over. Sparse ranking functions can be defined only for those points in the execution of a program which are semantically important, while all of the other intermediate steps can be ignored. Yet, they are no less efficacious for proving AST (Thm. IV.5): every sparse ranking function is a restriction of a ranking function. This also makes the ranking function method of proving AST more compatible with syntactic sugar, because the intermediate reduction steps implicit in the simplified notation can be ignored.

A ranking function (or sparse ranking function) provides a bound on the expected number of Y-reduction steps before the program terminates, therefore ranking functions cannot be constructed for terms whose expected number of Y-reduction steps is infinite, such as the simplest implementation of the 1D unbiased random walk. This restriction can be removed by generalising ranking functions to *antitone ranking functions*, which rather than having to decrease by a constant amount for each Y-reduction step, may decrease by a variable amount, depending on the value of the ranking function. Thanks to this feature, the antitone ranking function method is capable of handling programs which terminate arbitrarily slowly. We show that this method is also sound for proving AST (Thm. V.6), moreover it also enjoys a sparse function theorem (Thm. V.8).

Basic (non-random) lambda calculus has the very useful property of confluence, which implies (among other things) that even if execution of a program starts in a different order, it will still reach the same normal form eventually (assuming it does reach a normal form). Probabilistic lambda calculus does not have this property, because random choices may be duplicated, and evaluating the same subterm multiple times can yield different results. However, with a restricted set of reduction strategies, confluence may be regained. We introduce a novel addressing scheme for the possible random choices in a program's execution, which ensures that the same random choices are taken at corresponding positions in alternative reduction sequences, so that the same eventual result can be reached. This is then used to prove yet another extension to the ranking function theorem, that (Thm. VI.15) ranking functions may be defined with respect to alternative reduction strategies (which in some cases may lead to a considerably simpler execution and ranking function), and (Cor. VI.12) rankability in this sense still imples almost sure termination. The confluent trace semantics is of wider interest, and has other possible applications as well, for example in Bayesian inference algorithms.

By combining these methods we can prove AST of a variety of PPCF programs that are beyond the reach of methods in the literature, including non-affine recursion[1] (Ex. V.12 and V.13) and recursions that define higher-order functions (Ex. V.14).

*Outline:* We present the syntax and trace semantics of PPCF in Sec. II. In Sec. III, we show that ranking functions on terms induce supermartingales, which form the basis of a sound method for proving AST. We introduce sparse ranking functions in Sec. IV and antitone ranking functions in Sec. V, and illustrate how they can be used to prove AST via examples. In Sec. VI, we present a confluent trace semantics and demonstrate its usefulness. We discuss further applications of the confluent semantics in Sec. VII; and conclude with comments on related work and further directions in Sec. VIII.

*Additional materials:* Further details of some of the examples and all missing proofs can be found in the appendix.

## II. PROBABILISTIC PCF

### A. Syntax of PPCF

The language PPCF is a call-by-value (CBV) version of PCF with sampling of real numbers from the closed interval $[0, 1]$ [12, 28, 29]. Types and terms are defined as follows, where $r$ is a real number, $x$ is a variable, $f : \mathbb{R}^n \to \mathbb{R}$ is any measurable function, and $\Gamma$ is an environment:

$$\begin{aligned} \text{types } A, B &::= \mathsf{R} \mid A \to B \\ \text{values } V &::= \lambda x.M \mid \underline{r} \\ \text{terms } M, N &::= V \mid x \mid M_1\, M_2 \mid \underline{f}(M_1, \ldots, M_n) \mid \mathsf{Y}\, M \\ &\quad \mid \mathsf{if}(M < 0, N_1, N_2) \mid \mathsf{sample} \end{aligned}$$

The typing rules are standard (see Fig. 1). Terms are identified up to $\alpha$-equivalence, as usual. The set of all terms is denoted $\Lambda$, and the set of closed terms is denoted $\Lambda^0$.

To define the CBV reduction relation, let *evaluation contexts* be of the form:

$$\begin{aligned} E &::= [\,] \mid E\, M \mid V\, E \mid \underline{f}(\underline{r_1}, \ldots, \underline{r_{k-1}}, E, M_{k+1}, \ldots, M_n) \\ &\quad \mid \mathsf{Y}\, E \mid \mathsf{if}(E < 0, M_1, M_2) \end{aligned}$$

then a term reduces if it is formed by substituting a redex in a context i.e.

$$\begin{aligned} E[(\lambda x.M)V] &\to E[M[V/x]] \\ E[\underline{f}(\underline{r_1}, \ldots, \underline{r_n})] &\to E[\underline{f(r_1, \ldots, r_n)}] \\ E[\mathsf{Y}\lambda x.M] &\to E[\lambda z.M[(\mathsf{Y}\lambda x.M)/x]z] \\ &\qquad \text{where } z \text{ is not free in } M \\ E[\mathsf{if}(\underline{r} < 0, M_1, M_2)] &\to E[M_1] \text{ where } r < 0 \\ E[\mathsf{if}(\underline{r} < 0, M_1, M_2)] &\to E[M_2] \text{ where } r \geq 0 \\ E[\mathsf{sample}] &\to E[\underline{r}] \text{ where } r \in [0, 1]. \end{aligned}$$

We write $\to^*$ for the reflexive, transitive closure of $\to$. Every closed term either is a value or reduces to another term.

---

[1] Non-affine recursive programs are recursive programs that can, during the evaluation of the recursive body, make multiple recursive calls (of a first-order function) from distinct call sites.

$$\frac{}{\Gamma; x : A \vdash x : A} \qquad \frac{\Gamma; x : A \vdash M : B}{\Gamma \vdash \lambda x.M : A \to B} \qquad \frac{\Gamma \vdash M : A \to B \quad \Gamma \vdash N : A}{\Gamma \vdash M\,N : B}$$

$$\frac{\Gamma \vdash M : (A \to B) \to (A \to B)}{\Gamma \vdash \mathsf{Y}M : (A \to B)} \qquad \frac{\Gamma \vdash M : \mathsf{R} \quad \Gamma \vdash N_1 : A \quad \Gamma \vdash N_2 : A}{\Gamma \vdash \mathsf{if}(M < 0, N_1, N_2) : A}$$

$$\frac{}{\underline{r} : \mathsf{R}} \qquad \frac{}{\Gamma \vdash \mathsf{sample} : \mathsf{R}} \qquad \frac{\Gamma \vdash M_1 : \mathsf{R} \quad \ldots \quad \Gamma \vdash M_n : \mathsf{R}}{\Gamma \vdash \underline{f}(M_1, \ldots, M_n) : \mathsf{R}} \; (f : \mathbb{R}^n \to \mathbb{R})$$

Figure 1. Typing rules of SPCF

## B. Trace semantics

The relation $\to$ allows sample to reduce to any number in $[0, 1]$. This defines which results are possible, but to specify the exact probabilities, an additional argument is needed to determine the outcome of random samples. Let $I := [0, 1] \subset \mathbb{R}$, and define $\mathbb{S} := I^{\mathbb{N}}$, with the Borel $\sigma$-algebra $\Sigma_{\mathbb{S}}$. Equivalently, a basis of measurable sets is $\prod_{i=0}^{\infty} X_i$ where $X_i$ are all Borel and all but finitely many are $I$, and the probability measure $\mu_{\mathbb{S}}$ is given by $\mu_{\mathbb{S}}(\prod_{i=0}^{\infty} X_i) := \prod_{i=0}^{\infty} \mathrm{Leb}(X_i)$, writing Leb for the Lebesgue measure. The maps $\pi_h : \mathbb{S} \to I$ (projecting to the first element) and $\pi_t : \mathbb{S} \to \mathbb{S}$ (popping the first element) are then measurable. Following [30], we call the probability space $(\mathbb{S}, \Sigma_{\mathbb{S}}, \mu_{\mathbb{S}})$ the *entropy space*, and elements of $\mathbb{S}$ *traces*.

Define a *skeleton* to be a term but, instead of having real constants $\underline{r}$, it has a placeholder $\mathsf{X}$, so that each term $M$ has a skeleton $Sk(M)$, and each skeleton $S$ can be converted to a term $S[r]$ given a vector $r$ of $n$ real numbers to substitute in, where $n$ is the number of occurrences of $\mathsf{X}$ in $S$. Following [29, 31, 32], the $\sigma$-algebra on $\Lambda$ is defined by identifying $\Lambda$ with $\bigcup_{m \geq 0} (\mathsf{Sk}_m \times \mathbb{R}^m)$, where $\mathsf{Sk}_m$ is the set of skeletons containing $m$ occurrences of $\mathsf{X}$. Thus identified, we give $\Lambda$ the countable disjoint union topology of the product topology of the discrete topology on $\mathsf{Sk}_m$ and the standard topology on $\mathbb{R}^m$, and take the corresponding Borel $\sigma$-algebra. Note that the connected components of $\Lambda$ have the form $\{S\} \times \mathbb{R}^m$, with $S$ ranging over $\mathsf{Sk}_m$, and $m$ over $\mathbb{N}$.

The one-step reduction of the *trace-based* (or *sampling-style*) operational semantics [30, 31, 33] is given by the function $\mathrm{red} : \Lambda^0 \times \mathbb{S} \to \Lambda^0 \times \mathbb{S}$ where

$$\mathrm{red}(M, s) := \begin{cases} (E[N], s) & \text{if } M = E[R], R \to N \\ & \text{and } R \neq \mathsf{sample} \\ (E[\pi_h(s)], \pi_t(s)) & \text{if } M = E[\mathsf{sample}] \\ (M, s) & \text{if } M \text{ is a value} \end{cases}$$

The result after $n$ steps is then simply

$$\mathrm{red}^n(M, s) = \underbrace{\mathrm{red}(\ldots \mathrm{red}(}_{n} M, s) \ldots)$$

and the limit $\mathrm{red}^\infty$ can then be defined as a partial function as $\lim_{n \to \infty} \mathrm{red}^n(M, s)$ whenever that sequence becomes constant by reaching a value. A term $M$ terminates for a sample sequence $s$ if the limit $\mathrm{red}^\infty(M, s)$ is defined.

The reduction function is measurable, and the set of values is measurable, therefore the set of $s$ such that $M$ terminates at $s$ within $n$ steps is measurable for any $n$, therefore $\{s \mid M \text{ terminates at } s\}$ is measurable [12, 31]. We say that a term $M$ is *almost surely terminating* (*AST*) just if $\mu_{\mathbb{S}}(\{s \mid M \text{ terminates at } s\}) = 1$.

**Example II.1** (Geometric distribution)**.** The term

$$(\mathsf{Y}\lambda fn.\mathsf{if}(\mathsf{sample} - 0.5 < 0, n, f(n+1)))\,\underline{0},$$

which generates a geometric distribution, terminates on the set $\mathbb{S} \setminus [0.5, 1]^{\mathbb{N}}$, which has measure 1, therefore it terminates almost surely; whereas $\mathsf{if}(\mathsf{sample} - 0.5 < 0, \underline{0}, (\mathsf{Y}\lambda x.x)\,\underline{0})$, which terminates on the set $\pi_h^{-1}[[0, 0.5]]$, has probability 0.5 of failing to terminate.

*Remark* II.2. Our definition of AST is equivalent to that given in [12] (although the program semantics is stated in a slightly different way), except for the presence of a score construct for soft conditioning. The score construct is irrelevant to termination except that it fails if its argument is negative, thus allowing computations to fail after finitely many steps. If such a thing were added, it would merely require an additional side condition to our main theorems, that no failing term was reachable (or, in the case of the confluent semantics based results, no term where the reduction strategy is not defined is reachable via the reduction strategy unless it's a value).

## III. SUPERMARTINGALES

One approach to proving that a term terminates almost surely is to find some variant that is bounded below and, on average, decreases sufficiently quickly that it must eventually reach 0, similarly to the approach taken by [21] and others for imperative programs.

These variants are defined as functions from reachable terms (i.e. possible states of the program's execution) to real numbers. Specifically, let the set of reachable terms from a given closed starting term $M$ be $Rch(M) := \{N \in \Lambda \mid M \to^* N\}$, with the $\sigma$-algebra induced as a subset of $\Lambda$.

**Definition III.1.** A *ranking function on* $M \in \Lambda^0$ is a measurable function $f : Rch(M) \to \mathbb{R}$ such that $f(N) \geq 0$

for all $N$,[2] and

(i) $f(E[\mathsf{Y}\lambda x.N]) \geq 1 + f(E[\lambda z.N[(\mathsf{Y}\lambda x.N)/x]z])$ where $z$ is not free in $N$

(ii) $f(E[\mathsf{sample}]) \geq \int_I f(E[\underline{x}])\,\mathrm{Leb}(\mathrm{d}x)$

(iii) $f(E[R]) \geq f(E[R'])$ for any other redex $R$ with $R \to R'$

We say that the ranking function $f$ is *strict* if there exists $\epsilon > 0$ such that for all $E$ and $R \to R'$, $f(E[R']) \leq f(E[R]) - \epsilon$.

Any closed term for which a ranking (respectively, strict ranking) function exists is called *rankable* (respectively, *strictly rankable*). For example, $(\mathsf{Y}\lambda x.x)\,\underline{0}$ is not rankable.

It will be demonstrated later that for any rankable term $M$, if $(M_n)_{n \geq 0}$ is the reduction sequence starting from $M$ (considered as a stochastic process), then $(f(M_n))_{n \geq 0}$ is a supermartingale, and $M$ terminates almost surely, but first, some preliminaries about supermartingales (see e.g. [34, 35] for details).

Fix a probability space $(\Omega, \mathcal{F}, \mathbb{P})$ and a filtration $(\mathcal{F}_n)_{n \geq 0}$ (i.e. $\mathcal{F}_n \subseteq \mathcal{F}$ is a $\sigma$-algebra, and $\mathcal{F}_n \subseteq \mathcal{F}_{n+1}$ for all $n$). Let $T$ be a r.v. that takes values in $\mathbb{N} \cup \{\infty\}$. We call $T$ a *stopping time adapted to* $(\mathcal{F}_n)_{n \geq 0}$ just if $\{T = n\} \in \mathcal{F}_n$, for all $n$.

**Definition III.2.** (i) A sequence of r.v.s $(Y_n)_{n \geq 0}$ adapted to a filtration $(\mathcal{F}_n)_{n \geq 0}$ is a *supermartingale* if for all $n \geq 0$, $Y_n$ is integrable (i.e. $\mathbb{E}[|Y_n|] < \infty$), and $\mathbb{E}[Y_{n+1} \mid \mathcal{F}_n] \leq Y_n$ a.s. (i.e. for all $A \in \mathcal{F}_n$, $\int_A Y_{n+1}(\omega)\,\mathbb{P}(\mathrm{d}\omega) \leq \int_A Y_n(\omega)\,\mathbb{P}(\mathrm{d}\omega)$).

(ii) Let $\epsilon > 0$. Given a stopping time $T$ and a supermartingale $(Y_n)_{n \geq 0}$, both adapted to filtration $(\mathcal{F}_n)_{n \geq 0}$, we say that $(Y_n)_{n \geq 0}$ is a $\epsilon$-*ranking supermartingale w.r.t.* $T$ if for all $n$, $Y_n \geq 0$ and $\mathbb{E}[Y_{n+1} \mid \mathcal{F}_n] \leq Y_n - \epsilon \cdot \mathbf{1}_{\{T > n\}}$.[3] A *ranking supermartingale w.r.t.* $T$ is a $\epsilon$-ranking supermartingale w.r.t. $T$ for some $\epsilon > 0$.

*Remark* III.3. Our notion of ranking supermartingale is a slight generalisation of the original definition [21, 24], which does not involve an arbitrary stopping time.

Intuitively $Y_n$ gives the rank of the program after $n$ steps of computation, and $T$ is the time at which it reaches a value (which may be infinite if it fails to terminate). In a $\epsilon$-ranking supermartingale, each computation step causes a strict decrease in rank of at least $\epsilon$, provided the term in question is not a value.

**Lemma III.4.** *Let $(Y_n)_{n \geq 0}$ be a $\epsilon$-ranking supermartingale w.r.t. the stopping time $T$, then $T < \infty$ a.s., and $\mathbb{E}[T] \leq \frac{\mathbb{E}[Y_0]}{\epsilon}$.*

*Remark* III.5. Lem. III.4 is a slight extension of [21, Lem. 5.5], which asserts the same result but for the specific stopping time $T : \omega \mapsto \min\{n \mid Y_n(\omega) = 0\}$.

Let $T$ and $T'$ be stopping times adapted to $(\mathcal{F}_n)_{n \geq 0}$. Recall

---

[2] The fact that the ranking function can be 0 in some cases before it reaches a value is necessary to get Thm. IV.2 to work neatly.

[3] For $A \in \mathcal{F}$, we write $\mathbf{1}_A$ for the *indicator function* of $A$, i.e., the random variable defined by $\mathbf{1}_A(\omega) := 1$ if $\omega \in A$, and 0 otherwise.

the $\sigma$-algebra (consisting of measurable subsets "prior to $T$")

$$\mathcal{F}_T := \{A \in \mathcal{F} \mid \forall i \geq 0\,.\,A \cap \{T \leq i\} \in \mathcal{F}_i\}$$

and if $T \leq T'$, then $\mathcal{F}_T \subseteq \mathcal{F}_{T'}$.

The following is an iterated version of Doob's well-known Optional Sampling Theorem (see e.g. [21, 36]).

**Theorem III.6** (Optional Sampling). *Let $(X_n)_{n \geq 0}$ be a supermartingale, and $(T_n)_{n \geq 0}$ a sequence of increasing stopping times, all adapted to filtration $(\mathcal{F}_n)_{n \geq 0}$, then $(X_{T_n})_{n \geq 0}$ is a supermartingale adapted to $(\mathcal{F}_{T_n})_{n \geq 0}$ if one of the following conditions holds:*

*(i) each $T_n$ is bounded i.e. $T_n < c_n$ where $c_n$ is a constant*

*(ii) $(X_n)_{n \geq 0}$ is uniformly integrable.*

*A. Ranking functions and supermartingales*

Henceforth, fix the probability space $(\mathbb{S}, \Sigma_\mathbb{S}, \mu_\mathbb{S})$, and a closed PPCF term $M$. For $n \geq 0$, define the random variables

$$M_n(s) := \pi_0(\mathrm{red}^n(M, s))$$
$$T_M(s) := \min\{n \mid M_n(s) \text{ is a value}\}$$

and the filtration $(\mathcal{F}_n)_{n \geq 0}$ where $\mathcal{F}_n := \sigma(M_1, \cdots, M_n)$. Thus $T_M$ is the *runtime* of $M$ (and $M$ is AST iff $\mu_\mathbb{S}(T_M < \infty) = 1$); we say that $M$ is *positively almost surely terminating* (*PAST*) if $\mathbb{E}[T_M] < \infty$.

Our first result is the following theorem.

**Theorem III.7** (Deriving supermartingales). *If a closed PPCF term $M$ is rankable (respectively, strictly rankable) by $f$ then $(f(M_n))_{n \geq 0}$ is a supermartingale (respectively, ranking supermartingale w.r.t. stopping time $T_M$) adapted to $(\mathcal{F}_n)_{n \geq 0}$.*

The ranking supermartingale condition is satisfied essentially by a case analysis on the type of redex (Lem. A.1).

*B. Soundness of rankability*

In the rest of this section we show that if a PPCF term is rankable, then it is AST.

Let $f$ be a ranking function on $M \in \Lambda^0$. Define random variables on the probability space $(\mathbb{S}, \Sigma_\mathbb{S}, \mu_\mathbb{S})$:

$$T_{-1}(s) := -1$$
$$T_{n+1}(s) := \min\{k \mid k > T_n(s), M_k(s) \text{ a value}$$
$$\qquad\qquad\qquad \text{ or of form } E[\mathsf{Y}\lambda x.N]\}$$
$$Y_n(s) := f(M_{T_n(s)}(s))$$
$$T_M^{\mathsf{Y}}(s) := \min\{n \mid M_{T_n(s)}(s) \text{ is a value}\} \qquad (1)$$

We first state a useful property about r.v.s $T_0, T_1, T_2, \ldots$.

**Lemma III.8.** *$(T_n)_{n \geq 0}$ is an increasing sequence of stopping times adapted to $(\mathcal{F}_n)_{n \geq 0}$, and each $T_i$ is bounded.*

The random variable $T_M^{\mathsf{Y}}$, which we call the Y-*runtime* of $M$, can equivalently be defined as the number of Y-reduction steps in the reduction sequence of $M$. Note that, as the type system ensures that the reduction relation excluding Y-reduction is strongly normalising, only finitely many reductions can occur in a row without one of them being a Y-reduction, therefore $T_M^Y < \infty$ a.s. iff $M$ is AST. We say that $M$ is Y-*PAST* if $\mathbb{E}[T_M^Y] < \infty$.

**Lemma III.9.** $T_M^Y$ *is a stopping time adapted to* $(\mathcal{F}_{T_n})_{n \geq 0}$.

**Theorem III.10** (Soundness of rankability). *(i) If a closed PPCF term $M$ is rankable, then $M$ is AST and Y-PAST.*

*(ii) If a closed PPCF term $M$ is strictly rankable, then $M$ is PAST.*

*Proof.* Let $f$ be a ranking function on $M$. For (i), since $(T_n)_{n \geq 0}$ is an increasing sequence of stopping times, each is adapted to $(\mathcal{F}_n)_{n \geq 0}$ and bounded (Lem. III.8), and $(f(M_n))_{n \geq 0}$ is a supermartingale also adapted to $(\mathcal{F}_n)_{n \geq 0}$ (Thm. III.7), it follows from the Optional Sampling Thm. III.6 that $(Y_n)_{n \geq 0}$ is a supermartingale adapted to $(\mathcal{F}_{T_n})_{n \geq 0}$. Notice the stopping time $T_M^Y$ is also adapted to $(\mathcal{F}_{T_n})_{n \geq 0}$ (Lem. III.9); and we have that $(Y_n)_{n \geq 0}$ is a 1-ranking supermartingale. Therefore, by Lem. III.4, $T_M^Y < \infty$ a.s. and $\mathbb{E}[T_M^Y] < \infty$. Statement (ii) follows at once from Thm. III.7 and Lem. III.4. □

Thus the method of (strict) ranking function is sound for proving (positive) a.s. termination of PPCF programs. It is in fact also complete in a sense: if $\mathbb{E}[T_N^Y] < \infty$ for all $N \in Rch(M)$ then $M$ is rankable (Thm. IV.2).

## IV. CONSTRUCTING RANKING FUNCTIONS

Although rankability implies almost sure termination, the converse does not hold in general. For example,

$$\text{if}(-\text{sample} < 0, \underline{0}, (Y\lambda x.x)\underline{0})$$

terminates in 3 steps with probability 1, but isn't rankable because $(Y\lambda x.x)\underline{0}$ is reachable, although that has probability 0. Not only is this counterexample AST, it's PAST.

A ranking function can be constructed under the stronger assumptions that, for every $N$ reachable from $M$, the expected number of Y-reduction steps from $N$ to a value is finite. In particular, the expected number of Y-reduction steps from each reachable term is a ranking function. However, a finite number of expected Y-reduction steps does not necessarily imply a finite number of expected total reduction steps.

**Example IV.1.** The term $M = \Xi(\lambda x.x + 1)$ where

$$\Xi := Y\lambda f\, n.\text{if}(\text{sample} - 0.5 < 0, n\,\underline{0}, f(\lambda x.n(n\,x)))$$

is Y-PAST and rankable, terminating with only 2 Y-reductions on average, i.e., $\mathbb{E}[T_M^Y] = 2$, but applies the increment function $2^n$ times with probability $2^{-n-1}$ for $n \geq 0$, which diverges, i.e., $\mathbb{E}[T_M] = \infty$.

**Theorem IV.2.** *Given a closed term $M$, the function $f : Rch(M) \to \mathbb{R}$ given by*

$f(N) := \mathbb{E}[\text{number of Y-reduction steps from } N \text{ to a value}]$

*if it exists, is the least of all possible ranking functions of $M$.*

### A. Sparse ranking functions

Even in the case of reasonable simple terms, explicitly constructing a ranking function would be a lot of work, and Thm. IV.2 makes even stronger assumptions than almost sure termination, so it isn't useful for proving it.

**Example IV.3** (Geometric distribution). Let

$$\Theta := \lambda f, n.\text{if}(\text{sample} - 0.5 < 0, n, f(n+1)).$$

The term $(Y\Theta)\underline{0}$ generates a geometric distribution. Despite its simplicity, its $Rch$ contains all the terms in Fig. 2, for each $i \in \mathbb{N}, r \in I$.

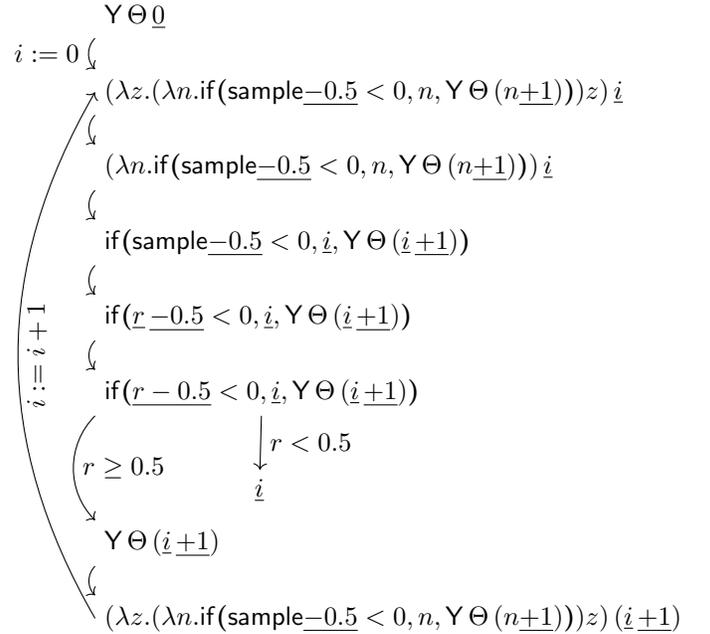

Figure 2. The reachable terms of $Y\Theta\underline{0}$

Even in this simple case, defining a ranking function explicitly is awkward because of the number of cases, although in most cases, because the value need only be greater than or equal to that of the next term in sequence, it suffices to take the ranking function as having the same value as the next term, so that overall it takes only 3 distinct values. (We will explain why later.)

The definition of rankability is also inconvenient for syntactic sugar. It could be useful, for example, to define $M \oplus_p N := \text{if}(\text{sample} - p < 0, M, N)$, where $M \oplus_p N$ reduces to $M$ or $N$, depending on the first value of $s \in \mathbb{S}$, with probability $p$ resp. $(1-p)$. Technically though, it reduces first to $\text{if}(\underline{r} - p < 0, M, N)$ for all $r \in I$, so those terms all need values of the ranking function too.

In both of these cases, there are only some values of the ranking function that are semantically important.

**Definition IV.4.** Define a *sparse ranking function* on a closed term $M$ to be a partial function $f : Rch(M) \rightharpoonup \mathbb{R}$ such that (i) $f(N) \geq 0$ for all $N \in \text{dom}(f)$, (ii) $M \in \text{dom}(f)$, (iii) for any $N \in \text{dom}(f)$, evaluation of $N$ will eventually reach some $O$ which is either a value or in $\text{dom}(f)$, and $f(N) \geq \mathbb{E}[f(O) +$ the number of Y-reduction steps from $N$ to $O$] (where $f(O)$ is taken to be 0 if $O$ is a value outside of $\text{dom}(f)$).

A sparse ranking function that is total is just a ranking function. Providing a sparse ranking function is essentially part way between providing a ranking function and directly proving almost sure termination.

**Theorem IV.5** (Sparse function). *Every sparse ranking function is a restriction of a ranking function.*

As a corollary, any term which admits a sparse ranking function terminates almost surely.

*B. Examples*

Let $M \oplus_p N = \text{if}(\text{sample} - p < 0, M, N)$, for $p \in (0, 1]$, then there are the pseudo-reduction relations

$$E[M \oplus_p N] \to^3 E[M] \qquad E[M \oplus_p N] \to^3 E[N]$$

$$\text{red}^3(E[M \oplus_p N], s) = \begin{cases} (E[M], \pi_t(s)) & \text{if } \pi_h(s) < p \\ (E[N], \pi_t(s)) & \text{if } \pi_h(s) \geq p. \end{cases}$$

A sparse ranking function could be defined with respect to this shortcut reduction simply by replacing $\to$ and $\text{red}$ in the definition of a sparse ranking function by a version that goes straight from $N \oplus_p O$ to $N$ or $O$. Such a pseudo-sparse ranking function would then be a partial function from a subset of $Rch(M)$, so it could also be considered as a partial function from all of $Rch(M)$, and it would in fact also be an actual sparse ranking function. It is therefore possible to prove rankability directly using the shortcutted reductions.

A similar procedure would work for other forms of syntactic sugar. If a closed term $N$ eventually reduces to one of a set of other terms $\{N_i \mid i \in I\}$ with certain probabilities, a sparse ranking function defined w.r.t. a reduction sequence that skips straight from $N$ to $N_i$ is also a valid sparse ranking function for the original reduction function, and therefore its existence implies almost sure termination. There is a caveat, however, that Y-reduction steps skipped over in the shortcut still need to be counted for the expected number of Y-reduction steps.

With this abbreviation, the geometric distribution example from earlier can be written as $(\mathsf{Y}\lambda fn.\ n \oplus_{0.5} f(n+1))\ \underline{0}$. It is then easy to see that the following is a sparse ranking function:

$$(\mathsf{Y}\lambda fn.\ n \oplus_{0.5} f(n+1))\ N \mapsto 2$$
$$\underline{i} \oplus_{0.5} (\mathsf{Y}\lambda fn.\ n \oplus_{0.5} f(n+1))\ (\underline{i}+1) \mapsto 1$$
$$\underline{i} \mapsto 0.$$

In fact, even the partial function $\mathsf{Y}\Theta N \mapsto 2$ alone is a sparse ranking function for this term.

## V. ANTITONE RANKING FUNCTIONS

**Example V.1** (Random walk). (i) One-dimensional (1D) biased (towards 0) random walk

$$M_1 = \bigl(\mathsf{Y}\ \lambda fn.\ \text{if}(n=0, 0, f\,(n-1) \oplus_{2/3} f\,(n+1))\bigr)\ \underline{10}$$

(ii) 1D unbiased random walk

$$M_2 = \bigl(\mathsf{Y}\ \lambda fn.\ \text{if}(n=0, 0, f\,(n-1) \oplus_{1/2} f\,(n+1))\bigr)\ \underline{10}$$

(iii) 1D biased (away from 0) random walk

$$M_3 = \bigl(\mathsf{Y}\ \lambda fn.\ \text{if}(n=0, 0, f\,(n-1) \oplus_{1/3} f\,(n+2))\bigr)\ \underline{10}$$

The term $M_1$ is rankable, terminating in 31 Y-steps on average, and $M_3$ only has a $1/1024$ chance of terminating, but in between, $M_2$ is AST, but isn't Y-PAST, therefore it isn't rankable and Thm. III.10 is insufficient to prove its termination. Thus we seek a generalised notion of ranking function so that $M_2$ becomes rankable, and then prove it sound i.e. rankable in this generalised sense implies AST.

**Definition V.2.** The definition of *antitone ranking function* $f$ for $M \in \Lambda^0$ is the same as that of ranking function except that in the case of Y-redex, we require the existence of an antitone (meaning: $r < r'$ implies $\epsilon(r) \geq \epsilon(r')$) function $\epsilon : \mathbb{R}_{\geq 0} \to \mathbb{R}_{>0}$ such that the ranking function $f : Rch(M) \to \mathbb{R}_{\geq 0}$ satisfies

$$f(E[R']) \leq f(E[R]) - \epsilon(f(E[R]))$$

where $R \to R'$ is the Y-redex rule. Any closed term for which an antitone ranking function exists is called *antitone rankable*.

Note that antitone ranking functions are actually a generalisation of ranking functions, even though the way we reference them may suggest that they are a type of ranking functions.

**Definition V.3.** Given a probability space $(\Omega, \mathcal{F}, \mathbb{P})$, and a supermartingale $(Y_n)_{n \geq 0}$ and a stopping time $T$ adapted to filtration $(\mathcal{F}_n)_{n \geq 0}$, we say that $(Y_n)_{n \geq 0}$ is an *antitone strict supermartingale w.r.t.* $T$ if for all $n \geq 0$, we have $Y_n \geq 0$, and there exists an antitone function $\epsilon : \mathbb{R}_{\geq 0} \to \mathbb{R}_{>0}$ satisfying $\mathbb{E}[Y_{n+1} \mid \mathcal{F}_n] \leq Y_n - \epsilon(Y_n) \cdot \mathbf{1}_{\{T>n\}}$.

**Theorem V.4.** *Let $(Y_n)_{n \geq 0}$ be an antitone strict supermartingale w.r.t. stopping time $T$. Then $T < \infty$ a.s.*

It is essential in this theorem that $\epsilon$ be defined for all $\mathbb{R}_{\geq 0}$, not just the values that $(Y_n)_{n \geq 0}$ actually takes (or at least if $(Y_n)_{n \geq 0}$ is uniformly bounded, $\epsilon$ must be positive at the supremum as well as the realized values).

**Example V.5.** Take the probability space $(\Omega, \mathcal{F}, \mathbb{P})$ where $\Omega$ is the closed interval of reals $[-1, 1]$, $\mathcal{F}$ the Borel $\sigma$-algebra, and $\mathbb{P}$ the corresponding Lebesgue probability measure. Let $\omega \in [-1, 1]$. Define random variables $T$ and $(Y_k)_{k \geq 0}$:

$$T(\omega) := \begin{cases} \min\{n \in \mathbb{N} \mid \omega > 2^{-n}\} & \text{if } \omega \in (0, 1] \\ \infty & \text{otherwise} \end{cases}$$

$$Y_k(\omega) := \begin{cases} 4 - 2^{-k} & \text{if } \omega \in [-1, 2^{-k}] \text{ (equivalently } k < T(\omega)) \\ 0 & \text{otherwise} \end{cases}$$

Plainly, $T$ is a stopping time, and $(Y_n)_{n \geq 0}$ is a supermartingale, adapted to the filtration $(\mathcal{F}_n)_{n \geq 0}$ where $\mathcal{F}_n = \mathcal{F}$ for all $n$. In this case, $(Y_n)_{n \geq 0}$ either tends to 4 as $n \to \infty$, or drops to 0 at some point, with an exponentially decreasing probability. With respect to the antitone function $\epsilon(x) = \frac{4-x}{4}$ and stopping time $T$, we have that $(Y_n)_{n \geq 0}$ is an antitone strict supermartingale, except that $\epsilon(4) = 0$, and $T = \infty$ with probability $\frac{1}{2}$, even though 4 is larger than any value $(Y_n)_{n \geq 0}$ can actually take, and $\epsilon(Y_n)$ never actually reaches 0.

**Theorem V.6** (Antitone ranking function soundness). *If a closed PPCF term $M$ is antitone rankable, then $T_M^{\mathsf{Y}} < \infty$ a.s. (equivalently, $M$ is AST).*

*Proof.* As in Thm. III.10, take the probability space $(\mathbb{S}, \Sigma_{\mathbb{S}}, \mu_{\mathbb{S}})$, and define the same r.v. $T_n, X_n, Y_n, T_M^{\mathsf{Y}}$. Thanks to Thm. III.7, $(Y_n)_{n \geq 0}$ is a supermartingale; and because $M$ is now assumed to be antitone rankable, it is an antitone strict supermartingale w.r.t. stopping time $T_M^{\mathsf{Y}}$. Thus, by Thm. V.4, $T_M^{\mathsf{Y}} < \infty$ a.s. □

As before, constructing antitone ranking fuctions completely is not necessary, and there is a corresponding notion of an antitone sparse ranking function.

**Definition V.7.** Define an *antitone sparse ranking function* on a closed term $M$ to be a partial function $F : Rch(M) \rightharpoonup \mathbb{R}$ such that for some antitone function $\epsilon : \mathbb{R}_{\geq 0} \to \mathbb{R}_{>0}$: (i) $f(N) \geq 0$ for all $N \in \text{dom}(f)$, (ii) $M \in \text{dom}(f)$, (iii) for any $N \in \text{dom}(f)$, evaluation of $N$ will eventually reach some $O$ which is either a value or in $\text{dom}(f)$, and $f(N) \geq \mathbb{E}[f(O) + \epsilon(f(O)) \times$ the number of Y-reduction steps from $N$ to $O]$ (where $f(O)$ is taken to be 0 if $O$ is a value outside of $\text{dom}(f)$).

**Theorem V.8** (Sparse function). *Every antitone sparse ranking function is a restriction of an antitone ranking function.*

As a corollary, any term which admits an antitone sparse ranking function terminates almost surely.

*Ex. V.1 revisited:* Programs $M_1$ and $M_2$ are AST. For $M_1$, the sparse ranking function

$$\bigl(\mathsf{Y}\, \lambda fn.\, \mathsf{if}(n = 0, 0, f(n-1) \oplus_{3/2} f(n+1))\bigr) \underline{x} \mapsto 3x + 1$$

suffices to prove its termination (and could equivalently be considered an antitone sparse ranking function with the constant antitone function $\epsilon_1(x) = 1$).

**Example V.9** (Unbiased random walk). For $M_2$, define the function $g_2 : \mathbb{R}_{\geq 0} \to \mathbb{R}_{\geq 0}$ by $g_2(x) = \ln(x+1) + 1$. Using shorthand $\Theta_2 = \mathsf{Y}\, \lambda fn.\, \mathsf{if}(n = 0, 0, f(n-1) \oplus_{1/2} f(n+1))$, we can define an antitone sparse ranking function $f_2 : Rch(M_2) \rightharpoonup \mathbb{R}_{\geq 0}$ by $f_2 : \Theta_2\, \underline{n} \mapsto g_2(n)$ and $\underline{0} \mapsto 0$ for $n \in \mathbb{N}$. For $n \geq 1$, $\Theta_2\, \underline{n}$ reduces in several steps to either $\Theta_2\, (\underline{n-1})$ or $\Theta_2\, (\underline{n+1})$,[4] each with probability $1/2$, with one Y-reduction. Now

$$g_2(n) - \frac{g_2(n-1) + g_2(n+1)}{2}$$
$$= \ln\left(\frac{n+1}{\sqrt{n(n+2)}}\right) = \frac{1}{2}\ln\left(\frac{(n+1)^2}{n(n+2)}\right)$$
$$= \frac{1}{2}\ln\left(1 + \frac{1}{n(n+2)}\right) > \frac{1}{n(n+2)+1}$$
$$> \frac{1}{2}\frac{1}{(n+1)^2} + \frac{1}{2}\frac{1}{(n+3)^2}$$

Moreover $\Theta_2\, \underline{0}$ reduces to $\underline{0}$ with one Y-reduction with a reduction in $g_2$ of 1. Therefore by setting $\epsilon_2(x) = \frac{1}{(e^{x-1}+1)^2}$, the condition on how much $g_2$ must decrease is met. It is also defined at $M_2 = \Theta_2\, \underline{10}$, and is non-negative, therefore it is an antitone sparse ranking function, and $M_2$ is AST.

*More challenging examples:* All of the following examples also have antitone ranking functions, provided in Appendix C.

**Example V.10** (Continuous random walk). In PPCF we can construct a function whose argument changes by a random amount at each recursive call: $\Theta\, \underline{10}$ where

$$\Theta := \mathsf{Y}\, \lambda fx.\, \mathsf{if}(x \leq 0, \underline{0}, f(x - \mathsf{sample}))$$

For a more complex (not Y-PAST) example, consider the following continuous random walk: $\Xi\, \underline{10}$ where

$$\Xi := \mathsf{Y}\, \lambda fx.\, \mathsf{if}(x \leq 0, \underline{0}, f(x - \mathsf{sample} + 1/2))$$

**Example V.11** (Fair-in-the-limit random walk). [25, §5.3]

$$\bigl(\mathsf{Y}\, \lambda fx.\, \mathsf{if}(x \leq 0, \underline{0}, f(x-1) \oplus_{\frac{x}{2x+1}} f(x+1))\bigr) \underline{10}$$

**Example V.12** (Non-affine recursion).

$$(\mathsf{Y}\lambda fx.x \oplus_{2/3} f(f(x+1)))\, \underline{1}$$

**Example V.13** (More complex non-affine recursion).

$(\mathsf{Y}\, \lambda fx.\, (\lambda e.X[e])\, \mathsf{sample})\, \underline{0}$ where
$X[e] := \mathsf{if}(e \leq p - x^{-2}, x, f^2(x+1) \oplus_e f(x+1))$

**Example V.14** (Higher-order recursion). Consider the higher-order function $\Xi : (\mathsf{R} \to \mathsf{R} \to \mathsf{R}) \to \mathsf{R} \to \mathsf{R} \to \mathsf{R}$ recursively defined[5] by

$\Xi := \mathsf{Y}\lambda \varphi\, f^{\mathsf{R} \to \mathsf{R} \to \mathsf{R}}\, s^{\mathsf{R}}\, n^{\mathsf{R}}$.
 $\quad \mathsf{if}(n \leq 0, s, f\, n\, (\varphi\, f\, s\, (n-1)) \oplus_p f\, n\, (\varphi\, f\, s\, (n+1)))$

For any $F : \mathsf{R} \to \mathsf{R} \to \mathsf{R}$ such that $F\, \underline{n}\, \underline{m}$ terminates with no Y reductions for all $m$ and $n$, $\Xi\, F\, \underline{x}\, \underline{y}$ is antitone rankable.

---
[4]Technically this is not quite correct, because of the distinction between $\underline{n} + 1$ and $\underline{n+1}$, but this ranking function can still be made to work by appealing to the Confluent Ranking Thm. VI.15, as can the others in this section with similar issues. See Rem. VI.18 for a fuller account.

[5]Inspired by examples in [37, 38].

*Completeness:* It seems very likely that this method is complete in the case that every reachable term is AST (Conj. VIII.1), but we have been unable to actually prove this. It is certainly at least capable of proving the termination of terms which terminate arbitrarily slowly (in the sense that there is no similar limitation to Thm. III.10, which can only prove termination of Y-PAST terms).

The following theorem does not prove completeness, but is suggestive in that direction:

**Theorem V.15.** *For any stopping time $T$ which is almost surely finite, if $(\mathcal{F}_n)_n$ is the coarsest filtration to which $T$ is adapted, then there is a supermartingale $(Y_n)_n$ adapted to $(\mathcal{F}_n)_n$ and an antitone function $\epsilon$ such that $(Y_n)_n$ is an antitone ranking supermartingale with respect to $T$ and $\epsilon$.*

## VI. Confluent trace semantics

When proving almost sure termination in this way, it is necessary to consider all of the terms a given term may reduce to, and organise them into a manageable number of cases to assign ranking function values. Sometimes however, the reduction that the programmer has in mind may not be strictly the call-by-value order defined so far; or the number of cases necessary for defining a sparse ranking function is excessive because the reduction does not proceed in a neat and orderly manner. In cases such as these (e.g. Ex. VI.17), it may be much more convenient to instead consider ranking functions with respect to alternative reduction strategies constructed to match the individual term in question, thereby reducing the number of cases needed and simplifying the reachable terms.

In order to be able to use alternative reduction strategies to prove properties of the program defined using the usual reduction strategy, we first prove the relation between these different strategies, providing a confluent version of the semantics. It is therefore necessary to have some way of relating the samples that will be used in runs of the program using different reduction strategies. Positions are defined as a way of addressing sample statements within a program independently of the reduction order. Next, a version of the reduction relation is presented that is nondeterministic, so that it allows a choice of what order to perform reductions in. The notion of positions is extended to potential positions, for samples which may appear later in the reduction sequence but not necessarily in the initial term. In order to allow potential positions in different reduction sequences to be considered equivalent, a relation $\sim^*$ is defined, and finally, all of these are used to construct a confluent version of the trace semantics, $\Rightarrow$, that is still nondeterministic in reduction order, but does specify the outcome of random choices. The desired properties of this semantics (confluence and equivalence to the standard semantics) are proved (Lem. VI.9 and Thm. VI.10), then it is used to give extended versions of ranking functions and the related AST results, Thm. VI.15.

Non-probabilistic lambda calculi are generally confluent, i.e. if a term $A$ reduces to both $B_1$ and $B_2$, there is some $C$ to which both $B_1$ and $B_2$ reduce, so the reduction order mostly doesn't matter. In the probabilistic case, this may not be true, because $\beta$-reduction can duplicate samples, so the outputs of the copies of the sample may be identical or independent, depending on whether the sample is taken before or after $\beta$-reduction. There are, however, some restricted variations on the reduction order that do not have this problem.

Even with such a restriction, a trace semantics in the style of the one already defined would not be entirely confluent, as, for example, $\text{red}^3(\text{sample} - \text{sample}, (1, 0, \dots))$ would be either $1$ or $-1$ depending on the order of evaluation of the samples, as that determines which sample from the pre-selected sequence is used for each one. To fix this, rather than pre-selecting samples according to the order they'll be drawn in, select them according to the position in the term where they'll be used instead.

A *position* is a finite sequence of steps into a term, defined inductively as

$$\alpha ::= \cdot \mid \lambda; \alpha \mid @_1; \alpha \mid @_2; \alpha \mid \underline{f}_i; \alpha$$
$$\mid \mathsf{Y}; \alpha \mid \mathsf{if}_1; \alpha \mid \mathsf{if}_2; \alpha \mid \mathsf{if}_3; \alpha.$$

The *subterm of $M$ at $\alpha$*, denoted $M \mid \alpha$, is defined as

$$M \mid \cdot = M$$
$$\lambda x.M \mid \lambda; \alpha = M \mid \alpha$$
$$M_1\, M_2 \mid @_i; \alpha = M_i \mid \alpha \quad \text{for } i = 1, 2$$
$$\underline{f}(M_1, \dots, M_n) \mid \underline{f}_i; \alpha = M_i \mid \alpha \quad \text{for } i \le n$$
$$\mathsf{Y}M \mid \mathsf{Y}; \alpha = M \mid \alpha$$
$$\mathsf{if}(M_1 < 0, M_2, M_3) \mid \mathsf{if}_i; \alpha = M_i \mid \alpha \quad \text{for } i = 1, 2, 3$$

so that every subterm is located at a unique position, but not every position corresponds to a subterm (e.g. $x\,y \mid \lambda$ is undefined). A position such that $M \mid \alpha$ does exist is said to *occur* in $M$. *Substitution* of $N$ at position $\alpha$ in $M$, written $M[N/\alpha]$, is defined similarly. For example, let $M = \lambda x\, y.y\,(\mathsf{if}(x < 0, y\,(\underline{f}(x)), \underline{3}))$ and $\alpha = \lambda; \lambda; @_2; \mathsf{if}_2; @_2$ then $M[\mathsf{sample}/\alpha] = \lambda x\, y.y\,(\mathsf{if}(x < 0, y\,\mathsf{sample}, \underline{3}))$.

Two subterms $N_1$ and $N_2$ of a term $M$, corresponding to positions $\alpha_1$ and $\alpha_2$, can overlap in a few different ways. If $\alpha_1$ is a prefix of $\alpha_2$ (written as $\alpha_1 \le \alpha_2$), then $N_2$ is also a subterm of $N_1$. If neither $\alpha_1 \le \alpha_2$ nor $\alpha_1 \ge \alpha_2$, the positions are said to be *disjoint*. The notion of disjointness is mostly relevant in that if $\alpha_1$ and $\alpha_2$ are disjoint, performing a substitution at $\alpha_1$ will leave the subterm at $\alpha_2$ unaffected.

Thus we can define a more general reduction relation[6] $\to$.

**Definition VI.1.** The binary relation $\to$ is defined by the following rules, each is conditional on a redex occurring at

---

[6] In this section $\to$ always means the relation defined in Def. VI.1.

position $\alpha$ in the term $M$:

$$\text{if } M \mid \alpha = (\lambda x.N)V,\ M \to M[N[V/x]/\alpha]$$
$$\text{if } M \mid \alpha = \underline{f}(\underline{r_1},\ldots,\underline{r_n}),\ M \to M[\underline{f(r_1,\ldots,r_n)}/\alpha]$$
$$\text{if } M \mid \alpha = \mathsf{Y}\lambda x.N,\ M \to M[\lambda z.N[(\mathsf{Y}\lambda x.N)/x]z/\alpha]$$
$$\text{where } z \text{ is not free in } N$$

if $M \mid \alpha = \mathsf{if}(\underline{r} < 0, N_1, N_2),\ M \to M[N_1/\alpha]$ where $r < 0$
if $M \mid \alpha = \mathsf{if}(\underline{r} < 0, N_1, N_2),\ M \to M[N_2/\alpha]$ where $r \geq 0$
if $M \mid \alpha = \mathsf{sample}$ and $\lambda$ does not occur after $@_2$ or $\mathsf{Y}$ in $\alpha$,
$$M \to M[\underline{r}/\alpha] \text{ where } r \in [0,1].$$

In each of these cases, $M \mid \alpha$ is the *redex*, and the reduction *takes place at* $\alpha$. Each subterm can be a redex in at most one way, but there can be multiple redexes at different positions.

The argument of a $\beta$ redex and the body of a $\mathsf{Y}$ redex may be duplicated by those reductions. It is therefore these cases that need to be handled carefully to avoid duplicating samples at the wrong time. In both cases, the potentially duplicated part must already be a value, which excludes terms like sample or sample$+\underline{1}$, which should be evaluated before being duplicated. In the other direction, if a sample occurs inside of a $\lambda$, it may need to be duplicated before being evaluated, which is why a sample reduction isn't allowed inside a $\lambda$ inside a $\mathsf{Y}$ or the right side of an application. These restrictions are in some cases unnecessarily strict, for example, in $(\lambda x.x)((\lambda y.\mathsf{sample})\underline{0})$, it would be fine to evaluate the sample first, but they are at least sufficient to ensure confluence in terms of the distribution of results. Getting individual traces to behave correctly will take more work though.

Labelling the pre-chosen samples by the positions in the term by using $I^{\{\alpha \mid (M\mid \alpha) = \mathsf{sample}\}}$ as the trace space would not be sufficient to solve the issue of different samples being used in corresponding locations in different reduction sequences because in some cases, a sample will be duplicated before being reduced, for example, in $(\lambda x.x\underline{0} \pm x\underline{0})(\lambda y.\mathsf{sample})$, both of the sample redexes that eventually occur originate at $@_2;\lambda$. It is therefore necessary to consider possible positions that may occur in other terms reachable from the original term. Even this is itself inadequate because some of the positions in different reachable terms need to be considered the same, and the number of reachable terms is in general uncountable, which leads to measure-theoretic issues.

*Reduction relation on skeletons:* We are thus led to consider the reduction relation on skeletons (and positions in a skeleton), which can be extended from the definitions on terms in the obvious way, with $\mathsf{if}(\mathsf{X} < 0, A, B)$ reducing nondeterministically to both $A$ and $B$, sample reducing to $\mathsf{X}$, and $\mathsf{X}$ considered as (the skeletal equivalent of) a value, so that $(\lambda x.A)\mathsf{X}$ reduces to $A[\mathsf{X}/x]$. For example, we have $(\lambda x.\mathsf{if}(x < 0, x, \mathsf{X}))\mathsf{sample} \to (\lambda x.\mathsf{if}(x < 0, x, \mathsf{X}))\mathsf{X} \to \mathsf{if}(\mathsf{X} < 0, \mathsf{X}, \mathsf{X}) \to \mathsf{X}$.

Given a closed term $M$, let $L_0(M)$ be the set of pairs, the first element of which is a $\to$-reduction sequence of skeletons starting at $Sk(M)$, and the second of which is a position in the final skeleton of the reduction sequence. As with the traces from $I^\mathbb{N}$ used to pre-select samples to use in the standard trace semantics, modified traces, which are elements of $I^{L_0(M)}$ (with one more caveat introduced after Def. VI.5), will be used to pre-select a sample from $I$ for each element of $L_0(M)$, which will then be used if a sample reduction is ever performed at that position.

A (skeletal) reduction sequence is assumed to contain the information on the locations of all of the redexes as well as the actual sequence of skeletons that occurs. For example, $(\lambda x.x)((\lambda x.x)\mathsf{X})$ could reduce to $(\lambda x.x)\mathsf{X}$ with the redex at either $\cdot$ or $@_2$, and these give different reduction sequences.

**Example VI.2.** Consider the terms

$$A[M] = \mathsf{if}(\mathsf{if}(M > 0, I, I)(\lambda y.\mathsf{sample})\,\underline{0} - \underline{0.5} > 0, \underline{0}, \Omega)$$
$$B = \mathsf{if}(\mathsf{sample} - \underline{0.5} > 0, \underline{0}, \Omega)$$

If terms rather than skeletons were used to label samples, the set of modified traces where $A[\mathsf{sample}]$ terminates would be

$$\bigcup_{r\in[0,1]} \{s \mid s(A[\mathsf{sample}], \mathsf{if}_1;\underline{\phantom{-}}_1;@_1;@_1;\mathsf{if}_1) = r,$$
$$s(A[\mathsf{sample}] \to A[\underline{r}] \to^* B, \mathsf{if}_1;\underline{\phantom{-}}_1) > 0.5\}.$$

This is a rather unwieldy expression, but the crucial part is that $r$ occurs twice in the conditions on $s$: once as the value a sample must take, and once in the location of a sample. As this set is unmeasurable, the termination probability would not even be well-defined. Labelling samples by skeletons instead, this problem does not occur because there are only countably many skeleton, and at each step in a reduction sequence, only finitely many could have occurred yet. Although skeletal reduction sequences omit the information on what the results of sampling were, they still contain all the necessary information on how many, and which, reductions took place.

For this particular term, $Sk(A[\underline{r}])$ does not depend on the value of $r$, therefore the set where it terminates becomes simply the following, which is measurable.

$$\{s \mid s(Sk(A[\mathsf{sample}]) \to Sk(A[\underline{0}]) \to^* Sk(B), \mathsf{if}_1;\underline{\phantom{-}}_1) > 0.5\}$$

Reduction sequences are used rather than reachable skeletons because if the same skeleton is reached twice, different samples may be needed.

**Example VI.3.** Consider the term $M = \mathsf{Y}(\lambda fx.\mathsf{if}(\mathsf{sample} - \underline{0.5} < 0, fx, x))\,\underline{0}$, which reduces after a few steps to $N = \mathsf{if}(\mathsf{sample} - \underline{0.5} < 0, M, \underline{0})$. If we label samples by just skeletons and positions, and the pre-selected sample for $(Sk(N), \mathsf{if}_1;\underline{\phantom{-}}_1)$ is less than $0.5$, $N$ reduces back to $M$, then $N$ again, then the same sample is used the next time, therefore it's an infinite loop, whereas if samples are labelled by reduction sequences, the samples for $M \to^* N$ are independent from the samples for $M \to^* N \to M \to^* N$, and so on.

The reduction sequences of skeletons will often be discussed as though they were just skeletons, identifying them with their final skeletons. With this abuse of notation, a reduction sequence $N$ (actually $N_1 \to^* N_n = N$) may be said to reduce to a reduction sequence $O$, where the reduction sequence implicitly associated with the final skeleton $O$ is $N_1 \to^* N_n \to O$.

This is still not quite sufficient to attain confluence because sometimes the same samples must be used at corresponding positions in different reduction sequences.

**Example VI.4.** The term $M = \mathsf{sample} + \mathsf{sample}$ has the reachable skeletons $N_1 = \mathsf{X} + \mathsf{sample}$, $N_2 = \mathsf{sample} + \mathsf{X}$, $O = \mathsf{X} + \mathsf{X}$ and $\mathsf{X}$, with reductions $M \to N_1 \to O \to \mathsf{X}$ and $M \to N_2 \to O \to \mathsf{X}$. In the reduction $M \to N_1$, the sample labelled $(M, \underline{+}_1)$ is used, and in the reduction $N_2 \to O$, the sample labelled $(M \to N_2, \underline{+}_1)$ is used. Each of these samples becomes the value of the first numeral in $O$ in their respective reduction sequences, therefore in order for confluence to be attained, they must be the same. Which elements of $L_0(M)$ must match can be described by the relation $\sim^*$:

**Definition VI.5.** The relation $\sim$ is defined as the union of the minimal symmetric relations $\sim_p$ ("p" for parent-child) and $\sim_c$ ("c" for cousin) satisfying

(i) If $N$ reduces to $O$ with the redex at position $\alpha$, and $\beta$ is a position in $N$ disjoint from $\alpha$, then $(N, \beta) \sim_p (O, \beta)$.

(ii) If $N$ $\beta$-reduces to $O$ at position $\alpha$, $\beta$ is a position in $N \mid \alpha; @_1; \lambda$ and $N \mid \alpha; @_1; \lambda; \beta$ is not the variable involved in the reduction, $(N, \alpha; @_1; \lambda; \beta) \sim_p (O, \alpha; \beta)$

(iii) If $N$ if-reduces to $O$ at position $\alpha$, with the first resp. second branch being taken, and $\alpha; \mathsf{if}_i; \beta$ occurs in $N$ (where $i = 2$ resp. 3), $(N, \alpha; \mathsf{if}_i; \beta) \sim_p (O, \alpha; \beta)$

(iv) If $N$, $O_1$ and $O_2$ match any of the following cases:

a) $N$ contains redexes at disjoint positions $\alpha_1$ and $\alpha_2$, $O_1$ is $N$ reduced first at $\alpha_1$ then $\alpha_2$ and $O_2$ is $N$ reduced first at $\alpha_2$ then at $\alpha_1$.
b) $N \mid \alpha = \mathsf{if}(\underline{r} < 0, N_1, N_2)$, where $r < 0$ (or, respectively, $r \geq 0$), $(N_2$ resp. $N_1) \mid \beta$ is a redex, and $O_1$ is $N$ reduced at $\alpha$ and $O_2$ is $N$ reduced first at $\alpha; (\mathsf{if}_3$ resp. $\mathsf{if}_2); \beta$ then at $\alpha$
c) $N \mid \alpha = \mathsf{if}(\underline{r} < 0, N_1, N_2)$, where $r < 0$ (or, respectively, $r \geq 0$), $(N_1$ resp. $N_2) \mid \beta$ is a redex, and $O_1$ is $N$ reduced first at $\alpha$ then at $\alpha; \beta$ and $O_2$ is $N$ reduced first at $\alpha; (\mathsf{if}_2$ resp. $\mathsf{if}_3); \beta$ then at $\alpha$
d) $N \mid \alpha = (\lambda x.A)B$, there is a redex in $A$ at position $\beta$, $O_1$ is $N$ reduced first at $\alpha$ then at $\alpha; \beta$, and $O_2$ is $N$ reduced first at $\alpha; @_1; \lambda; \beta$ then at $\alpha$
e) $N \mid \alpha = (\lambda x.A)B$, $B \mid \beta$ is a redex, $(\gamma_i)_i$ is a list of all the positions in $A$ where $A \mid \gamma = x$, ordered from left to right, $O_1$ is $N$ reduced first at $\alpha; @_2; \beta$ then at $\alpha$, and $O_2$ is $N$ reduced first at $\alpha$ then at $\alpha; \gamma_i; \beta$ for each $i$ in order.
f) $N \mid \alpha = \mathsf{Y}(\lambda x.A)$, $A$ reduced at $\beta$ is $A'$, $(\gamma_i)_i$ is a list of all the positions where $A' \mid \gamma = x$, ordered from left to right, $O_1$ is $N$ reduced first at $\alpha; \mathsf{Y}; \lambda; \beta$ then at $\alpha$, and $O_2$ is $N$ reduced first at $\alpha$ then at $\alpha; \lambda; @_1; \gamma_i; \mathsf{Y}; \lambda; \beta$ for each $i$ in order where $\gamma_i$ is left of $\beta$ then at $\alpha; \lambda; @_1; \beta$ then at $\alpha; \lambda; @_1; \gamma_i; \mathsf{Y}; \lambda; \beta$ for the remaining values of $i$.

(in which case $O_1$ and $O_2$ are equal as skeletons, but with different reduction sequences), $O'_1$ and $O'_2$ are the results of applying some reduction sequence to each of $O_1$ and $O_2$ (the same reductions in each case, which is always possible because they're equal skeletons), and $\delta$ is a position in $O'_1$ (or equivalently $O'_2$), then $(O'_1, \delta) \sim_c (O'_2, \delta)$.

The $\sim_c$-rules are illustrated in Fig. 3.

**Example VI.6.** In Ex. VI.4, $(M, \underline{+}_1) \sim_p (M \to N_2, \underline{+}_1)$ by case i of $\sim_p$ (because the reduction $M \to N_2$ occurs at $\underline{+}_2$, which is disjoint from $\underline{+}_1$), and similarly, $(M, \underline{+}_2) \sim_p (M \to N_1, \underline{+}_2)$.

If we extend it to have three samples, $\sim_c$ becomes necessary as well: Let $M_{sss} = \mathsf{sample} + \mathsf{sample} + \mathsf{sample}$ (taking the three-way addition to be a single primitive function), $M_{Xss} = \mathsf{X} + \mathsf{sample} + \mathsf{sample}$, and so on. There are then reduction sequences $M_{sss} \to M_{Xss} \to M_{XXs} \to M_{XXX} \to \mathsf{X}$ and $M_{sss} \to M_{sXs} \to M_{XXs} \to M_{XXX} \to \mathsf{X}$. For the first two reductions, these reduction sequences take the same samples by $\sim_p$, case i, as in Ex. VI.4. The next reduction uses the samples labelled by $(M_{sss} \to M_{Xss} \to M_{XXs}, \underline{+}_3)$ and $(M_{sss} \to M_{sXs} \to M_{XXs}, \underline{+}_3)$, which are related by $\sim_c$, case a, therefore when these reduction sequences reach $M_{XXX}$, they still contain all the same numbers, as desired.

The reflexive transitive closure $\sim^*$ of this relation is used to define the set of *potential positions* $L(M) = L_0(M)/\sim^*$, and each equivalence class can be considered as the same position as it may occur across multiple reachable skeletons. If $(N, \alpha) \sim^* (O, \beta)$, then $N \mid \alpha$ and $O \mid \beta$ both have the same shape (i.e. they're either both the placeholder $\mathsf{X}$, both variables, both applications, both samples etc.), therefore it's well-defined to talk of the set of *potential positions where there is a* sample, $L_s(M)$. The new sample space is then defined as $I^{L_s(M)}$, with the Borel $\sigma$-algebra and product measure. Since $I^{L_s(M)}$ is a countable product, the measure space is well-defined [36, Cor. 2.7.3].

Before defining the new version of the reduction relation red, the following lemma is necessary for it to be well-defined.

**Lemma VI.7.** *The relation $\sim$ is defined on $L_0(M)$ with reference to a particular starting term $M$, so different versions, $\sim_M$ and $\sim_N$, can be defined starting at different terms. If $M \to N$, then $\sim_N^*$ is equal to the restriction of $\sim_M^*$ to $L_0(N)$.*

At each reduction step $M \to N$, the sample space must be restricted from $I^{L_s(M)}$ to $I^{L_s(N)}$. The injection $L_0(N) \to L_0(M)$ is trivial to define by appending $Sk(M) \to Sk(N)$ to each path, and using Lem. VI.7, this induces a corresponding injection on the quotient, $L(N) \to L(M)$. The corresponding map $L_s(N) \to L_s(M)$ is then denoted $i(M \to N)$.

**Definition VI.8** ($\Rightarrow$ reduction). Unlike in the purely call-by-value case, the version of the reduction relation that takes into account samples is still a general relation rather than a function, so it is denoted "$\Rightarrow$" instead of "red", and it relates $\biguplus_{M \in \Lambda_0} I^{L_s(M)}$ to itself. We write an element of $\biguplus_{M \in \Lambda_0} I^{L_s(M)}$ as $(M', s)$ where the *term* $M' \in \Lambda^0$ and $s \in I^{L_s(M')}$.

$(M, s) \Rightarrow (N, s \circ i(M \to N))$ if $M \to N$ at $\alpha$ and either

the redex is not sample, or

$M \mid \alpha = \mathsf{sample}$ and $N = M[\underline{s(Sk(M), \alpha)}/\alpha]$

This reduction relation now has all of the properties required of it. In particular, it can be considered an extension of the standard trace semantics (as will be seen later in Thm. VI.10), and also:

**Lemma VI.9.** *The relation $\Rightarrow$ is confluent.*

*Reduction strategy determinising $\Rightarrow$:* The reduction relation $\Rightarrow$ is nondeterministic, so it admits multiple possible reduction strategies. A *reduction strategy* starting from a closed term $M$ is a measurable partial function $f$ from $Rch(M)$ to positions, such that for any reachable term $N$ where $f$ is defined, $f(N)$ is a position of a redex in $N$, and if $f(N)$ is not defined, $N$ is a value. Using a reduction strategy $f$, a subset of $\Rightarrow$ that isn't nondeterministic, $\Rightarrow_f$, can be defined by $(N, s) \Rightarrow_f (N', s')$ just if $(N, s) \Rightarrow (N', s')$ and $N$ reduces to $N'$ with the redex at $f(N)$.

The usual call-by-value semantics can be implemented as one of these reduction strategies, given by (with $V$ a value and $T$ a term that isn't a value and $M$ a general term)

$$\mathrm{cbv}(TM) = @_1; \mathrm{cbv}(T)$$
$$\mathrm{cbv}(VT) = @_2; \mathrm{cbv}(T)$$
$$\mathrm{cbv}(\underline{f}(V_1, \ldots, V_{k-1}, T, M_{k+1}, \ldots, M_n)) = \underline{f}_k; \mathrm{cbv}(T)$$
$$\mathrm{cbv}(\mathsf{Y}T) = \mathsf{Y}; \mathrm{cbv}(T)$$
$$\mathrm{cbv}(\mathsf{if}(T < 0, M_1, M_2)) = \mathsf{if}_1; \mathrm{cbv}(T)$$
$$\mathrm{cbv}(V) \text{ is undefined}$$
$$\mathrm{cbv}(T) = \cdot \text{ otherwise}$$

(this last case covers redexes at the root position).

A closed term $M$ terminates with a given reduction strategy $f$ and samples $s$ if there is some natural number $n$ such that $(M, s) \Rightarrow_f^n (N, s')$ where $f$ gives no reduction at $N$. The term *terminates almost surely w.r.t. $f$* if it terminates with $f$ for almost all $s$.

With these definitions, it is now possible to relate the confluent trace semantics back to the standard trace semantics.

**Theorem VI.10.** *A closed term $M$ is AST with respect to $\mathrm{cbv}$ iff it is AST.*

**Theorem VI.11.** *If $M$ terminates with some reduction strategy $f$ and trace $s$, it terminates with $\mathrm{cbv}$ and $s$.*

**Corollary VI.12** (Reduction strategy independence)**.** *If $M$ is AST with respect to any reduction strategy, it is AST.*

*Proof.* Suppose $M$ is AST w.r.t. $f$. Let the set of samples with which it terminates with this reduction strategy be $X$. By Thm. VI.11, $M$ also terminates with $\mathrm{cbv}$ and every element of $X$, and $X$ has measure 1, by assumption, therefore $M$ is AST with respect to $\mathrm{cbv}$ therefore by Thm. VI.10 it is AST. $\square$

All of the theorems on the termination of rankable terms therefore extend to other reduction strategies too. The proofs of Thm. III.10, IV.5, V.6 and V.8 are all sufficiently generic with respect to what the reduction relation actually is that they can be directly applied to other reduction strategies. They only require that the number of reductions that can occur without any of them being a Y-reduction is bounded for any starting term (which is true, because Thm. A.2 applies equally to any reduction strategy). For a reduction strategy $r$ and term $M$, just substitute $N[r(N)]$ being a Y-redex for $N$ being of the form $E[\mathsf{Y}\lambda x.O]$, and $r(N)$ being undefined for $N$ being a value.

The domain of definition of the ranking functions also needs to be changed from the reachable terms $Rch(M)$ to the *reachable terms with respect to the reduction strategy $r$*,

$$Rch_r(M) := \{N \mid \exists n, (N_i)_{0 \leq i \leq n} : \\ N_0 = M, N_n = N, N_i \to N_{i+1} \text{ at } r(N_i)\}.$$

More explicitly, the modified forms of the theorems are:

**Definition VI.13.** An *antitone ranking function on $M$ w.r.t. a reduction strategy $r$* is a measurable function $f : Rch_r(M) \to \mathbb{R}$ such that $f(N) \geq 0$ for all $N$, and there exists an antitone function $\epsilon : \mathbb{R}_{\geq 0} \to \mathbb{R}_{>0}$ such that

- $f(N) \geq \epsilon(f(N)) + f(N')$ if $N[r(N)]$ is a Y-redex and $N$ reduces to $N'$ at $r(N)$
- $f(N) \geq \int_I f(N[\underline{x}/r(N)]) \mathrm{Leb}(\mathrm{d}x)$ if $N[r(N)] = \mathsf{sample}$
- $f(N) \geq f(N')$ if $r(N)$ is any other redex, where $N \to N'$ at $r(N)$.

Any closed term for which an antitone ranking function with respect to a reduction strategy $r$ exists is called *antitone rankable w.r.t. $r$*.

**Definition VI.14.** An *antitone sparse ranking function on $M$ w.r.t. a reduction strategy $r$* is a partial function $f : Rch_r(M) \rightharpoonup \mathbb{R}$ such that there exists some antitone function $\epsilon : \mathbb{R}_{\geq 0} \to \mathbb{R}_{>0}$ such that (i) $f(N) \geq 0$ for all $N \in \mathrm{dom}(f)$, (ii) $M \in \mathrm{dom}(f)$, (iii) for any $N \in \mathrm{dom}(f)$, evaluation of $N$ at the positions specified by $r$ will eventually reach some $O$ such that either $r(O)$ isn't defined or $f(O)$ is, and $f(N) \geq \mathbb{E}[f(O) + \epsilon(f(O)) \times$ the number of Y-reduction steps from $N$ to $O]$ (where $f(O)$ is taken to be 0 if $O$ is a value outside of $\mathrm{dom}(f)$).

If the antitone function is just the constant function 1 instead, then it is called a *(sparse) ranking function on $M$ w.r.t. $r$* instead, and the term is called *rankable w.r.t. $r$*.

**Theorem VI.15** (Confluent ranking). *If a closed PPCF term $M$ has a ranking function, sparse ranking function, antitone ranking function or antitone sparse ranking function w.r.t. a reduction strategy $r$, then $M$ is AST w.r.t. $r$, and AST.*

When constructing an (antitone) sparse ranking function with respect to an alternative reduction strategy, it is intuitively simplest to define the sparse ranking function and the reduction strategy together. For each key reachable term where the sparse ranking function is defined, simply decide which sub-term should be reduced next, and how far before the next key reachable term. The only restrictions to bear in mind on evaluating sub-terms are that a sample may not be evaluated inside a $\lambda$ that's inside a Y or on the right of an application, and a $\beta$ redex may not be reduced until its argument is a value. This can be seen as applied to Ex. VI.17 at the end of Sec. D.

If, as seems likely, Thm. V.6 is complete for terms from which every reachable term is AST, then by the following theorem, Thm. VI.15 can't actually prove the termination of any terms not already provable by Thm. V.6 and Thm. V.8.

**Theorem VI.16.** *For any closed term $M$ and reduction strategy $r$ on $M$, if every term in $Rch_r(M)$ is AST, then every term in $Rch(M)$ is AST.*

*Custom reduction strategy for AST analysis:* However, there are some terms which terminate more simply and directly via some alternative reduction strategy. For these examples, a custom reduction strategy is specified for each term under consideration.

**Example VI.17** (Sum of powers). The following term picks a random natural number $n$, then computes $\sum_{k=0}^{n} k^n$.

$$M = (\lambda n.P[n])\lfloor \mathsf{sample}^{-1/2} \rfloor$$
$$P[n] = (\lambda p.\Xi[p]\,n)(\lambda x.\Theta[x]\,n))$$
$$\Xi[p] = \mathsf{Y}\lambda fn.\mathsf{if}(n=0,\underline{0},p\,n+f(n-1))$$
$$\Theta[x] = \mathsf{Y}\lambda fn.\mathsf{if}(n=0,\underline{1},x\times f(n-1)).$$

With the CBV reduction strategy, the recursion in $\Theta[x]$ (that defines the function $k \mapsto k^x$) is executed separately for every term in the sum in $\Xi$. Not only is this much more complicated than simplifying $\Theta[x]\,\underline{n}$ first, it requires considerably more Y-reduction steps, so many so that $M$ is Y-PAST with respect to a reduction strategy that simplifies $\Theta[x]\,\underline{n}$ first (call it "$r$"), but isn't Y-PAST with respect to the standard reduction strategy, so that $M$ is rankable with respect to $r$, but isn't rankable. It is still antitone rankable, but this is much more complicated. Details of the ranking function construction with and without using the confluent semantics are given in Sec. D.

*Remark* VI.18. Even some of the examples given earlier implicitly use slightly non-standard reduction strategies to make the definitions of their ranking functions slightly simpler. For example, the antitone sparse ranking function in Ex. V.9 was defined at $\Theta_2\,\underline{n}$ (where $\Theta_2 = \mathsf{Y}\,\lambda fn.\mathsf{if}(n=0,0,f\,(n-1)\oplus_{1/2} f\,(n+1)))$, but $\Theta_2\,\underline{10}$ actually reduces to $\Theta_2\,(\underline{10}-1)$ or $\Theta_2\,(\underline{10}+1)$, and (taking the second of these), it then reduces to $(\lambda z.(\lambda n.\mathsf{if}(n=0,0,\Theta_2\,(n-1)\oplus_{1/2}\Theta_2\,(n+1)))z)\,(\underline{10}+1)$, then $(\lambda z.(\lambda n.\mathsf{if}(n=0,0,\Theta_2\,(n-1)\oplus_{1/2}\Theta_2\,(n+1)))z)\,\underline{11}$, bypassing $\Theta_2\,\underline{11}$ entirely. The antitone sparse ranking function in the form given can be justified anyway, by considering it to be using a reduction strategy which is the same as cbv except that in $\Theta_2\,(f\,\underline{n})$, it evaluates the redex at $@_2$ first.

Ex. V.14 also makes use of an alternative reduction strategy. In this case, it is to allow the antitone sparse ranking function to skip over all of those terms where the unknown function $F$ is part way through evaluation.

## VII. APPLICATIONS

Although the definition of potential positions in a term in Sec. VI was intended simply to define a variant of trace semantics that has a restricted version of confluence, it is likely to also be applicable to inference algorithms (for probabilistic programs) as well.

In Lightweight Metropolis-Hastings (LMH) or Single-site MH (originally suggested by [39]; see also [3]), the program is executed, with a random value being chosen each time a sample redex is evaluated. The trace of samples used is recorded, then for the next step, it is modified slightly before the program is run again. This time, when sample redexes are evaluated, they may be taken from the pre-determined trace instead of chosen randomly. The change to the trace may be accepted or rejected, depending on the probability weight associated with each execution of the program (based on a conditioning language construct), and after this process is repeated sufficiently many times, the distribution of results of the program tends towards the true posterior distribution.

It is essential to the efficiency of this algorithm that after modifying the trace, the weight of the resultant program execution is likely to be similar to the weight of the previous execution. This is satisfied by the values of the samples being similar to their original values, which (for a sufficiently well-behaved program) will tend to make the entire execution proceed similarly. In the simplest version of this algorithm, the samples in the trace are used one at a time, in the order they are encountered. However, if as a result of some random choice, more or fewer samples are taken during some stage of the program, subsequent samples in the trace will be displaced, and end up being used for different purposes than previously, thereby decreasing the correlation between the program runs.

It is possible to reduce this problem by labelling the samples in the trace by something other than the order in which they are used. If the labels of samples in the trace correspond closely to the roles that the samples play in the program execution, this will increase the efficiency of the inference algorithm. In [39], for example, something rather like a stack trace is used, although this addressing scheme turns out to have various problems [40, 41]. Potential positions could also be used for a similar purpose, although in practice, a more compact representation than an entire skeletal reduction sequence would be necessary. Selecting an appropriate addressing scheme is also important for variational inference [42].

The $\sim$ relation as it has been defined is not exactly what would be necessary for inference (for example, $\sim_c$ only becomes relevant when there are multiple different reduction strategies being considered), but something much like it is still likely to be useful.

Another possible application for the confluent trace semantics would be as a justification, within a trace-style semantics, of some forms of equational reasoning. By introducing reduction steps that go backwards to terms not reachable from the original term, it may in some cases be simplified, or equivalences between terms may be proved, by appealing to confluence results such as Lem. VI.9 and Thm. VI.11.

## VIII. Related work and further directions

Most methods for proving AST of programs in the literature do not support continuous distributions [16, 18–20, 43, 44]. The few that do [21–23] are for first-order imperative programs.

A fortiori, our results specialise to a method for proving termination of non-random lambda calculus. Thus restricted, our approach is closest in spirit to Jones' work on size-change termination [45, 46]. Dal Lago and Grellois [43] extend this to programs with discrete distributions, but their method is limited to affine recursion of first-order functions. We think that the supermartingales of our paper are a good basis for giving a unifying account. Breuvart and Dal Lago [47] develop systems of intersection types from which the termination probability of programs (with discrete distributions) can be inferred from (infinitely many) type derivations.

Kobayashi et al. [44] show that the termination probability of order-$n$ probabilistic recursion schemes ($n$-PHORS) can be obtained as the least fixpoint of suitable order-$(n-1)$ fixpoint equations, which are solvable by Kleene fixpoint iteration. By contrast, our approach works for programs with continuous distributions. Note that $n$-PHORS is strictly less expressive then order-$n$ call-by-name PPCF: for example, the PPCF terms in Ex. V.10, V.13 and V.14 are not definable as PHORS.

The main result by McIver et al. [25] is similar to our Thm. V.6, but in an imperative language. While we require that the antitone ranking function decreases for every Y-reduction step, they require that it decreases for every iteration of a certain while loop (the ranking function in that case is defined in the context of a particular loop), which is similar to Y but limited to tail recursion. The difference in the exact progress condition is insignificant. A ranking function satisfying either progress condition can easily be converted to satisfy the other.

The confluent semantics, and the results based on it, is new in the setting of a functional language, because an imperative language does not have any equivalent of other redexes, or a similar nondeterministic structure. The sparse ranking function construction is also more useful in a functional language (although it would be possible to give an equivalent in an imperative language), because in an imperative language where the order of execution is more rigid, the iterations of the while loop provide a natural set of checkpoints where it is reasonable to give values of the ranking function.

*Further directions:* We have not yet been able to prove Conj. VIII.1 which would imply that Thm. V.6 is almost complete.

**Conjecture VIII.1** (Completeness). *If every term reachable from a certain PPCF term is AST, then that term is antitone rankable.*

The Def. VI.1 of redexes and which positions are acceptable to reduce at is sufficiently restrictive to guarantee confluence, but is also a little more restrictive in some cases than is necessary for this purpose. For example, if the argument of a function is not yet a value, but its reduction to a value would be deterministic, or the function is affine, then applying the function before reducing its argument would not cause any problematic duplication of random samples. Similarly, if sample occurs at position $\alpha; @_2; \beta; \lambda; \gamma$, but the function at $\alpha; @_1$ is affine, evaluating the sample may in some cases also work fine. A more complete characterisation of which redexes can be reduced without breaking confluence could be interesting.

Although this is a broadly applicable method of proving almost sure termination of probabilistic functional programs, and Thm. V.8 and VI.15 make it more convenient to use, some method of automating the construction and checking of (antitone) ranking functions, even partially, could make it considerably more practically useful, especially in cases where almost sure termination is merely a side-condition for some other theorem or algorithm to be applicable.

*Conclusions:* We have presented the first application of martingales to probabilistic lambda calculus, and the first version of trace semantics that's capable of satisfying a restricted form of confluence. Though the construction of ranking functions is inherently difficult, using sparse ranking functions, antitone ranking functions and ranking functions w.r.t. an alternative reduction strategy, we have shown that a great variety of programs can be proved to be AST, including some that are beyond the reach of methods in the literature.

# Appendix A
## Supplementary Materials for Sec. III

**Lemma III.4.** *Let $(Y_n)_{n \geq 0}$ be a $\epsilon$-ranking supermartingale w.r.t. the stopping time $T$, then $T < \infty$ a.s., and $\mathbb{E}[T] \leq \frac{\mathbb{E}[Y_0]}{\epsilon}$.*

*Proof.* We first prove (by induction on $n$): for all $n \geq 0$
$$\mathbb{E}[Y_n] \leq \mathbb{E}[Y_0] - \epsilon \cdot \left( \sum_{i=0}^{n-1} \mathbb{P}[T > i] \right).$$
It follows that $\sum_{i=0}^{\infty} \mathbb{P}[T > i]$ converges; hence we have $\lim_{i \to \infty} \mathbb{P}[T > i] = 0$ and so $\mathbb{P}[T < \infty] = 1$. It then remains to observe: if $T < \infty$ a.s., then $\mathbb{E}[T] = \sum_{i=0}^{\infty} \mathbb{P}[T > i]$. □

**Theorem III.7** (Deriving supermartingales). *If a closed PPCF term $M$ is rankable (respectively, strictly rankable) by $f$ then $(f(M_n))_{n \geq 0}$ is a supermartingale (respectively, ranking supermartingale w.r.t. stopping time $T_M$) adapted to $(\mathcal{F}_n)_{n \geq 0}$.*

We shall obtain the theorem as a corollary of a technical lemma (Lem. A.1).

We say that a given PPCF term is: *type 1* if it has the shape $E[\mathsf{Y}\lambda x.N]$, *type 2* if it has the shape $E[\mathsf{sample}]$; *type 3* if it has the shape $E[R]$ where $R$ is any other redex, and *type 4* if it is a value. Henceforth fix an $n \geq 0$, and define $\mathbf{T}_i := \{s \mid M_n(s) \text{ is type } i\}$. It is straightforward to see that each $\mathbf{T}_i \in \mathcal{F}_n$, and $\{\mathbf{T}_1, \mathbf{T}_2, \mathbf{T}_3, \mathbf{T}_4\}$ is a partition of $\mathbb{S}$. Hence it suffices to prove the following lemma.

**Lemma A.1** (Technical). *For all $i \in \{1, 2, 3, 4\}$ and $A \in \mathcal{F}_n$*
$$\int_A \mu_\mathbb{S}(\mathrm{d}s)\, f(M_{n+1})[s \in \mathbf{T}_i] \leq \int_A \mu_\mathbb{S}(\mathrm{d}s)\, f(M_n)[s \in \mathbf{T}_i]\,[7]$$
*Hence $\mathbb{E}[f(M_{n+1}) \mid \mathcal{F}_n] \leq f(M_n)$ a.s.*

First, some notation.

Given an $\mathbb{N}$-indexed sequence (e.g. $s \in \mathbb{S}$) and $m \geq 0$, we write $s_{\leq m} \in I^m$ to mean the prefix of $s$ of length $m$. For $n \geq 0$, define
$$\#\mathrm{draw}_n(s) := |\{k < n \mid \exists E : M_k(s) = E[\mathsf{sample}]\}|$$
so that $\pi_1(\mathrm{red}^n(M, s)) = \pi_t(\cdots(\pi_t(s)))$, with $\pi_t$ applied $\#\mathrm{draw}_n(s)$ times. The $\mathcal{F}_n$-measurability of $M_n$ (and hence of $\#\mathrm{draw}_n$) follows from [31]. Take $s \in A \in \mathcal{F}_n$ with $\#\mathrm{draw}_n(s) = l$. For any $s' \in \mathbb{S}$, if $s_{\leq l} = s'_{\leq l}$ then $s' \in A$. It follows that $\{s_{\leq l}\} \cdot I^\mathbb{N} \subseteq A$.

*Proof.* We show the non-trivial case of $i = 2$. First we express
$$f(M_{n+1})[s \in \mathbf{T}_2] = \sum_{i \in \mathcal{I}} f(E_i[\sigma_i(s)][\underline{\rho(s)}])[s \in U_i] \quad (2)$$
where
- $\mathcal{I}$ is a countable indexing set
- $E_i[\cdot][\mathsf{sample}] \in \mathsf{Sk}_{j_i}$, and $\sigma_i : \mathbb{S} \to \mathbb{R}^{j_i}$, and $\rho(s) := \pi_h(\pi_1(\mathrm{red}^n(M, s))) = \pi_h(\pi_t^{l_i}(s)) \in \mathbb{R}$
- $\{U_i\}_{i \in \mathcal{I}}$ is a partition of $\mathbf{T}_2$, where each $U_i$ is determined by a skeletal environment $E_i$ and a number (of draws) $l_i \leq n$, so that $U_i$ is the set of traces where after $n$ reduction steps, $l_i$ samples have been used, and the term has reached $E_i[r][\mathsf{sample}]$ for some $r \in \mathbb{R}^{j_i}$. (We use the (measurable) function $\sigma_i : \mathbb{S} \to \mathbb{R}^{j_i}$ to skolemise the existentially quantified $r$.) Equivalently

$$U_i := \#\mathrm{draw}_n^{-1}[l_i] \cap M_n^{-1}[\{E_i[r][\mathsf{sample}] \mid r \in \mathbb{R}^{j_i}\}] \in \mathcal{F}_n$$

---

[7]where (the Iverson bracket) $[P] := 1$ if the statement $P$ hold, and 0 otherwise.

Observe that if $s \in U_i$ then $\{s_{\leq l_i}\} \cdot I^{\mathbb{N}} \subseteq U_i$; in fact $(U_i)_{\leq l_i} \cdot I^{\mathbb{N}} = U_i$. This means that for any measurable $g : \mathbb{S} \to \mathbb{R}_{\geq 0}$, if $g(s)$ only depends on the prefix of $s$ of length $(l_i + 1)$, then, writing $\widehat{g} : I^{l_i+1} \to \mathbb{R}_{\geq 0}$ where $g(s) = \widehat{g}(s_{\leq l_i+1})$, we have: for any $A \in \mathcal{F}_n$

$$\int_{A \cap U_i} \mu_{\mathbb{S}}(\mathrm{d}s)\, g(s) = \int_{(A \cap U_i)_{\leq l_i+1}} \mathrm{Leb}_{l_i+1}(\mathrm{d}t)\, \widehat{g}(t) \quad (3)$$

Take $s \in U_i$, and set $l = l_i$. Plainly $\sigma_i(s)$ depends on $s_{\leq l}$, and $\rho(s)$ depends on $s_{\leq l+1}$. Take $u \in (U_i)_{\leq l}$. It then follows from the definition of ranking function that

$$\int_I \mathrm{Leb}(\mathrm{d}r)\, f(E_i[\widehat{\sigma}_i(u)][\underline{r}])) \leq f(E_i[\widehat{\sigma}_i(u)][\mathsf{sample}])).$$

Take $A \in \mathcal{F}_n$, and integrating both sides, we get

$$\int_{(A \cap U_i)_{\leq l}} \mathrm{Leb}_l(\mathrm{d}u) \int_I \mathrm{Leb}(\mathrm{d}r)\, f(E_i[\widehat{\sigma}_i(u)][\underline{r}]))$$
$$\leq \int_{(A \cap U_i)_{\leq l}} \mathrm{Leb}_l(\mathrm{d}u)\, f(E_i[\widehat{\sigma}_i(u)][\mathsf{sample}])).$$

Since $\mathrm{Leb}_{l+1}$ is the (unique) product measure satisfying $\mathrm{Leb}_{l+1}(V \times B) = \mathrm{Leb}_l(V) \cdot \mathrm{Leb}(B)$, and $(U_i)_{\leq l_i} \cdot I^{\mathbb{N}} = U_i$, we have

$$\int_{(A \cap U_i)_{\leq l+1}} \mathrm{Leb}_{l+1}(\mathrm{d}u')\, f(E_i[\widehat{\sigma}_i(u')][\underline{\widehat{\rho}(u')}]))$$
$$\leq \int_{(A \cap U_i)_{\leq l}} \mathrm{Leb}_l(\mathrm{d}u')\, f(E_i[\widehat{\sigma}_i(u')][\mathsf{sample}]))$$

and so, by (3)

$$\int_{A \cap U_i} \mu_{\mathbb{S}}(\mathrm{d}s)\, f(E_i[\sigma_i(s)][\underline{\rho(s)}]))$$
$$\leq \int_{A \cap U_i} \mu_{\mathbb{S}}(\mathrm{d}s)\, f(E_i[\sigma_i(s)][\mathsf{sample}])). \quad (4)$$

Now, integrating both sides of (2), we have

$$\int_A \mu_{\mathbb{S}}(\mathrm{d}s)\, f(M_{n+1})[s \in \mathbf{T}_2]$$
$$= \int_A \mu_{\mathbb{S}}(\mathrm{d}s) \sum_{i \in \mathcal{I}} f(E_i[\sigma_i(s)][\underline{\rho(s)}])[s \in U_i]$$
$$= \sum_{i \in \mathcal{I}} \int_{A \cap U_i} \mu_{\mathbb{S}}(\mathrm{d}s)\, f(E_i[\sigma_i(s)][\underline{\rho(s)}])$$
$$\leq \sum_{i \in \mathcal{I}} \int_{A \cap U_i} \mu_{\mathbb{S}}(\mathrm{d}s)\, f(E_i[\sigma_i(s)][\mathsf{sample}]) \quad \because (4)$$
$$= \int_A \mu_{\mathbb{S}}(\mathrm{d}s) \sum_{i \in \mathcal{I}} f(E_i[\sigma_i(s)][\mathsf{sample}])[s \in U_i]$$
$$= \int_A \mu_{\mathbb{S}}(\mathrm{d}s) f_n(M)[s \in \mathbf{T}_2]$$

$\square$

As an immediate corollary of Lem. A.1, each $f(M_n)$ is integrable. This concludes the proof of Thm. III.7.

**Lemma III.8.** $(T_n)_{n \geq 0}$ *is an increasing sequence of stopping times adapted to* $(\mathcal{F}_n)_{n \geq 0}$, *and each* $T_i$ *is bounded.*

We first show that $T_1$ is bounded i.e. $T_1 \leq n$, for some $n \in \mathbb{N}$.

A first proof idea is to construct a reduction argument to the strong normalisation of the simply-typed lambda-calculus. There is a classical transform of conditionals into the pure lambda-calculus. However, each evaluation of $\mathsf{sample}$ (almost surely) returns a different number, which complicates such a transform.

Given a closed PPCF term $M$, we transform it to a term $\ulcorner M \urcorner$ of the nondeterministic simply-typed lambda-calculus as follows. (We may assume that the nondeterministic calculus is generated from a base type $\iota$ with a $\iota$-type constant symbol $r$, and a function symbol $\bot : A$ for each function type $A$.)

$$\ulcorner \mathsf{sample} \urcorner := r$$
$$\ulcorner y \urcorner := y$$
$$\ulcorner \lambda y.N \urcorner := \lambda y.\ulcorner N \urcorner$$
$$\ulcorner M_1\, M_2 \urcorner := \ulcorner M_1 \urcorner \ulcorner M_2 \urcorner$$
$$n \geq 0, \ulcorner f\, M_1 \cdots M_n \urcorner := (\lambda z_1 \cdots z_n.r)\ulcorner M_1 \urcorner \cdots \ulcorner M_n \urcorner$$
$$\ulcorner \mathsf{if}(B, M_1, M_2) \urcorner := \bigl(\lambda z.(\ulcorner M_1 \urcorner + \ulcorner M_2 \urcorner)\bigr)\ulcorner B \urcorner$$
$$\ulcorner \mathsf{Y}N \urcorner := (\lambda y.\bot)\ulcorner N \urcorner$$

In the above, variables $z, z_1, \cdots, z_n$ are assumed to be fresh; $f$ ranges over primitive functions and numerals.

The idea is that $\ulcorner M \urcorner$ captures the initial, non-recursive operational behaviour of $M$, so that $\ulcorner M \urcorner$ simulates the reduction of $M$ until the latter reaches a value, or a term of the form $E[\mathsf{Y}\lambda x.N]$.

It is straightforward to see that for every trace $s \in \mathbb{S}$, there is a reduction sequence of $\ulcorner M \urcorner$ that simulates (an initial subsequene of) the reduction of $M$ under $s$.

Finally thanks to the following

**Theorem A.2** (de Groote [48]). *The simply-typed nondeterministic lambda calculus is strongly normalising.* $\square$

we have:

(i) Every $\ulcorner M \urcorner$-reduction terminates on reaching $r$, or a term of the shape $E[\bot N_1 \cdots N_n]$.
(ii) Further there is a finite bound (say $l$) on the length of such $\ulcorner M \urcorner$-reduction sequences; and $l$ bounds the stopping time $T_1$.

This concludes the proof of Lem. III.8.

**Lemma III.9.** $T_M^{\mathsf{Y}}$ *is a stopping time adapted to* $(\mathcal{F}_{T_n})_{n \geq 0}$.

*Proof.* To see $\{T_M^{\mathsf{Y}} = n\} \in \mathcal{F}_{T_n}$ for a given $n$, we need to show that $\{T_M^{\mathsf{Y}} = n\} \cap \{T_n \leq i\} \in \mathcal{F}_i$, for all $i \in \mathbb{N}$. Let $\mathcal{V}$ be the set of values in $\Lambda^0$ (which is measurable), and let $\mathcal{W}$ be the non-values. Then, it suffices to observe that $\{T_M^\mathsf{Y} = n\} \cap \{T_n \leq i\} = \bigcup_{l=n}^{i} \left(M_l^{-1}[\mathcal{V}] \cap M_{l-1}^{-1}[\mathcal{W}] \cap \{T_n = l\}\right)$, where each of $M_l^{-1}[\mathcal{V}]$, $M_{l-1}^{-1}[\mathcal{W}]$, and $\{T_n = l\}$ is in $\mathcal{F}_i$. $\square$

## APPENDIX B
## SUPPLEMENTARY MATERIALS FOR SEC. IV

**Theorem IV.2.** *Given a closed term $M$, the function $f : Rch(M) \to \mathbb{R}$ given by*

$$f(N) := \mathbb{E}[\textit{number of } \mathsf{Y}\textit{-reduction steps from } N \textit{ to a value}]$$

*if it exists, is the least of all possible ranking functions of $M$.*

*Proof.* Let $f$ be the candidate least ranking function defined above, and suppose $g$ is another ranking function such that $f(N) > g(N)$ for some $N \in Rch(M)$. The restrictions of $f$ and $g$ to $Rch(N)$ have the same properties assumed of $f$ and $g$, so assume w.l.o.g. that $N = M$. The difference $g - f$ is then a supermartingale (with the same setup as in Thm. III.7); therefore $\mathbb{E}[g(M_n)] \leq \mathbb{E}[f(M_n)] + g(M) - f(M)$, for all $n$. Now $\mathbb{E}[f(M_n)] = \sum_{k=n}^{\infty} \mathbb{P}[M_k = E[\mathsf{Y}N]$ for some $E, N] \to 0$ as $n \to \infty$; therefore as $g(M) - f(M) < 0$, eventually $\mathbb{E}[g(M_n)] < 0$, which is impossible. It follows that $g \geq f$ as required.

In order for $f$ to be the least ranking function of $M$, it also has to actually be a ranking function itself. Each of the conditions on a ranking function is easily verified from the definition of $f$. $\square$

**Theorem IV.5** (Sparse function). *Every sparse ranking function is a restriction of a ranking function.*

*Proof.* Take a closed term $M$ and sparse ranking function $f$ on $M$. Define $f_1 : Rch(M) \rightharpoonup \mathbb{R}$ by

$$f_1(N) := f(N) \text{ whenever } f(N) \text{ is defined,}$$
$$f_1(V) := 0 \text{ for values } V \text{ not in the domain of } f.$$

Define $(\text{next}(N, s), \_) = \text{red}^n(N, s)$ for the least $n \geq 0$ such that it's in the domain of $f_1$, and $g(N, s) := |\{m < n \mid \text{red}^m(N, s) \text{ is of the form } (E[\mathsf{Y} \lambda x.N'], s')\}|$. The function next is well-defined (i.e. $n$ is finite) for all $N \in Rch(M)$ by induction on the path from $M$ to $N$, by the third condition on sparse ranking functions. Define $f_2(N) = \int_\mathbb{S} f_1(\text{next}(N, s)) + g(N, s) \mu_\mathbb{S}(\text{d}s)$. The (total) function $f_2$ agrees with $f$ on $f$'s domain, and it is a ranking function on $M$ (in fact, the least ranking function of which $f$ is a restriction, by the same argument as Thm. IV.2). $\square$

## APPENDIX C
## SUPPLEMENTARY MATERIALS FOR SEC. V

**Theorem V.4.** *Let $(Y_n)_{n \geq 0}$ be an antitone strict supermartingale w.r.t. stopping time $T$. Then $T < \infty$ a.s.*

*Proof.* First, as $(Y_n)$ is a supermartingale, $\mathbb{E}[Y_n] \leq \mathbb{E}[Y_0]$. Therefore

$$\mathbb{E}[Y_n \mid T > n]$$
$$= \quad \{ \text{ rearranging terms } \}$$
$$\frac{\mathbb{P}[T > n] \mathbb{E}[Y_n \mid T > n] + \mathbb{P}[T \leq n] \mathbb{E}[Y_n \mid T \leq n] - \mathbb{P}[T \leq n] \mathbb{E}[Y_n \mid T \leq n]}{\mathbb{P}[T > n]}$$
$$= \quad \{ \text{ definition of conditional expectation } \}$$
$$\frac{\mathbb{E}[Y_n] - \mathbb{P}[T \leq n] \mathbb{E}[Y_n \mid T \leq n]}{\mathbb{P}[T > n]}$$
$$\leq \quad \{ Y_n \geq 0 \text{ always } \}$$
$$\frac{\mathbb{E}[Y_n]}{\mathbb{P}[T > n]}$$
$$\leq \quad \{ (Y_n)_n \text{ is a supermartingale } \}$$
$$\frac{\mathbb{E}[Y_0]}{\mathbb{P}[T > n]}$$

*Claim*: For all $0 < x \leq 1$, $\mathbb{P}[T > B_x] \leq x$, where $B_x = \left\lceil \frac{\mathbb{E}[Y_0] + 1}{x \, \epsilon(\mathbb{E}[Y_0] \, x^{-1})} \right\rceil$ and $\epsilon : \mathbb{R}_{\geq 0} \to \mathbb{R}_{>0}$ is the antitone function.

As the convex hull of $\epsilon$ (the greatest convex function less than or equal to it) satisfies all the conditions assumed of $\epsilon$, in addition to being convex, assume wlog that $\epsilon$ is convex.

Assume for a contradiction that $\mathbb{P}[T > B_x] > x$. Then, take $n \leq B_x$. We have

$$\mathbb{E}[Y_n - Y_{n+1}]$$
$$= \quad \{ \mathcal{F}_n \subseteq \mathcal{F}_{n+1}, \text{ def. \& linearity of cond. expectation } \}$$
$$\mathbb{E}[Y_n - \mathbb{E}[Y_{n+1} \mid \mathcal{F}_n]]$$
$$\geq \quad \{ \text{ antitone strict assumption } \}$$
$$\mathbb{E}[\epsilon(Y_n) \cdot \mathbf{1}_{\{T > n\}}]$$
$$= \quad \{ \text{ definition of expectation conditioning on an event } \}$$
$$\mathbb{P}[T > n] \, \mathbb{E}[\epsilon(Y_n) \mid T > n]$$
$$\geq \quad \{ \text{ Jensen's inequality } \}$$
$$\mathbb{P}[T > n] \, \epsilon(\mathbb{E}[Y_n \mid T > n])$$
$$\geq \quad \{ \text{ proved earlier } \}$$
$$\mathbb{P}[T > n] \, \epsilon\left(\frac{\mathbb{E}[Y_0]}{\mathbb{P}[T > n]}\right)$$
$$> \quad \{ \text{ assumption, } \mathbb{P}[T > n] \geq \mathbb{P}[T > B_x] > x \}$$
$$x \, \epsilon\left(\mathbb{E}[Y_0] \, x^{-1}\right).$$

Therefore, by a telescoping sum

$$\mathbb{E}[Y_{B_x}] = \mathbb{E}[Y_{B_x} - Y_0 + Y_0] \leq \mathbb{E}[Y_0] - B_x \, x \, \epsilon(\mathbb{E}[Y_0] \, x^{-1})$$
$$\leq -1 < 0$$

which is a contradiction, therefore the claim must be true, therefore $P[T > n] \to 0$ as $n \to \infty$, therefore $T < \infty$ a.s. $\square$

**Theorem V.8** (Sparse function). *Every antitone sparse ranking function is a restriction of an antitone ranking function.*

*Proof.* Take a closed term $M$ and an antitone sparse ranking function $f$ on $M$, with a corresponding antitone function $\epsilon$. Assume wlog that $\epsilon$ is convex (as if it isn't, we can just take its convex hull instead). As in Theorem IV.5, define $f_1 : Rch(M) \rightharpoonup \mathbb{R}$ by

$$f_1(N) = f(N) \text{ whenever } f(N) \text{ is defined,}$$
$$f_1(V) = 0 \text{ for values } V \text{ not in the domain of } f.$$

Define $(\text{next}(N, s), \_) = \text{red}^n(N, s)$ for the least $n \geq 0$ such that it's in the domain of $f_1$, and $g(N, s) := |\{m < n \mid \text{red}^m(N, s) \text{ is of the form } (E[\mathsf{Y}\,\lambda x.N'], s')\}|$. The function $\text{next}$ is well-defined (i.e. $n$ is finite) for all $N \in Rch(M)$ by induction on the path from $M$ to $N$, by the third condition on antitone partial ranking functions. Define $f_2(N) := \int_\mathbb{S} f_1(\text{next}(N, s)) + \epsilon(f_1(\text{next}(N, s)))g(N, s)\,\mu_\mathbb{S}(\mathrm{d}s)$. For any term $N$ where $f$ is defined, $f_2(N) = f(N)$, and the value that $f_2$ would have at $N$ if $f$ were not defined at $N$ is $\leq f(N)$, by the third condition on antitone partial ranking functions. In order to show that $f_2$ is an antitone partial ranking function, it therefore suffices to show that the value that $f_2$ would have had at each term if $f$ were not defined at that term is at least the expectation of $f_2$ after one reduction step (plus $\epsilon(f_2(N))$ if the reduction step is a $t\mathsf{Y}$-reduction). For any term $N$ which is not of the form $E[R]$ for some $\mathsf{Y}$-redex $R$, this is trivial. If $R$ is a $\mathsf{Y}$-redex, then $\epsilon(f_2(N)) \leq \int_\mathbb{S} \epsilon(f_1(\text{next}(N, s)))\,\mu_\mathbb{S}(\mathrm{d}s)$ by the convexity of $\epsilon$, because $n$ is bounded. Therefore, the (total) function $f_2$, which agrees with $f$ on $f$'s domain, is an antitone ranking function on $M$ (with the same function $\epsilon$ if it's convex). $\square$

*Ex. V.10* (Continuous random walk)

Let $\Theta := \mathsf{Y}\,\lambda fx.\mathsf{if}(x \leq 0, 0, f(x - \mathsf{sample}))$. We can construct a sparse ranking function $f$ for $\Theta\,\underline{10}$ as follows:

$$\Theta\,\underline{l} \mapsto l + 2$$
$$\mathsf{if}(\underline{l} \leq 0, 0, \Theta\,(\underline{l} - \mathsf{sample})) \mapsto l + 1$$
$$\underline{0} \mapsto 0.$$

For a more complex (not Y-PAST) example, consider the following "continuous random walk": $\Xi\,\underline{10}$ where

$$\Xi := \mathsf{Y}\,\lambda fx.\mathsf{if}(x \leq 0, 0, f(x - \mathsf{sample} + 1/2))$$

Let $g(x) := 2 + \ln(x + 1)$, and let $\epsilon$ be the function specified by

$$\epsilon(g(x - 1/2)) = g(x) - \int_{x-1/2}^{x+1/2} g(y)\,\mathrm{d}y.$$

The limit of this as $x \to \infty$ and $g(x + 1/2) \to \infty$ is 0, and $\frac{\mathrm{d}}{\mathrm{d}x}\epsilon(g(x + 1/2)) = \frac{1}{x+1} + \ln(1 - \frac{1}{x+1/2}) < 0$, and $g$ is monotonic increasing; therefore $\epsilon$ is antitone and bounded below by 0. We define an antitone sparse ranking function by:

$$\Xi\,\underline{l} \mapsto g(l), \qquad \underline{0} \mapsto 0$$

The value of $g$ after one Y-reduction step is at least $g(l - 1/2)$, therefore the expectation of $\epsilon$ after one Y-reduction step is at most $\epsilon(g(l - 1/2)) = g(x) - \int_{x-1/2}^{x+1/2} g(y)\,\mathrm{d}y$. Thus

- in case $l > 0$, $g$ decreases by the required amount
- in case $l \leq 0$, $g(\Xi l) \geq 2 + \ln(1/2) > \ln(\frac{3\sqrt{3}}{2}) - 1 = \epsilon(g(0 - 1/2))$ as well.

Hence this is a valid antitone sparse ranking function and the term is AST.

*Ex. V.11* (Fair-in-the-limit random walk) [25, §5.3]

$$\underbrace{\left(\mathsf{Y}\,\lambda fx.\,\mathsf{if}(x \leq 0, 0, f(x-1) \oplus_{\frac{x}{2x+1}} f(x+1))\right)}_{\Xi}\,\underline{10}$$

To construct an antitone ranking function, we solve the recurrence relation:

$$z_0 = 0$$
$$n \geq 1, \quad z_n > \tfrac{n}{2n+1}\,z_{n-1} + \tfrac{n+1}{2n+1}\,z_{n+1}.$$

For $n > 2$, $z_n = \ln(n - 1)$ works. The expected decrease is $\frac{1}{2}(\ln(1 + \frac{1}{(n-1)^2-1}) - \frac{1}{2n+1}\ln(1 + \frac{2}{n-2}))$, and using the fact that $\frac{x}{x+1} \leq \ln(1+x) \leq x$ for $x > 0$, this is at least $\frac{1}{2}(\frac{1}{(n-1)^2} - \frac{1}{2n+1}\frac{2}{n-2}) = \frac{n^2}{2(n-1)^2(2n+1)(n-2)}$, which (again for $n > 2$) is positive and antitone. For $n = 0, 1, 2$, we take $\epsilon$ to be 9/40 (the same as its value at 3), then set $z_2 = 2\ln 2 - \ln 3 - 1$, $z_1 = 3\ln 2 - 2\ln 3 - 3$, $z_0 = 4\ln 2 - 3\ln 3 - 6$. Some of those values are negative, but it's still bounded below, so by adding a constant offset it can be corrected, and the term is AST.

**Example C.1** (Escaping spline). [25, §5.4]

$$\underbrace{\left(\mathsf{Y}\,\lambda fx.\,0 \oplus_{\frac{1}{x+1}} f(x+1)\right)}_{\Xi}\,\underline{10}$$

In this case, the fact that the ranking function must decrease at each Y-step (in an antitone sparse ranking function) by the expected value of $\epsilon$ not at the current term, but at the next term where the ranking function is defined, is a little harder to deal with, because the variable $x$ can change all the way to 0 in one step, therefore simply adding a small offset doesn't suffice to compensate for this fact.

Consider the candidate ranking function $\Xi\,\underline{n} \mapsto n + 1$. For each $n$, $\Xi\,\underline{n}$ reduces to either $\underline{0}$ (with probability $\frac{1}{n+1}$) or $\Xi\,\underline{n+1}$, therefore the expected value of the ranking function is $\frac{n(n+2)}{n+1} = n + 1 - \frac{1}{(n+1)^2}$, and the required decrease is $\frac{\epsilon(0)+n\epsilon(n+2)}{n+1}$, therefore eventually, the expected decrease isn't enough, whatever the value of $\epsilon(0)$ is.

If instead, we take the ranking function to be defined after the Y-reduction but before the sample-reduction as well, this can be resolved. Letting $\Theta[n] = \underline{0} \oplus_{\frac{1}{n}} \Xi\,n$:

$$\Theta[\underline{n}] \mapsto n$$
$$\Xi\,\underline{10} \mapsto 12.$$

For each $n$, $\Theta[n]$ reduces to either 0 or $\Theta[n+1]$, with a Y-reduction only in the latter case, and the condition that this is an antitone sparse ranking function is that $n \geq \frac{(n-1)(n+1)}{n} + \frac{n-1}{n}\epsilon(n+1)$ and $12 \geq 11 + \epsilon(11)$. These are satisfied by setting $\epsilon(x) = \min(1, 1/x)$, therefore this term is AST.

*Non-affine recursion*

Many of the recent advances in the development of AST verification methods [19, 21, 22, 24, 25, 49–53] are concerned with loop-based programs. We can view such loops as tail-recursive programs that are, in particular, *affine recursive*, i.e., in each evaluation (or run) of the body of the recursion, recursive calls are made from at most one call site [43, §4.1]. (Note that whether a program is affine recursive cannot be checked by just counting textual occurrences of variables.) Termination analysis of *non-affine recursive* probabilistic programs does not seem to have received much attention. Methods such as those presented in [43] are explicitly restricted to affine programs, and are unsound otherwise. By contrast, many probabilistic programming languages allow for richer recursive structures [5, 7, 54].

*Ex. V.12 (Non-affine recursion)*

Let
$$\Xi := \mathsf{Y}\lambda f x. x \oplus_{2/3} f(f(x+1))$$

A suitable sparse ranking function for $\Xi \underline{1}$ would be $\Xi^n \underline{i} \mapsto 3n$ (for $n \geq 0$), because $\Xi^n \underline{i}$ reduces to either $\Xi^{n+1} \underline{i+1}$ or $\Xi^{n-1} \underline{i+1}$, with probabilities $1/3$ and $2/3$. The value of $i$ does not actually matter for the progress of this recursion. It is basically another variant of the biased random walk, except that the relevant variable is the number of copies of $\Xi$, instead of a real number in the term.

*Ex. V.13 (More complex non-affine recursion)*

Let
$$\Xi = \mathsf{Y}\,\lambda f x.\,(\lambda e. X[e])\,\mathsf{sample}$$
$$X[e] = \mathsf{if}(e \leq p - x^{-2}, x+1, f^2(x+1) \oplus_e f(x+1))$$

This example is a little more complex because of the use of a random sample, $e$, as a first class value, which cannot be modelled via discrete distributions.

We can use the ranking function method (coupled with the solution of linear recurrence relations) to show that provided $p \geq \frac{5-\sqrt{21}}{2}$, the program $\Xi \underline{3}$ is AST. Consider the edge case, that $p = \frac{5-\sqrt{21}}{2}$ exactly.

The term $\Xi^n \underline{x}$ (for $n > 0$, $x^{-2} < p$) reduces to either $\Xi^{n-1}\,\underline{x+1}$, $\Xi^n\,\underline{x+1}$ or $\Xi^{n+1}\,\underline{x+1}$, with probabilities $p - x^{-2}$, $(1-p+x^{-2})\frac{p-x^{-2}}{2}$ and $(1-p+x^{-2})(1-\frac{p-x^{-2}}{2})$ respectively. Let $a(\Xi^n \underline{x}) = n + 2\,x^{-1}$. This is a supermartingale (because the decrease in $2\,x^{-1}$ as $x$ increases is enough to offset the average increase in $n$), but it does not satisfy the antitone-strict progress condition. It does however have a bounded-below variance, so the usual method of using $\ln a$ instead of just $a$ works. Calculating the exact amount that $\ln a$ decreases is not necessary, because it can be bounded as follows: $a$ changes by at least $1/2$ with probability at least $1/11$ (assuming $x \geq 3$), therefore $\ln(a)$ decreases in expectation by at least $\frac{1}{11}(\ln(a+\frac{1}{2}) - \ln(a) - \frac{1}{2a})$, using the fact that a linear approximation to $\ln$, applied to $a$, at least doesn't increase, then adding the deviation of $\ln$ from its linear approximation $\ln(a) + \frac{x-a}{a}$ (and using the fact that the linear approximation is everywhere an overestimate), we obtain a sufficiently strong bound on the decrease of $\ln(a)$ that it's an antitone sparse ranking function. (It can obviously be extended to $\Xi\,\underline{1}$ too, which reduces to $\Xi^n\,\underline{3}$ or a value after a bounded number of steps, but that complicates the analysis a little.)

*Higher-order recursion*

There is an important formalism for constructing (non-random) higher-order recursive functions, viz., *higher-order recursion schemes* (HORS) (see e.g. [55, 56]). Recently [44] have extended HORS to *probabilistic higher-order recursion schemes* (PHORS), which are HORS augmented with probabilistic (binary) branching $\oplus_p$. As HORS are in essence the $\lambda\mathsf{Y}$-calculus [57] (i.e. pure simply-typed lambda calculus with recursion, generated from a finite base type), PHORS are definable in PPCF: (order-$n$) PHORS is encodable as (order-$n$) call-by-name PPCF, but the former is strictly less expressive (because the underlying HORS is not Turing complete). For example, the PPCF term in Ex. V.14 is not definable in PHORS.

A relevant result here is that the AST problem is decidable for order-1 PHORS (by reduction to the PSPACE-hard solvability of finite systems of polynomial equations with real coefficients; see [58]), but undecidable for PHORS with order greater than 1 [44].

*Ex. V.14 (Higher-order recursion)*

Consider the higher-order function $\Xi : (\mathsf{R} \to \mathsf{R} \to \mathsf{R}) \to \mathsf{R} \to \mathsf{R} \to \mathsf{R}$ recursively defined by
$$\Xi := \mathsf{Y}\lambda\varphi\,f^{\mathsf{R}\to\mathsf{R}\to\mathsf{R}}\,s^{\mathsf{R}}\,n^{\mathsf{R}}.$$
$$\mathsf{if}(n \leq 0, s, f\,n\,(\varphi\,f\,s\,(n-1)) \oplus_p f\,n\,(\varphi\,f\,s\,(n+1)))$$

Let $F : \mathsf{R} \to \mathsf{R} \to \mathsf{R}$ be a function such that $F\,\underline{n}\,\underline{m}$ terminates with no $\mathsf{Y}$ reductions for all $m$ and $n$. Then $\Xi\,F\,\underline{x}\,\underline{y}$ has the antitone sparse ranking function
$$F\,\underline{n_0}\,(F\,\underline{n_1}\,(\ldots(\Xi\,F\,\underline{x}\,\underline{n})\ldots)) \mapsto g_2(n),$$
where $g_2$ is as in Ex. V.9, using the same antitone function too.

The inequalities required for this to be an antitone partial supermartingale are satisfied with just the same reasoning as in Ex. V.9, therefore this term too is antitone rankable.

Again, this is not actually the correct reduction order, in that $F$ should be applied to its argument $n_i$ and reduced to a value before its second argument is expanded, but this would complicate the presentation, and this version can be justified by Thm. VI.15 instead. The reduction strategy implied here should be clear enough (assuming that, where it isn't specified, it just matches cbv), but to be more precise, let

$$r\left(F\,\underline{n_0}\,(\ldots(F\,\underline{n_{k-1}}\,X)\ldots)\right) = @_2^k; \mathrm{cbv}(X),$$

where $X$ is not a value or of the form $F\,\underline{n_k}\,Z$ for some non-value $Z$. Within $X$, there is necessarily a redex at $\mathrm{cbv}(X)$. The only conditions that $@_2^k; \mathrm{cbv}(X)$ is also a position of a redex in the whole term are that if the redex is a sample, it is not inside a $\lambda$ that's inside a Y or on the right of an application, which is true because cbv never selects a redex inside of a $\lambda$. After $n$ reaches 0, the term is $F\,\underline{n_0}\,(\ldots(F\,\underline{n_k}\,\underline{x})\ldots)$, then the innermost $F$ is evaluated completely in cbv order, therefore (by assumption) it terminates. Because there are no Y-reductions in the evaluation of $F$, it can't reach any other term of the form $F\,\underline{n'_k}\,X$, therefore $k$ never changes until this whole subexpression reaches a number, at which point $k$ decreases by 1 and the next $F$ is evaluated.

**Theorem V.15.** *For any stopping time $T$ which is almost surely finite, if $(\mathcal{F}_n)_n$ is the coarsest filtration to which $T$ is adapted, then there is a supermartingale $(Y_n)_n$ adapted to $(\mathcal{F}_n)_n$ and an antitone function $\epsilon$ such that $(Y_n)_n$ is an antitone ranking supermartingale with respect to $T$ and $\epsilon$.*

*Proof.* If $T$ is bounded by $b$ a.s. (for $n$ constant), the statement is trivially true by taking $\epsilon$ to be constantly 1, and $Y_n = b-n$. Otherwise, let $(t_n)_{n\geq 0}$ be defined recursively such that $t_0 = 0$, $t_{n+1} > t_n$ and $\mathbb{P}[T > t_{n+1}] \leq \frac{1}{4}\mathbb{P}[T > t_n]$. We will then define a supermartingale $(Y_n)_{n\geq 0}$ and nonrandom sequence $(y_n)_{n\geq 0}$ such that $Y_n = 0$ iff $T < n$, $Y_n = y_n$ iff $T \geq n$, and $y_n \geq 2^k$ for $n \geq t_k$.

The antitone function $\epsilon$ is defined piecewise and recursively in such a way as to force all the necessary constraints to hold:

$$\text{for } x \in [0,1),\; \epsilon(x) = 1$$
$$\text{for } x \in [2^k, 2^{k+1}),\; \epsilon(x) = \min\left(\epsilon(2^{k-1}), \frac{2^k}{t_{k+1}-t_k}\right).$$

This ensures that

- $\epsilon$ is weakly decreasing.
- $\epsilon$ is bounded below by 0.
- For $x \geq 2^k$, $\epsilon(x) \leq \frac{2^k}{t_{k+1}-t_k}$.

The sequence $(y_n)_{n\geq 0}$ is then defined recursively by

$$y_0 = 2, \quad y_{n+1} = (y_n - \epsilon(y_n))\frac{\mathbb{P}[T > n]}{\mathbb{P}[T > n+1]}.$$

The fact $y_{t_k} \geq 2^{k+1}$ is proven by induction on $k$. For base case, $2 \geq 2$, as required. For the inductive case $k+1$, we first do another induction to prove that $y_n \geq 2^k$ for all $t_k \leq n \leq t_{k+1}$. The base case of this inner induction follows from the outer induction hypothesis a fortiori. Take the greatest $k$ such that $n > t_k$. By the induction hypothesis, $y_{t_k} \geq 2^k$.

$$y_n = y_{t_k}\frac{\mathbb{P}[T > t_k]}{\mathbb{P}[T > n]} - \sum_{m=t_k}^{n-1}\epsilon(y_m)\frac{\mathbb{P}[T > m+1]}{\mathbb{P}[T > n]}$$
$$\geq \frac{\mathbb{P}[T > t_k]}{\mathbb{P}[T > n]}(y_{t_k} - \sum_{m=t_k}^{n-1}\epsilon(y_m))$$
$$\geq \frac{\mathbb{P}[T > t_k]}{\mathbb{P}[T > n]}(y_{t_k} - \sum_{m=t_k}^{n-1}\frac{2^k}{t_{k+1}-t_k})$$
$$= \frac{\mathbb{P}[T > t_k]}{\mathbb{P}[T > n]}(y_{t_k} - \frac{2^k(n-t_k)}{t_{k+1}-t_k})$$
$$\geq \frac{\mathbb{P}[T > t_k]}{\mathbb{P}[T > n]}(y_{t_k} - 2^k)$$
$$\geq \frac{\mathbb{P}[T > t_k]}{\mathbb{P}[T > n]}(2^{k+1} - 2^k)$$
$$= \frac{\mathbb{P}[T > t_k]}{\mathbb{P}[T > n]}2^k$$
$$\geq 2^k$$

as required. Substituting $n = t_{k+1}$ and using the fact that $\frac{\mathbb{P}[T > t_k]}{\mathbb{P}[T > t_{k+1}]} > 4$, the same reasoning gives $y_{t_{k+1}} \geq 2^{k+2}$, as required.

The fact $y_n \geq 2^k$ for some $k$ implies that $y_n > 0$ for all $n$, which implies that $Y_n \geq 0$. The other condition for $(Y_n)$ to be an antitone-strict supermartingale is

$$\mathbb{E}[Y_{n+1} \mid \mathcal{F}_n] = \mathbb{E}[y_{n+1}\mathbf{1}_{\{T\geq n+1\}} \mid \mathcal{F}_n]$$
$$= y_{n+1}\mathbb{E}[\mathbf{1}_{\{T\geq n+1\}} \mid \mathcal{F}_n]$$
$$= y_{n+1}\mathbf{1}_{\{T\geq n\}}\frac{\mathbb{P}[T > n+1]}{\mathbb{P}[T > n]}$$
$$= (y_n - \epsilon(y_n))\mathbf{1}_{\{T\geq n\}}$$
$$= Y_n - \epsilon(y_n)\mathbf{1}_{\{T\geq n\}}$$

therefore $(Y_n)_n$ is an antitone strict supermartingale with respect to $(T, \epsilon)$. □

The assumption that $(\mathcal{F}_n)_{n\geq 0}$ is the coarsest filtration to which $T$ is adapted is used in the equality between the third and fourth lines here. This condition is the main reason that this proof does not extend directly to a completeness result for antitone ranking functions.

## Appendix D
### Supplementary materials for Sec. VI

The rules defining $\sim_c$ can be more intuitively understood as diagrams. In the following diagrams, a circle represents a term (or skeleton), and the leaves on the trees are subterms at the positions indicated by the labels in the other nodes. The labels on the arrows between trees are the positions of the reductions, and dashed arrows represent some number of reductions in a row. If a reduction sequence includes one branch of one of these cases at some point in it, then the positions in the reduction sequence obtained by substituting in the other branch are all related by $\sim_c$ to the corresponding positions in the original reduction sequence.

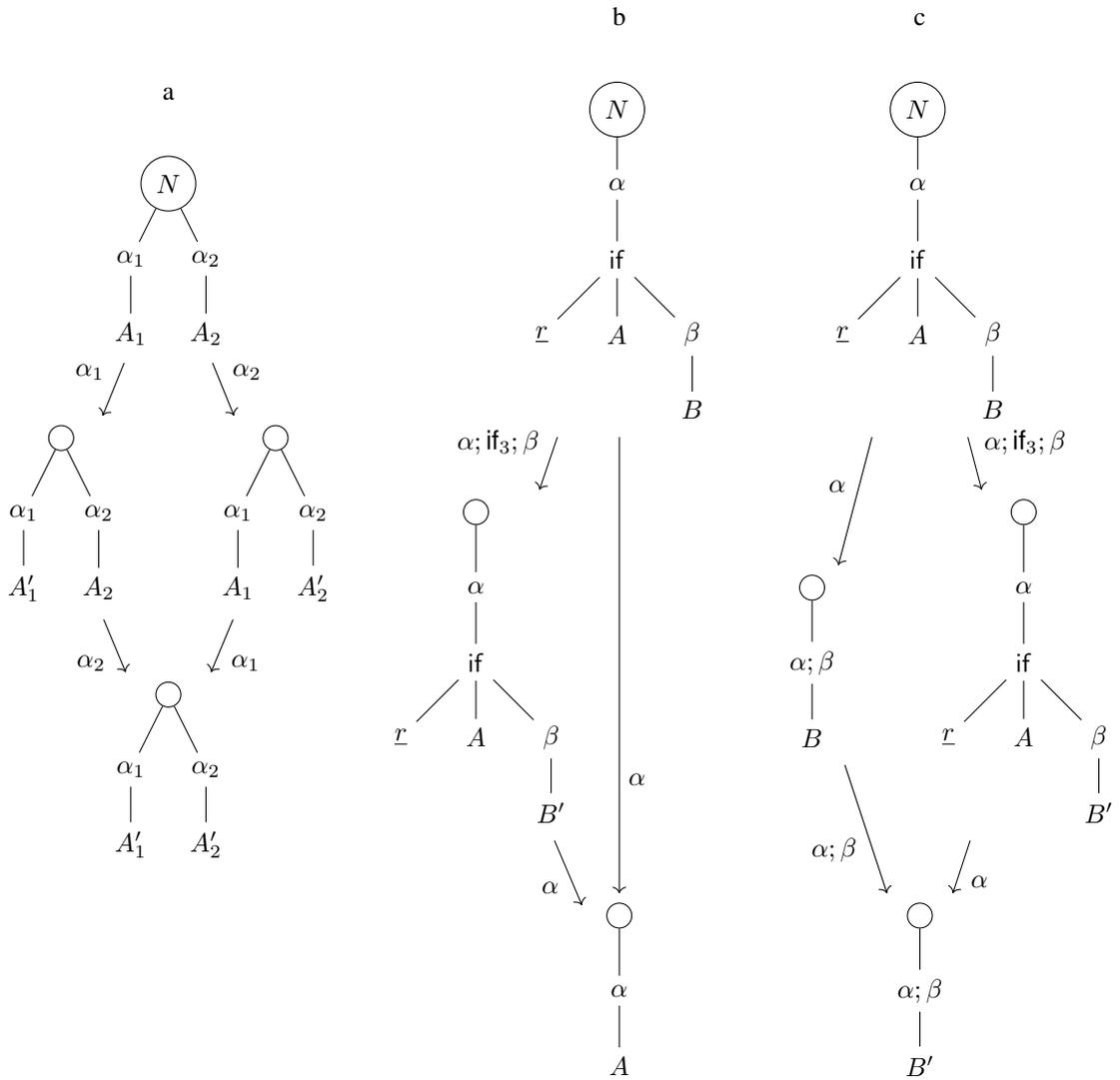

d 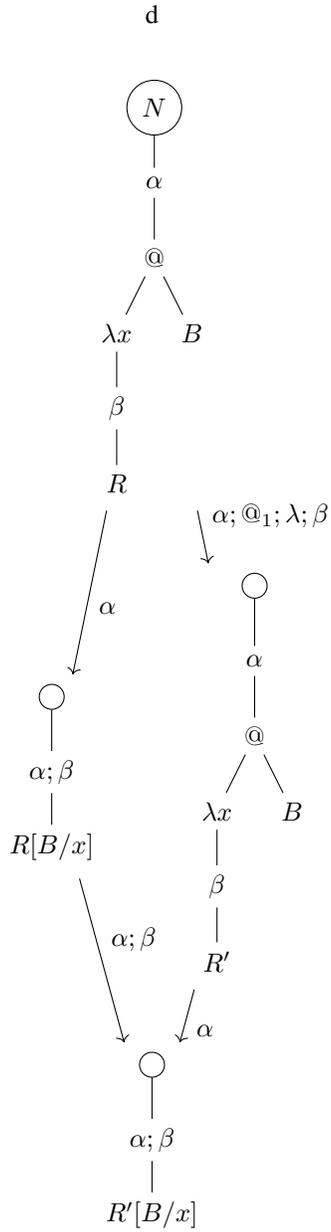 e 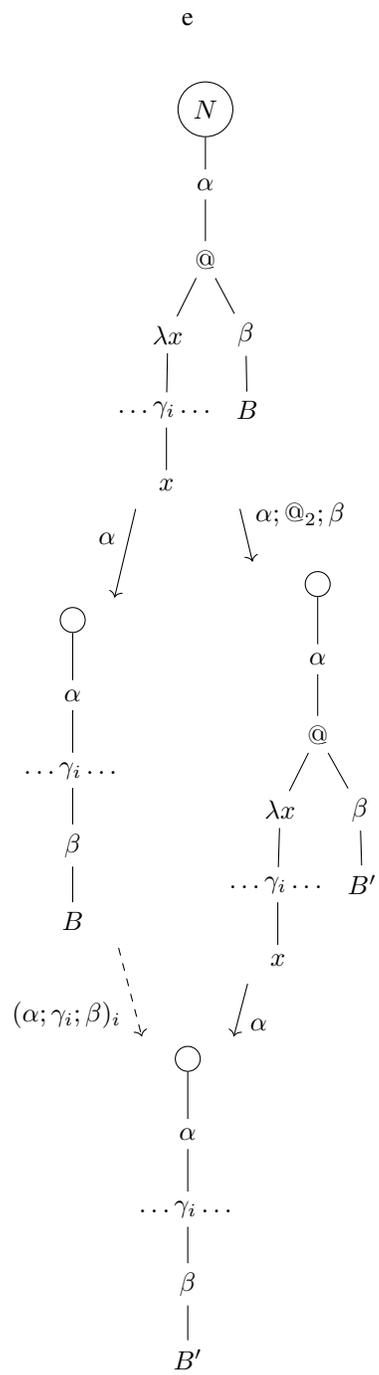

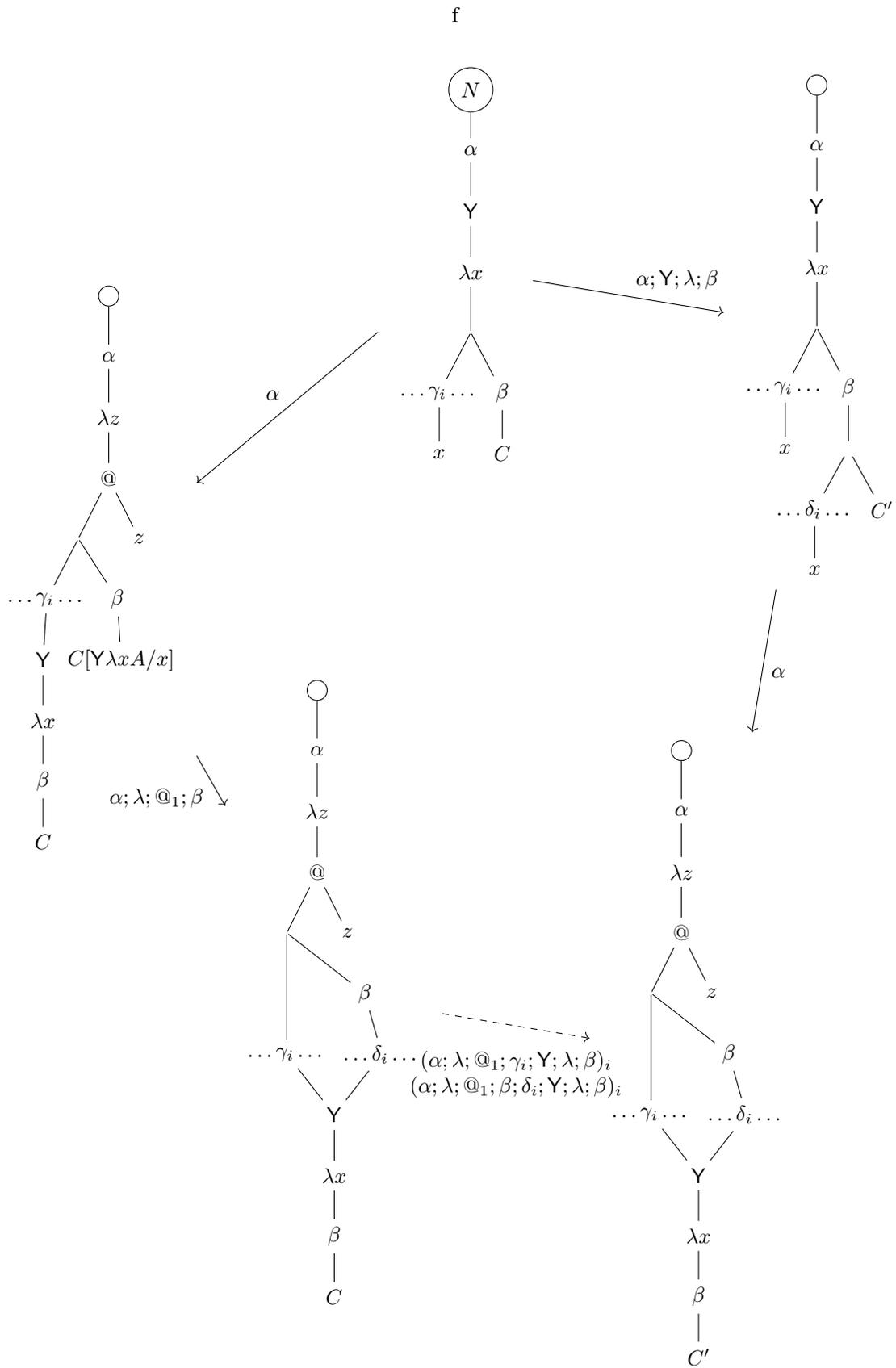

Figure 3. Illustration of the $\sim_c$ rules.

*a)* : Although the six cases in the definition of $\sim_c$ are defined separately, there are some commonalities worth noting. Each case (except a, which is symmetrical) has a right branch and a left branch (the branches of case b are unfortunately displayed the wrong way around in the diagram). The right branch has reduction steps at two positions, $\beta$ then $\alpha$, where $\beta > \alpha$ (i.e. $\beta$ is inside $\alpha$), and in the left branch, the order of these reductions are swapped, so the reductions are at $\alpha$ then $\beta_1, \ldots, \beta_n$. The position $\alpha$ of one of the reductions is unchanged, and the other reduction may have multiple (in cases e and f) or 0 (in cases b and e) images in the left branch. The images $(\beta_i)$ are all still inside (greater than or equal to) $\alpha$, and although they are not generally equal to $\beta$, the subskeletons at these positions have a similar shape to the subskeleton (initially) at position $\beta$, so that the reductions at these positions are the same type of reduction (e.g. both $\beta$-reduction or both if-reduction). Case a is also similar, except that the relevant positions are all disjoint rather than contained in one another, and either branch could be considered the right branch.

**Lemma D.1.** *If $(N_1, \alpha_1) \sim^* (N_2, \alpha_2)$, there is some descendants $N_1', N_2'$ of $N_1$ resp. $N_2$, and a position $\alpha'$ in both of them such that $(N_1, \alpha_1) \sim_p^* (N_1', \alpha') \sim_c^* (N_2', \alpha') \sim_p^* (N_2, \alpha_2)$.*

*Proof.* The $\sim_p$ relation can be split into $\sim_\downarrow \cup \sim_\uparrow$, where $(A, \alpha) \sim_\downarrow (B, \beta)$ if $A \to B$ and $(A, \alpha) \sim_p (B, \beta)$, and similarly $(A, \alpha) \sim_\uparrow (B, \beta)$ if $B \to A$ and $(A, \alpha) \sim_p (B, \beta)$. At each stage of this proof, it will be assumed that the $\sim_c$ steps and the $\sim_p$ steps in the sequence from $(N_1, \alpha_1)$ to $(N_2, \alpha_2)$ are rearranged such that there is never a $\sim_c$ immediately before a $\sim_\downarrow$, and there is never a $\sim_c$ immediately after a $\sim_\uparrow$. This rearrangement is always possible, because if there is some subsequence $(A, \alpha) \sim_c (B, \alpha) \sim_\downarrow (C, \beta)$, then there is an alternate path $(A, \alpha) \sim_\downarrow (B', \beta) \sim_c (C, \beta)$, where the reduction and the $\sim_\downarrow$ from $A$ to $B'$ are the same as those from $B$ to $C$, and the $\sim_c$ step is the same, but with the reduction sequences $O_1 \to^* O_1'$ and $O_2 \to^* O_2'$ in the definition of $\sim_c$ extended by the reduction $B \to C$. With these rearrangements assumed, it is sufficient to prove that the $\sim_p$ steps can be rearranged so that all of the $\sim_\downarrow$ steps come before all of the $\sim_\uparrow$ steps (possibly introducing some $\sim_c$ steps in the process).

The rearrangement to put all of the $\sim_\downarrow$s before all of the $\sim_\uparrow$s proceeds by induction on the number of $\sim_\downarrow$s that occur after the first occurrence of $\sim_\uparrow$. If it's 0, all of the $\sim_p$s are already in the correct order, and we're done. Otherwise, take the subsequence from the first $\sim_\uparrow$ to the first $\sim_\downarrow$ after that. This must be rearranged to some $\sim_\downarrow$s followed by some $\sim_\uparrow$s, then once that's done, the number of $\sim_\downarrow$s after the first $\sim_\uparrow$ in the overall sequence will have decreased by 1.

Let this subsequence be $A \sim_\uparrow^n B \sim_\downarrow C$. Suppose for induction that $A \sim_\uparrow^k (D, \delta) \sim_\downarrow^* (E, \epsilon) \sim_\uparrow^* C$ for some $D, \delta, E, \epsilon$, and $0 \le k \le n$, where in the reduction sequence $D \to^* E$, for any pair of reductions at positions which aren't disjoint, the reduction at the innermost (greater) position occurs first. Reduction sequences with this property will be called "parallel", as it is analagous to the parallel reduction introduced by Tait and Martin-Löf in their proof of the Church-Rosser theorem [59, 60]. This induction is in reverse, with $k$ decreasing from $n$ to 0. If $k = n$, simply set $(D, \delta) = B$ and $(E, \epsilon) = C$. If $k = 0$, then $A$ is related to $B$ by a sequence of $\sim_\downarrow$s then $\sim_\uparrow$s, as desired. In the intermediate steps, $k$ must be decreased by one, so a subsequence of one $\sim_\uparrow$ followed by a parallel sequence of $\sim_\downarrow$s must be replaced by a parallel sequence of $\sim_\downarrow$s followed by some $\sim_\uparrow$s.

For any individual $\sim_\uparrow$ then $\sim_\downarrow$ pair, let $(A, \alpha) \sim_\uparrow (B, \beta) \sim_\downarrow (C, \gamma)$. The skeleton $B$ reduces to both $A$ and $C$. If these are the same reduction, then $(A, \alpha) = (C, \gamma)$ and this subsequence may be removed. If they are different, then the way they overlap corresponds to one of the cases in the definition of $\sim_c$, either case a if the reduction positions are disjoint, or one of the cases b–f if one of the reduction positions is inside the other. In any case, there is a reduction sequence from each of $A$ and $C$ to some common skeleton $D$ ($O_1$ and $O_2$ in the definition of $\sim_c$). For the same reason that $(B, \beta) \sim_\downarrow (C, \gamma)$, there is some $\delta$ such that $(A, \alpha) \sim_\downarrow^* (D, \delta)$, and similarly for the same reason that $(B, \beta) \sim_\downarrow (A, \alpha)$, for the same $\delta$, $(C, \gamma) \sim_\downarrow^* (D, \delta)$. There are a lot of cases to check for this statement, but all of them are rather simple, and similar to each other. There is therefore an alternative $\sim^*$ sequence from $(A, \alpha)$ to $(C, \gamma)$, of the form $(A, \alpha) \sim_\downarrow^* (D_1, \delta) \sim_c (D_2, \delta) \sim_\uparrow^* (C, \gamma)$, where $D_1$ and $D_2$ are the alternative reduction sequences leading to the same skeleton $D$. There are only multiple $\sim_\downarrow$ steps in the result if the position of the reduction $B \to C$ is inside the position of the reduction $B \to A$, and similarly, there are only multiple $\sim_\uparrow$ steps in the result if the position of the reduction $B \to A$ is inside the position of the reduction $B \to C$. If there are multiple $\sim_\downarrow$ steps, they are not necessarily parallel, but in the only case where they aren't (case f of $\sim_c$), the subsequence of $\sim_\downarrow$ steps with the reductions at $\alpha; \lambda; @_1; \beta$ then $\alpha; \lambda; @_1; \gamma_i; \mathsf{Y}; \lambda; \beta$ for each $i$ where $\gamma_i \ge \beta$ (in the notation of the definition of $\sim_c$) can be replaced by $\sim_\downarrow$s with the reductions at $\alpha; \lambda; @_1; \gamma_i'; \mathsf{Y}; \lambda; \beta$ then $\alpha; \lambda; @_1; \beta$, then another $\sim_c$, where $(\gamma_i')$ is the list of positions in $A$ (*before* the reduction at $\beta$ to $A'$, unlike $(\gamma_i)$) where $A|\gamma_i' = x$ and $\gamma_i' > \beta$.

The effect of this rearrangement is (ignoring the $\sim_c$s) to swap the order of the $\sim_\uparrow$ and the $\sim_\downarrow$, except that if the positions of one of these reductions is inside the other, that inner $\sim_p$ may be duplicated. If the positions are disjoint, both of them are unchanged (corresponding to case a of $\sim_c$), and if one is inside the other, then the outer position is unchanged and the other resultant positions, although they aren't equal to the original inner position, are still inside the outer position. Taking the sub-sequence of one $\sim_\uparrow$ followed by a parallel sequence of $\sim_\downarrow$s, this rearrangement can be repeatedly applied to move the $\sim_\uparrow$ further along the sequence. If at some point a $\sim_\uparrow$ and $\sim_\downarrow$ match (by relating the same (skeleton, position) pairs in opposite directions), they may be removed and the process may stop early. The $\sim_\uparrow$ may be duplicated if it passes a $\sim_\downarrow$ whose position is outside of the $\sim_\uparrow$'s position, but by the assumption that the sequence of $\sim_\downarrow$s is initially parallel, no position inside of this occurs later in the sequence, therefore all of the resultant $\sim_\uparrow$s pass the remaining $\sim_\downarrow$s without changing or duplicating them. The $\sim_\uparrow$s may be further duplicated, but the number of steps left in this process is bounded by the product, for all of the remaining $\sim_\downarrow$s, of 2+ the number of occurrences of the variable relevant to the reduction, because for each $\sim_\downarrow$, that is more than the number of duplicates it could produce for any $\sim_\uparrow$ that passes it. It is also possible (earlier) for some of the $\sim_\downarrow$s to be duplicated, but the only case where this process would not terminate, that both the $\sim_\downarrow$s and the $\sim_\uparrow$s continue being duplicated forever so that the number of switches left to do never decreases, is prevented by the parallelness condition, as all of the duplications of $\sim_\downarrow$s occur before all of the duplications of $\sim_\uparrow$s.

It still remains to be shown that the sequence of $\sim_\downarrow$s left at the end of this can be made parallel. The changes that may have occurred since the previous version of this sequence (which was known to be parrallel already), are that some of the $\sim_\downarrow$s may have been duplicated by passing the $\sim_\uparrow$, and if that $\sim_\uparrow$ matches one of the $\sim_\downarrow$s, that $\sim_\downarrow$ is removed. The position of the $\sim_\uparrow$'s reduction is the same for all of these duplications, because it is not changed until later in the sequence, when the $\sim_\downarrow$s with positions outside of that may occur. The resultant positions after a duplication are in the same position relative to each other, and therefore the order of the remaining $\sim_\downarrow$s is automatically parallel, except that (letting the position of the $\sim_\uparrow$'s reduction be $\alpha$) if there are $\sim_\downarrow$s with reductions at positions $\alpha; @_1; \lambda; \beta$ and $\alpha; @_2; \gamma$, or $\alpha; \mathsf{Y}; \lambda; \beta$ and $\alpha; \mathsf{Y}; \lambda; \gamma$, some of the resultant $\sim_\downarrow$s' positions may be inside each other where they were originally not. Similarly to the case where a $\mathsf{Y}$ reduction and a reduction at a position inside it are swapped though, the order can be fixed by swapping some of the $\sim_\downarrow$ steps with each other. In the first case, the reduction at position $\alpha; @_1; \lambda; \beta$ ends up at $\alpha; \beta$, and the reduction at position $\alpha; @_2; \gamma$ ends up at the positions $\alpha; \delta_i; \gamma$, where $(\delta_i)$ is the positions of the variable relevant to the $\sim_\uparrow$'s reduction inside its lambda. The positions of the variable in the lambda may be different before the reduction at $\beta$, but the positions of the redexes corresponding to the original redex at $\alpha; @_2; \gamma$ are changed similarly. Rather than reducing at $\alpha; \beta$ then at $\alpha; \delta_i; \gamma$, it is therefore possible to reach the same point by reducing at $\alpha; \delta'_i; \gamma$ then at $\alpha; \beta$, where $(\delta'_i)$ is the positions of the relevant variable within the lambda before the reduction at $\alpha; @_1; \lambda; \beta$ (and again, a $\sim_c$ of some sort must also be introduced). The other case, that the $\sim_\uparrow$'s reduction is a $\mathsf{Y}$-reduction, is similar. Each reduction which is duplicated has one image at $\alpha; \lambda; @_1$; its original relative position, and one for each occurrence of the variable. Those corresponding to variable occurrences may need to be moved to before those at $\lambda; @_1$.

In summary, to reach the desired order of $\sim_p$ steps, each $\sim_\downarrow$ may have to be moved past some $\sim_\uparrow$ steps, and it may get duplicated in the process, but the resultant sequence of $\sim_\downarrow$s can all be put in parallel order, and it has been shown that moving a $\sim_\uparrow$ past a sequence of $\sim_\downarrow$s is always possible if they're in parallel order, therefore the overall process terminates and reaches a point where all of the $\sim_\downarrow$s precede all of the $\sim_\uparrow$s, and all of the $\sim_c$s are in between them. All of these rearrangements leave the end points of the sequence unchanged, therefore this is an alternative sequence of $\sim$ steps between $(N_1, \alpha_1)$ and $(N_2, \alpha_2)$ of the form $(N_1, \alpha_1) \sim_\downarrow^* (N'_1, \alpha') \sim_c^* (N'_2, \alpha') \sim_\uparrow^* (N_2, \alpha_2)$. □

If $(X, \alpha) \sim_c (Y, \alpha)$ then the final skeletons in $X$ and $Y$ are equal, and $(X, \beta) \sim_c (Y, \beta)$ for any $\beta$ that occurs in this skeleton, therefore $\sim_c$ can be considered to apply to reduction sequences even without positions, $X \sim_c Y$ iff $(X, \cdot) \sim_c (Y, \cdot)$. The structure of $\sim_c$ can be seen more clearly by choosing a canonical example from each equivalence class. For this, the member of the class that's in call-by-value order is used. A reduction sequence is in call-by-value order if for every pair of reductions in the sequence at positions $\alpha$ and $\beta$, where the reduction at $\alpha$ happens first, either $\alpha \leq \beta$ or $\alpha$ and $\beta$ are disjoint, with $\alpha$ to the left of $\beta$ (i.e. the first elements of the sequences $\alpha, \beta$ where they differ are $@_1$ and $@_2$, $\underline{f}_i$ and $\underline{f}_j$ for $i < j$, or $\mathsf{if}_i$ and $\mathsf{if}_j$ for $i < j$, in that order). Note that although the reductions that occur in the sequence are in call-by-value order, it may not be the actual call-by-value reduction sequence starting from that point because some redexes may remain un-reduced even when other reductions that should happen afterwards occur. It is only required that the reductions that do occur occur in the correct order.

**Lemma D.2.** *For any reduction sequence $X$, there is a unique reduction sequence $X_{cbv}$ that is related to $X$ by $\sim_c^*$ and is in call-by-value order, and if $X \sim_c^* Y$, then $X_{cbv} = Y_{cbv}$.*

*Proof.* Define $X_n$ recursively so that $X_0 = X$ and if $X_n$ is not in CBV order, take the last pair of reductions in $X_n$ which aren't in CBV order. They are adjacent in the sequence, as the CBV order is a total order. Let the positions of these reductions be $\alpha$ and $\beta$ respectively. If they are disjoint, they can be considered as one of the branches of $\sim_c$ case a. Otherwise, $\beta < \alpha$, and the sub-skeleton at position $\beta$ at the appropriate point in the reduction sequence is a redex with non-trivial subterms, therefore it's of one of the forms $\mathsf{if}(\mathsf{X} > 0, A, B)$, $(\lambda x.A)B$ or $\mathsf{Y}\lambda x.A$, therefore the reductions $\alpha$ and $\beta$ form the $O_2$ branch of one of the cases b, c, d, e or f of $\sim_c$. In either case, define $X_{n+1}$ as equal to $X_n$ except taking the other (left) branch, so that the order of the reductions at $\alpha$ and $\beta$ are switched (although the equivalent(s) of the reduction at $\alpha$ may not actually be at that position).

This sequence of swaps rearranges the reductions in $X$ rather like insertion-sort (but in some cases duplicating or deleting the new element), and eventually reaches some $X_n$ that is in CBV order. This is defined to be $X_{cbv}$.

*b) :* If $Y \sim_c X \sim_c^* X_{cbv}$, and the reductions involved in $Y \sim_c X$ are not the last pair of reductions in $Y$ that aren't in CBV order, the order of the $\sim_c$s can be rearranged so that the sequence $Y \sim_c^* X_{cbv}$ is the canonical sequence given above, therefore $Y_{cbv} = X_{cbv}$. To be more precise about this rearrangement, consider the sequence of $\sim_c$ steps from $Y$ to $X_{cbv}$. The first step takes some sub-sequence of the reduction sequence $Y$ and swaps the order of those reductions. If $X = X_{cbv}$, then $X = Y_1 = Y_{cbv}$ already. Otherwise, consider the cases according to whether the image of this subsequence in $X$ occurs earlier than the next pair of reductions to be swapped (i.e. the last pair of reductions in $X$ that occur not in CBV order). If not, then either the step $Y \sim_c X$ changes this subsequence into CBV order, or out of it. If it is into, then it was the last out-of-order pair in $Y$ therefore the whole sequence $Y \sim_c^* X_{cbv}$ is already in the canonical order and $Y_{cbv} = X_{cbv}$. If it switches the pair to the wrong order, then the result of that switch is the last out-of-order pair in $X$ therefore $Y = X_1$, and the sequence $X_1 \sim_c^* X_{cbv}$ is just a suffix of the sequence $X \sim_c^* X_{cbv}$ therefore $(X_1)_{cbv} = X_{cbv}$ and again $Y_{cbv} = X_{cbv}$.

The remaining cases are when the subsequence of reduction steps involved in $Y \sim_c X$ occur before the last out-of-order pair in $X$, but possibly overlapping. If there is no overlap, then the $\sim_c$ steps do not interfere with each other at all, and can simply be performed in the other order. This gives a different sequence of $\sim_c$ steps $Y \sim_c Y_1 \sim_c X_1 \sim_c^* X_{cbv}$. As $Y_{cbv} = (Y_1)_{cbv}$, we can proceed by induction in this case and use the fact that $(Y_1)_{cbv} = (X_1)_{cbv} = X_{cbv}$ to acheive the result.

In the other case, there is some overlap between the regions involved in $Y \sim_c X$ and $X \sim_c X_1$, but the last reduction in $X \sim_c X_1$ is later than the last reduction in $Y \sim_c X$. There are only 2 reductions in $X$ involved in $X \sim_c X_1$, therefore the region involved in $Y \sim_c X$ ends on the first of the reductions involved in $X \sim_c X_1$. Now consider the specific sequence of positions of the reductions in $X$ involved in either of these steps. Let the last of these positions be $\alpha$, the first $\gamma$, and the second-last, third-last and so on $\beta_1, \beta_2, \ldots$ respectively, and let the positions of the redexes in $Y$ involved in $Y \sim_c X$ be $\beta$ then $(\gamma_i)$ in order. Then $X \sim_c X_1$ swaps $\beta_1$ and $\alpha$, and $\alpha$ comes before $\beta_1$ in CBV order (i.e. $\alpha < \beta_1$ or $\alpha$ is left of $\beta_1$), and the step $Y \sim_c X$ either proceeds forwards towards CBV order, swapping $\beta$ below $\gamma_1 = \gamma$ to form $\gamma$ then $(\beta_i)$, or proceeds backwards, taking $Y$ even farther from CBV order, swapping the $(\gamma_i)$ to above $\beta$, forming $\gamma$ then $\beta_1 = \beta$. In each of the cases below, it is required to show that the canonical sequence $(X_i)_i$ eventually reaches some term which is the same as one in $(Y_j)_j$, and from that point on, they match, therefore they reach the same end result: $X_{cbv} = Y_{cbv}$.

- $Y \sim_c X$ in the backwards direction (so that $Y$ is closer to CBV order than $X$ is): By swapping the roles of $X$ and $Y$, and $\beta$ and $\gamma$, this is equivalent to the $Y \sim_c X$ forwards case. The assumption that $\alpha$ is before $\beta_1$ in CBV order (so that $\beta_1, \alpha$ is the pair to be swapped in $X \sim_c X_1$) maps to the assumption that $\alpha$ is before $\gamma_1$ (so that the result doesn't just follow trivially by $Y_1$ being equal to $X$) and vice-versa. Then one of the cases below establishes that $X_i = Y_j$ for some $i$ and $j$, therefore after swapping back, $Y_i = X_j$.

- $Y \sim_c X$ forwards and $\alpha$ comes after $\gamma$ in CBV order: In this case, the $\gamma$ and $\alpha$ reduction steps are already in the correct order in $Y$, therefore $\beta$ and $\gamma$ are already the last out-of-order pair of reductions in $Y$, $Y_1 = X$, and $Y \sim_c X \sim_c^* X_{cbv}$ is already the canonical sequence from $Y$ therefore $Y_{cbv} = X_{cbv}$.

- $Y \sim_c X$ forwards and $\alpha$ is disjoint from $\gamma$ and $\beta$: In this case, the sequence $(X_i)_i$ starts by swapping $\alpha$ and $\beta_1$ by case a, then because $\gamma$ is disjoint from $\alpha$ and comes after is in CBV order, $\gamma$ is right of $\alpha$, therefore either all the remaining $\beta_i$s, are $\geq \gamma$ (if $\beta$ and $\gamma$ aren't disjoint) or there are no remaining $\beta_i$s (if $\beta$ and $\gamma$ are disjoint, in which case they swap by case a and there is only $\beta_1$). The sequence $(X_i)_i$ therefore proceeds to swap all the remaining $\beta_i$s then $\gamma$, in order, below $\alpha$, resulting in $X_i$ for some $i$ having as its relevant sub-reduction-sequence $\alpha, \gamma, \beta_k, \ldots, \beta_1$. The other sequence $(Y_j)_j$ starts by swapping $\gamma$ then $\beta$ below $\alpha$, then swapping $\gamma$ and $\beta$. Because $\alpha$ is disjoint from $\gamma$, the reduction there does not affect the sub-skeleton at position $\gamma$, therefore this last $\sim_c$ step proceeds identically to how it did in $Y \sim_c X$, resulting in $\gamma, \beta_k, \ldots, \beta_i$. Overall, this results in $Y_3$ having the relevant sub-reduction-sequence $\alpha, \gamma, \beta_k, \ldots, \beta_1$, therefore $Y_3 = X_i$ for the aforementioned value of $i$.

- $Y \sim_c X$ forwards, $\alpha \leq \gamma$ and $\alpha$ (and therefore also $\gamma$) is disjoint from $\beta$: In this case, $(X_i)_i$ proceeds by swapping $\beta_1 = \beta$ below $\alpha$ by case a, then swapping $\gamma$ below $\alpha$ by one of the other cases, and $(Y_j)_j$ proceeds by swapping $\gamma$ and $\alpha$, then swapping $\beta$ below $\alpha$ then all of the images of $\gamma$ (which are $\geq \alpha$ and therefore left of $\beta$). In both cases, the subterm at $\alpha$ when $\alpha$ and $\gamma$ are swapped is unaffected by the reduction at $\beta$, therefore it produces the same result in both cases, therefore the overall sequence in both cases is $\alpha$ then the images of $\gamma$ then $\beta$.

- Both $\gamma$ and $\beta$ are $> \alpha$ but they're disjoint from each other, and either the skeleton at position $\alpha$ is of the form $if(X > 0, A, B)$ or it is of the form $AB$ and either both $\gamma$ and $\beta$ are $> \alpha; @_1$ or they are both $> \alpha; @_2$: In this case, $(X_i)_i$ proceeds by swapping $\beta$ then $\gamma$ below $\alpha$, in each case possibly forming 0 or multiple images. The only way the number of images of each of these positions can be different is if the subskeleton at $\alpha$ is an if, and there are 0 of one of them and 1 of the other, in which case they're trivially in the correct order already. Otherwise, all of the images of $\beta$ and $\gamma$ are disjoint from one another, therefore none of the positions, or their relative order, change when they are swapped, therefore the next part of the sequence $(X_i)_i$ is just insertion-sort running on the images of $\beta$ and $\gamma$ with $\sim_c$ case a swaps until they're in the correct order. The sequence $(Y_j)_j$ starts by swapping $\gamma$ then $\beta$ below $\alpha$, then as before, sorting the images into the correct order by case a swaps. The images originally formed of $\gamma$ and $\beta$ are the same in both cases, therefore the final order is the same, so $X_i = Y_j$ for some $i$ and $j$.

- $Y \sim_c X$ forwards, $\alpha;@_1 < \gamma$, and $\alpha;@_2 < \beta$: Let $x$ be the variable involved in the $\beta$-reduction at $\alpha$, so that the sub-skeleton at $\alpha$ is $(\lambda x.A)B$ for some $A, B$, where $\gamma = \alpha;@_1;\lambda;\gamma'$ and $A \to A'$ at $\gamma'$, and $\beta = \alpha;@_2;\beta'$, and $B \to B'$ at $\beta'$. In this case, $(X_i)_i$ starts by swapping $\beta$ below $\alpha$ by case e, producing one image of $\beta$ for each $x$ in $A'$, then swapping $\gamma$ below $\alpha$ by case d, producing a single image $\alpha;\gamma'$. For each of the instances of $x$ in $A$ left of $\gamma'$, the reduction at $\alpha;\gamma'$ is then swapped below the corresponding image of $\beta$. The sequence $(Y_j)_j$ starts by swapping $\gamma$ below $\alpha$ then $\beta$ below $\alpha$, but in this case the images of $\beta$ produced may be different. In this case the swap happens earlier in the reduction sequence than $\gamma$, so that the body of the lambda is still $A$ rather than $A'$, and the reduction at $\gamma$ may rearrange or change the number of instances of $x$. Next, each of the images of $\beta$ to the right of $\alpha;\gamma'$ are swapped with it by case a. At this point, the next few reductions before $\alpha;\gamma'$ will in general be images of $\beta$ at positions inside $\alpha;\gamma'$. The subskeleton at position $\alpha;\gamma'$ before these reductions is $(A|\gamma')[B/x]$, then by the images of $\beta$ this reduces to $(A|\gamma')[B'/x]$, then it reduces at its root position ($\alpha;\gamma'$ in the overall skeleton) to $(A'|\gamma')[B'/x]$. For each position of an $x$ in $(A|\gamma')$, there are 0 or more corresponding positions of $x$ in $(A'|\gamma')$, depending on what type of reduction $A \to A'$ is. Because $B$ and $B'$ cannot contain any instances of the variable involved in the reduction $A \to A'$ (if it's a $\beta$-reduction or Y-reduction), each instance of $B$ in $(A|\gamma')[B/x]$ has the same set of images in $(A'|\gamma')[B/x]$ as the corresponding $x$ in $(A|\gamma')$. When these images of $\beta$ that overlap with the image of $\gamma$ are swapped below it, each of them has its own images, one at each position of $B$ in $(A'|\gamma')[B/x]$ corresponding to its original copy of $B$ in $(A|\gamma')[B/x]$. After all of them are swapped below $\alpha;\gamma'$ (and rearranged among themselves by case a), there is therefore one reduction at each copy of $B$ in $(A'|\gamma')[B/x]$, i.e. one position $\beta'$ relative to each $x$ in $(A'|\gamma')$. They are therefore the same as the images of $\beta$ in $X_1$ that overlap with $\alpha;\gamma'$, because those are also one reduction at position $\beta'$ relative to each $x$ in $(A'|\gamma)$. Both $(X_i)_i$ and $(Y_j)_j$ therefore eventually reach a point where the relevant sub-reduction-sequence is $\alpha$, the images of $\beta$ to the left of $\alpha;\gamma'$, $\alpha;\gamma'$ itself, the aforementioned images of $\beta$ that overlap with $\alpha;\gamma'$, then all the images of $\beta$ to the right of $\alpha;\gamma'$.

- $Y \sim_c X$ forwards, $\gamma$ and $\beta$ are disjoint and both $> \alpha$, and the reduction at $\alpha$ is a Y-reduction: This is similar to the previous case, but more complicated, so it will need some definitions to explain properly what's going on. Let the term at $\alpha$ before any of the reductions be $Y\lambda x.A$, let $\gamma = \alpha;Y;\lambda;\gamma'$, let $\beta = \alpha;Y;\lambda;\beta'$, let $A|\beta' = B \to B'$ with the reduction at $\cdot$, let $A|\gamma' = C \to C'$ with the reduction at $\cdot$, let the positions in $A$ where $x$ occurs that are left of $\gamma'$, between $\gamma'$ and $\beta'$ (but still disjoint from both), and right of $\beta'$ respectively be $(\delta_i^l)$, $(\delta_i^m)$ and $(\delta_i^r)$, let the positions in $B$, $C$, $B'$ and $C'$ respectively where $x$ occurs be $(\delta_i^B)$, $(\delta_i^C)$, $(\delta_i^{B'})$ and $(\delta_i^{C'})$ (so that all the positions in $A$ where $x$ occurs, in left-to-right order, are $(\delta_i^l)_i, (\delta_i^C)_i, (\delta_i^m)_i, (\delta_i^B)_i, (\delta_i^r)_i$). The sequence $(X_i)_i$ starts by swapping $\beta$ then $\gamma$ below $\alpha$. With the shorthand that $\gamma'(\epsilon) = \alpha;\lambda;@_1;\epsilon;Y;\lambda;\gamma'$, $\gamma'_0 = \alpha;\lambda;@_1;\gamma'$, and similarly for $\beta'$, the relevant portion of $X_2$ is then $\alpha$, $(\gamma'(\delta_i^l))_i$, $\gamma'_0$, $(\gamma'(\delta_i^{C'}))_i$, $(\gamma'(\delta_i^m))_i$, $(\gamma'(\delta_i^B))_i$, $(\gamma'(\delta_i^r))_i$, $(\beta'(\delta_i^l))_i$, $(\beta'(\delta_i^{C'}))_i$, $(\beta'(\delta_i^m))_i$, $\beta'_0$, $(\beta'(\delta_i^{B'}))_i$, $(\beta'(\delta_i^r))_i$. Next in $(X_k)_k$, for each $i$, $\gamma'(\delta_i^r)$ are swapped past all the images of $\beta$ until it is immediately before $\beta'(\delta_i^r)$ by case a, then for each $i$, $\gamma'(\delta_i^B)$ is swapped past all the images of $\beta$ until $\beta'_0$ by case a, then swapped with $\beta'_0$ by one of the other cases depending on what type of reduction $B \to B'$ is and the relative positions. Even if it's case d or f, all of the images of $\gamma'(\delta_i^B)$ are disjoint from each other (and from the positions of the other reductions after $\beta'_0$) because the redex, $C$, doesn't contain any instances of the variable involved in the reduction $B \to B'$. Expanding the definitions, this is swapping $\alpha;\lambda;@_1;\delta_i^B;Y;\lambda;\gamma'$ with $\alpha;\lambda;@_1;\beta'$ at a point in the reduction sequence where the subskeleton at $\alpha;\lambda;@_1;\beta'$ is $B$ with a mixture of $Y\lambda x.A$ and $Y\lambda x.A[C'/\gamma']$ substituted for its occurrences of $x$, and one of these occurrences of $x$ is at $\delta_i^B$. For each $\delta_i^B$, there are some corresponding $\delta_k^{B'}$, where the images of the $x$ at $\delta_j^B$ end up after the reduction $B \to B'$. These cover all the $\delta_k^{B'}$, and uniquely, and for each $\delta_i^B$, the images of $\zeta;\delta_i^B;\theta$ after swapping it with $\zeta;\beta'$ are precisely $(\zeta;\delta_k^{B'};\theta)$ for those same values of $k$, therefore after swapping all of the $(\gamma'(\delta_i^B))_i$ down past $\beta'_0$, their images are $(\gamma'(\delta_i^{B'}))_i$. After swapping with $\beta'_0$, each of these images than swaps by case a to take its place among the $(\beta'(\delta_i^{B'}))_i$. Next up in $(X_i)_i$, the $(\gamma'(\delta_i^m))_i$s and the $(\gamma'(\delta_i^{C'}))_i$s swap down past some images of $\beta$ they're disjoint from to take their place before their matching image of $\beta$, then $\gamma'_0$ swaps past the $(\beta'(\delta_i^l))_i$, stopping immediately before $\beta'(\delta_1^{C'})$, then the $(\gamma'(\delta_i^l))_i$ and the $(\beta'(\delta_i^l))_i$ mix. The end result of the rearrangement of (this subsequence of) the reduction sequence is therefore $\alpha$, $(\gamma'(\delta_i^l), \beta'(\delta_i^l))_i$, $\gamma'_0$, $(\gamma'(\delta_i^{C'}), \beta'(\delta_i^{C'}))_i$, $(\gamma'(\delta_i^m), \beta'(\delta_i^m))_i$, $\beta'_0$, $(\gamma'(\delta_i^{B'}), \beta'(\delta_i^{B'}))_i$, $(\gamma'(\delta_i^r), \beta'(\delta_i^r))_i$.

The other sequence, $(Y_j)_j$, proceeds similarly, but with the images of $\gamma$ starting out after the images of $\beta$. It swaps all the images of $\beta$ that are right of $\gamma_0$ to the appropriate places by case a, then swaps all the $(\beta'(\delta_i^C)_i$ first to then past $\gamma'_0$, resulting in $(\beta'(\delta_i^{C'}))_i$ for the same reason as with swapping those images of $\gamma$ that overlapped with $\beta'_0$ past it in the $(X_i)_i$ case above, then finally swapping the $(\beta'(\delta_i^l))_i$s and the $(\gamma'(\delta_i^l))_i$s into the correct order. As required, this produces the same result as in $(X_i)_i$.

- $Y \to X$ forwards, and $\alpha < \gamma < \beta$: Let $\gamma'$ be such

that $\gamma = \alpha; \mathsf{if}_i; \gamma'$, $\alpha; @_1; \lambda; \gamma'$, $\alpha; @_2; \gamma'$ or $\alpha; \mathsf{Y}; \lambda; \gamma'$, so that $\gamma'$ is the freely varying later part of $\gamma$ as in the definition of $\sim_c$, whichever case applies to swapping $\gamma$ and $\alpha$. Similarly, let $\beta'$ be the freely varying part of $\beta$ relative to $\gamma$. Let the images of $\gamma$ after swapping it with $\alpha$ (where this swap occurs later in the reduction sequence than $\beta$) then be $(\alpha; \delta_i; \gamma')_i$, and the images of $\beta$ after swapping it with $\gamma$ be $(\gamma; \epsilon_i; \beta')_i$. Depending on the skeleton at $\alpha$, and $\gamma$'s position within it, $(\delta_i)_i$ may be empty (case b), a singleton containing $\cdot$ (cases c and d), all the positions of the relevant variable in the lambda in the skeleton at $\alpha$ (case e), or $\lambda; @_1$ and $\lambda; @_1; \zeta; \mathsf{Y}; \lambda$ for each position $\zeta$ of the relevant variable (case f), but in all cases (except a, which is excluded because none of $\alpha, \gamma$ and $\beta$ are disjoint) the set of images has this general structure. Furthermore, the positions $(\delta_i)_i$ are determined only by the general position of $\gamma$ within $\alpha$ (the part excluded from $\gamma'$), and the positions of the relevant variable within the skeleton at $\alpha$. They do not depend on $\alpha$ or $\gamma$, so that if both initial positions were moved somewhere else the relative positions of the images of $\gamma$ would be unaffected, and similarly if some position within $\gamma$ were swapped with $\alpha$, the relative positions of its images would be the same (except to the extent that, in case f, the positions of the relevant variable are taken after the inner reduction takes place, so that some of those may still change). Because there are so many of them and they don't actually affect the multiset of positions where reductions take place, we will be ignoring swaps by case a in the proof of this case, and just assuming that all the final positions end up in the correct order.

The final set of reduction positions that we will be proving both $(X_i)_i$ and $(Y_j)_j$ reach is $\alpha, (\alpha; \delta_i; \gamma', (\alpha; \delta_i; \gamma'; \epsilon_j; \beta')_j)_i$, where as usual the sequences are expanded out to the full list. For the sequence $(X_i)_i$, first the images of $\beta$ in $X_0$ are $(\gamma, \epsilon_i, \beta')$. For each of these in turn, it is swapped with $\alpha$. By the fact mentioned above that sub-positions are mapped to the same set of images except in case f, the images of $\gamma, \epsilon_j, \beta'$ after this swap are $(\alpha; \delta_i^j; \gamma', \epsilon_j, \beta')_i$, where $(\delta_i^j)_i$ is the same as $(\delta_i)_i$ except that in the case where this is not the first image of $\beta$ swapped below $\alpha$ and the reduction at $\alpha$ is a Y-reduction (and therefore swaps with it proceed by case f), the positions of the relevant variable are taken after the reductions at $\gamma$ and $\gamma; \epsilon_k; \beta'$ for $k \leq j$ (this implies that the first of these, $\delta_i^n = \delta_i$). All the images $\alpha; \delta_i^j; \gamma'; \epsilon_j; \beta'$ where $\delta_i^j > \lambda; @_1; \gamma'; \epsilon_{j+1}; \beta'$ are then swapped with $\alpha; \lambda; @_1; \gamma'; \epsilon_{j+1}; \beta'$ (which is $\alpha; \delta_i^{j+1}; \gamma'; \epsilon_{j+1}; \beta'$ for some $i$). As in the case where $\alpha; \mathsf{Y} < \gamma, \beta$ and $\gamma$ and $\beta$ are disjoint, the effects of these swaps may duplicate or delete the reductions in such a way that all the remaining images of $\gamma; \epsilon_j; \beta'$ are $(\alpha; \delta_i^{j+1}; \gamma'; \epsilon_j; \beta')$. This is repeated with $\delta_i^{j+2}$ and so on until these images are all in the correct order, at which point they are $(\alpha; \delta_i; \gamma'; \epsilon_j; \beta')_i$. After all of these are done, $\gamma$ is swapped with $\alpha$, yielding $(\alpha; \delta_i^{-1}; \gamma')_i$. As with the images of $\beta$, the subset of these which are inside $\alpha; \lambda; @_1; \gamma'; \epsilon_j; \beta'$ for each $j$ in turn are swapped with it, until they are all in the correct order and the images left of $\gamma$ are $(\alpha; \delta_i; \gamma')_i$. At this point, $X_i$ has the desired value.

The sequence $(Y_j)_j$ proceeds similarly. First $\gamma$ swaps with $\alpha$, yielding $\alpha$ and $(\alpha; \delta_i; \gamma')_i$, then $\beta$ swaps with $\alpha$, yielding one image of $\beta$ for each $\delta_i$, except that, in case f, those images at positions inside $\alpha; \lambda; @_1; \gamma'$ may differ because they are earlier in the reduction sequence than any of the images of $\gamma$. Each of the images of $\beta$ in turn is moved to the corresponding image of $\gamma$, except that those overlapping with $\alpha; \lambda; @_1; \gamma'$ may be duplicated or deleted on the way. The images of $\beta$ after just this process (which does not actually correspond to any term in $(Y_j)_j$, but it's sufficiently independent that the order doesn't matter that much) are $(\alpha; \delta_i; \gamma'; \beta'')$ (where $\gamma; \beta'' = \beta$). The images of these for each $i$, after swapping with $\alpha; \delta_i; \gamma'$, is $(\alpha; \delta_i; \gamma'; \epsilon_j; \beta')_j$, because the reductions produced by a swap are unaffected by its overall location, and the skeleton at $\alpha; \delta_i; \gamma'$ at the relevant point in the sequence is equal to the skeleton at $\gamma$ initially, therefore this is equivalent to immediately swapping $\beta$ and $\gamma$ as in $Y \sim_c X$ (except that in the case where the reduction at $\alpha$ is a Y-reduction, the skeleton at $\alpha; \lambda; @_1; \gamma'$ is not actually equal to the skeleton initially at $\gamma$ because something was substituted in for the variable bound at $\alpha; \mathsf{Y}$, but this doesn't affect the variable involved in the reduction at $\gamma$, so this doesn't actually matter). Therefore also in this case, eventually the set of reductions in the relevant portion of $Y_j$ is $\alpha, (\alpha; \delta_i; \gamma', (\alpha; \delta_i; \gamma'; \epsilon_j; \beta')_j)_i$, therefore it is equal to some $X_i$.

In summary, if the swap in $Y \sim_c X$ doesn't overlap with the swap $X \sim_c X_0$, the sequence of swaps can be rearranged until it does, and in any other case, there is some $Y_j$ that's equal to some $X_i$, therefore these sequences reach the same end point and if $Y \sim_c X$, then $Y_{cbv} = X_{cbv}$, and by chaining these together, if $Y \sim_c^* X$, $Y_{cbv} = X_{cbv}$.

c) : If $Y$ is in CBV order and $Y \sim_c^* X$, then $Y_{cbv} = X_{cbv}$ but also $Y_{cbv} = Y$ because the sequence $(Y_i)_i$ terminates immediately therefore $X_{cbv}$ is the unique reduction sequence in call-by-value order that's related to $X$ by $\sim_c^*$. $\square$

**Lemma VI.7.** *The relation $\sim$ is defined on $L_0(M)$ with reference to a particular starting term $M$, so different versions, $\sim_M$ and $\sim_N$, can be defined starting at different terms. If $M \to N$, then $\sim_N^*$ is equal to the restriction of $\sim_M^*$ to $L_0(N)$.*

*Proof.* $\sim_N^*$ is trivially a subset of $\sim_M^*$ because $\sim_N$ is a subset of $\sim_M$.

In the other direction, suppose $(N_1, \alpha_1) \sim_M^* (N_2, \alpha_2)$ where both $N_1$ and $N_2$ are descendants of $N$. By Lem. D.1, take $(N_1, \alpha_1) \sim_{p,M}^* (N_1', \alpha') \sim_{c,M}^* (N_2', \alpha') \sim_{p,M}^* (N_2, \alpha_2)$.

The $\sim_p^*$ steps remain within $Rch(N)$, and $\sim_p$ does not depend on the history of the reduction sequences, therefore $(N_1, \alpha_1) \sim_{p,N}^* (N_1', \alpha') \sim_{c,M}^* (N_2', \alpha') \sim_{p,N}^* (N_2, \alpha_2)$.

Let $X$ be the call-by-value reduction sequence related to both $N_1'$ and $N_2'$ by $\sim_{c,M}$ given by Lem. D.2. As $M \to N$ is the first reduction in the sequence $N_1'$, it is the last to be affected by the $\sim_{c,M}$ sequence $N_1' \sim_{c,M}^* X$ given by Lem. D.2 therefore it can be split into $N_1' \sim_{c,M}^* Y_1 \sim_{c,M}^* X$ where $Y_1$ is in CBV order except possibly for its first reduction, which is still $M \to N$. Let the position of the reduction $M \to N$ be $\beta$. The rearrangement of the reduction sequence $Y_1 \sim_{c,M} X$ consists of moving the reduction at $\beta$ down past the other reductions in $Y_1$, possibly duplicating or deleting it in the process, but not affecting the positions or the correct order for any of the other reductions. The reductions derived from $M \to N$ can be identified as follows:

A position in some (reduction sequence of) term(s) is *derived from* another if it is related by the reflexive transitive closure of $\leadsto$, where $(V, \gamma) \leadsto (W, \delta)$ just if $V \to W$ and one of the following cases holds:

- $(V, \gamma) \sim_p (W, \delta)$
- $\gamma = \delta$, $V \to W$ at $\epsilon$, and $\epsilon > \gamma$
- $V \mid \epsilon = (\lambda x.U)Z, U \mid \zeta = x, \gamma = \epsilon; @_2; \theta$ and $\delta = \epsilon; \zeta; \theta$
- $V \mid \epsilon = \mathsf{Y}(\lambda x.U), \gamma = \epsilon; \mathsf{Y}; \lambda; \zeta$ and $\delta = \epsilon; \lambda; @_1; \zeta$
- $V \mid \epsilon = \mathsf{Y}(\lambda x.U), U \mid \zeta = x, \gamma = \epsilon; \theta$ and $\delta = \epsilon; \lambda; @_1; \zeta; \theta$

Crucially, the reductions in $X$ can be partitioned into 2 sets: those at positions derived from $(M, \beta)$, and those at positions equal to the reductions in $Y_1$ (in the same order as they occur in $Y_1$). Using the same construction for $N_2'$ shows that the positions of the reductions in $Y_2$ are also the positions of the reductions in $X$ other than those derived from $(M, \beta)$, therefore $Y_1 = Y_2$ therefore $N_1' \sim_{c,M}^* Y_1 \sim_{c,M}^* N_2'$ and every reduction sequence in this sequence starts with $M \to N$ therefore $N_1' \sim_{c,N}^* N_2'$.

Combining this with the $\sim_{p,N}^*$s at the beginning and end then yields the desired result that $(N_1, \alpha_1) \sim_N^* (N_2, \alpha_2)$, therefore the restriction of $\sim_M^*$ to $\sim_N$'s domain $(Rch(N))$ is a subset of $\sim_N^*$ therefore the two versions of $\sim$ match on this domain, as desired. $\square$

**Lemma VI.9.** *The relation $\Rightarrow$ is confluent.*

*Proof.* Suppose that $(M, s) \Rightarrow^* (M_1, s_1)$ and also $(M, s) \Rightarrow^* (M_2, s_2)$, then it is required to prove that there is some $(M', s')$ such that both $(M_1, s_1) \Rightarrow^* (M', s')$ and $(M_2, s_2) \Rightarrow^* (M', s')$. First consider the special case where $(M, s) \Rightarrow (M_1, s_1)$ and $(M, s) \Rightarrow (M_2, s_2)$, with only a single step in each case. Let the positions of the redexes in $M \to M_1$ and $M \to M_2$ be $\alpha_1$ and $\alpha_2$ respectively.

First consider the case that $\alpha_1$ and $\alpha_2$ are disjoint. Let $M_1 = M[X_1/\alpha_1]$ and similarly for $X_2$, then let $M' = M[X_1/\alpha_1][X_2/\alpha_2]$. As the positions are disjoint, the substitutions commute and both $M_1$ and $M_2$ reduce (with $\to$) to $M'$. Let $s' = s \circ i(M \to M_1) \circ i(M_1 \to M')$. The injection $i(M \to M_1) \circ i(M_1 \to M')$ consists of prepending $M \to M_1 \to M'$ to each reduction sequence, but by case a of $\sim_c$, this is equivalent to prepending $M \to M_2 \to M'$, which is $i(M \to M_2) \circ i(M_2 \to M')$. In the case that the reduction $M \to M_2$ isn't a sample-reduction, this is enough to establish that $(M_1, s_1) \Rightarrow (M', s')$ (and similarly if the redex of $M \to M_1$ isn't sample, $(M_2, s_2) \Rightarrow (M', s')$). If it is sample though, in order for it to be the case that $(M_1, s_1) \Rightarrow (M', s')$, it is additionally necessary that $X_2$, the result of the reduction at $\alpha_2$, be $s_1(M_1, \alpha_2)$, which follows from the fact that $(M, \alpha_2) \sim (M_1, \alpha_2)$ by case 1 of $\sim_p$. The case that the reduction $M \to M_1$ is a sample-reduction is similar.

The case that $\alpha_1 = \alpha_2$ is trivial, because there is at most one possible $\Rightarrow$ reduction at any given position, therefore $(M_1, s_1) = (M_2, s_2)$ already.

The remaining case is that $\alpha_1 < \alpha_2$ or $\alpha_1 > \alpha_2$. Assume without loss of generality that $\alpha_1 < \alpha_2$. For each possible case of what type of redex $M \mid \alpha_1$ is, and $\alpha_2$'s position within it, there is a corresponding case of $\sim_c$, and similarly to the case where $\alpha_1$ and $\alpha_2$ are disjoint, the term $O_1 = O_2$ from the definition of $\sim_c$ is a suitable value of $M'$. The term $M \mid \alpha_1$ can't be sample, because it has strict subterms, but the case that $M \mid \alpha_2 = $ sample is still somewhat more complicated. $\alpha_2$ can't be within $\alpha_1; @_2$ or $\alpha_1; \mathsf{Y}; \lambda$ (cases e and f of $\sim_c$) because if $\alpha_2 > \alpha_1; @_2$, $M \mid \alpha_1; @_2$ must be a value, therefore $\alpha_2 > \alpha_1; @_2; \lambda$ and those are not valid positions to reduce a sample. The case that $\alpha_2$ is within the branch of an if statement that is deleted by the reduction at $\alpha_1$ (case b) doesn't actually present a problem because there is no corresponding reduction to $M \to M_2$ in the other branch, which leaves cases c and d, that $\alpha_2$ is in the $\mathsf{if}(\cdot, \cdot, \cdot)$ branch that isn't deleted, and that $\alpha_2 < \alpha_1; @_1; \lambda$. The values that the samples take in these cases match because $(M, \alpha) \sim_p (M_1, \alpha; \beta)$ by cases 3 and 2 of $\sim_p$ respectively.

*d)* : In the case that $(M, s) \Rightarrow^* (M_1, s_1)$ or $(M, s) \Rightarrow^* (M_2, s_2)$ by multiple steps, it is possible to repeatedly replace a pair of the form $A \Leftarrow B \Rightarrow C$ by $A \Rightarrow^* D \Leftarrow^* C$ by the construction above, but it is not immediate that this process terminates, because each $\Rightarrow$ or $\Leftarrow$ may be replaced by multiple, so the sequence of $\Rightarrow$s and $\Leftarrow$s from $(M_1, s_1)$ to $(M_2, s_2)$ may get longer at some steps. However, termination can be proved by noting that the structure of this process is identical to the process of swapping $\sim_\uparrow$ and $\sim_\downarrow$ steps in Lem. D.1. To be more precise, in this case there is a sequence of $\Leftarrow$ and $\Rightarrow$ steps, each of which has an associated reduction position and initial and final terms, and if a $\Leftarrow$ immediately precedes a $\Rightarrow$, they may be swapped to produce some number of $\Rightarrow$s, followed by some number of $\Leftarrow$s. In the case of Lem. D.1, there is a sequence of $\sim_\uparrow$ and $\sim_\downarrow$ steps, each of which has an associated reduction position and initial and final skeletons, and if a $\sim_\uparrow$ immediately precedes a $\sim_\downarrow$, they may be swapped to produce some number of $\sim_\downarrow$s followed by some number of $\sim_\uparrow$s. In both cases, the number and reduction positions of the resultant steps are determined by the case of $\sim_c$ that matches the way the initial reduction positions overlap, and the initial skeleton (or the skeleton of the initial term). The same argument that the process in Lem. D.1 terminated is therefore applicable here too. At every stage, the sequence of reductions that results from one of the initial $\Rightarrow$s is a parallel sequence, and swapping a parallel sequence of $\Rightarrow$s with a $\Leftarrow$ always terminates, therefore each $\Rightarrow$ in turn can be moved past all of the $\Leftarrow$s, and the process as a whole will terminate in a state where all of the $\Rightarrow$s precede all of the $\Leftarrow$s, i.e. a pair of reduction sequences $(M_1, s_1) \Rightarrow^* (M', s') \Leftarrow^* (M_2, s_2)$. □

**Lemma D.3.** *If $A$ is some descendant of $M$ and $(A, \gamma) \sim^* (M, \delta)$, then $(A, \gamma) \sim_p^* (M, \delta)$, with the length of the reduction sequences decreasing by one each step from $A$ to $M$.*

*Proof.* This is a simple induction on $\sim^*$. In the base case, $(A, \gamma) = (M, \delta)$ therefore $(A, \gamma) \sim_p^* (M, \delta)$ trivially. Otherwise, suppose that $(M, \delta) \sim_p^* (B, \epsilon) \sim (A, \gamma)$. Either $(B, \epsilon) \sim_p (A, \gamma)$ or $(B, \epsilon) \sim_c (A, \gamma)$. In the first case, either $B \to A$, in which case the result follows directly, or $A \to B$, in which case the fact that each (reduction sequence of) skeleton(s) has only one parent implies that a $(M, \delta) \sim_p^* (A, \gamma)$ directly as a subsequence of the path to $(B, \epsilon)$.

In the $\sim_c$ case, consider the definition of $\sim_c$. Either $O_1' = A$ and $O_2' = B$ or vice-versa. As $M \to^* N \to^* O_1 \to^* B$ and $(M, \delta) \sim_p^* (B, \epsilon)$, and $\sim_p$ only relates positions in a term and its parent, there are some positions $\zeta, \theta$ such that $(M, \delta) \sim_p^* (N, \zeta) \sim_p^* (O_1, \theta) \sim_p^* (B, \epsilon)$. It follows that $(O_2, \theta) \sim_p^* (A, \gamma)$ by following the same path, therefore it suffices to provide the only missing portion of the path from $M$ to $A$, i.e. to prove that $(N, \zeta) \sim_p^* (O_2, \theta)$ given $(N, \zeta) \sim_p^* (O_1, \theta)$ (or vice-versa).

If $\zeta$ is disjoint from all the positions of reduction from $N$ to $O_1$ and $O_2$ (and consequently $\zeta = \theta$), this follows from case 1 of $\sim_p$. Otherwise, this can be proved by taking cases from the definition of $\sim_c$. This is rather long, but all of the cases are similar. The general idea is that the reductions from $N$ to $O_1$ correspond to the reductions from $N$ to $O_2$, so that if a position is related by $\sim_p$ across that reduction, it is related in the other branch for the same reason. Case d, where $B = O_1'$ rather than $O_2'$, is given here in more detail as an illustrative example:

Let $I$ be $N$ reduced at $\alpha$, and $J$ be $N$ reduced at $\alpha; @_1; \lambda; \beta$, so that $N \to I \to O_1$ and $N \to J \to O_2$. All of the reduction positions are $\geq \alpha$, and $\zeta$ is not disjoint from all of them, therefore $\zeta$ is not disjoint from $\alpha$. Let $\iota$ be the position such that $(N, \zeta) \sim_p (I, \iota) \sim_p (O_1, \theta)$. The fact that $(N, \zeta) \sim_p (I, \iota)$ implies that $\zeta > \alpha; @_1; \lambda$. Let $\zeta = \alpha; @_1; \lambda; \kappa$ and $\theta = \alpha; \kappa'$. If $\kappa$ is disjoint from $\beta$, then $(N, \zeta) \sim_p (J, \zeta) \sim_p (O_2, \alpha; \kappa') = (O_2; \theta)$ by cases 1 and 2 of $\sim_p$. In the other case, that $\kappa$ isn't disjoint from $\beta$, $\kappa > \beta$ because none of the positions $\leq \alpha; @_1; \lambda; \beta$ in $N$ are related to any position in $I$ by $\sim_p$. As $(I, \alpha; \kappa) \sim_p (O_1, \alpha; \kappa')$ (with the redex at $\alpha; \beta$), for exactly the same reason $(N, \alpha; @_1; \lambda; \kappa) \sim_p (J, \alpha; @_1; \lambda; \kappa')$ (with the redex at $\alpha; @_1; \lambda; \beta$). Because $(N, \alpha; @_1; \lambda; \kappa) \sim_p (J, \alpha; @_1; \lambda; \kappa')$, $N|\alpha; @_1; \lambda; \kappa = J|\alpha; @_1; \lambda; \kappa'$, and $N|\alpha; @_1; \lambda; \kappa \neq$ the variable of $N|\alpha; @_1$, therefore $J|\alpha; @_1; \lambda; \kappa'$ is also not the variable therefore $(J, \alpha; @_1; \lambda; \kappa') \sim_p (O_2, \alpha; \kappa') = (O_2, \theta)$ by case 2 of $\sim_p$. Combining these results, $(N, \zeta) \sim_p^* (O_2; \theta)$ as desired. □

**Lemma D.4.** *If $M \to N$, with the redex at position $\alpha$, then no position in any term reachable from $N$ is related by $\sim^*$ to $(M, \alpha)$.*

*Proof.* Suppose on the contrary that $(M, \alpha) \sim^* (A, \beta)$, where $N \to^* A$, then by Lem. D.3, $(M, \alpha) \sim_p^* (A, \beta)$. $M \neq A$ therefore $(M, \alpha) \sim_p$ some position in $N$, but in all the cases of the definition of $\sim_p$, no position in the child term is related to the position of the redex. □

Essentially what Lem. D.4 demonstrates is that the samples taken during any reduction sequence are independent of each other. This is made more precise in the following lemmas.

**Lemma D.5.** *For any skeletons $M \to N$, with the redex at position $\alpha$, and measurable set of samples $S \subset I^{L_s(N)}$, $\mu(S) = \mu(\{s \in I^{L_s(M)} \mid s \circ i(M \to N) \in S\})$, and furthermore, if $M|\alpha =$ sample, for any $S \subset I \times I^{L_s(N)}$, $\mu(S) = \mu(\{s \in I^{L_s(M)} \mid (s(M, \alpha), s \circ i(M \to N)) \in S\})$.*

*Proof.* If the result holds for all sets $S$ of the form $\{s \in I^{L_s(N)} \mid \forall j : s(j) \in x_i\}$, where $(x_j)_{j \in J}$ is a family of measurable subsets of $I$ indexed by some finite set $J \subset L_s(N)$ of potential positions, it also holds for all other $S$ by taking limits and disjoint unions.

Take such a $(x_j)_{j \in J}$, and define $K = i(M \to N)[J]$. Because $i(M \to N)$ is injective, it defines a bijection between $J$ and $K$. We can then calculate the measure $\mu(\{s \in I^{L_s(M)} \mid s \circ i(M \to N) \in S\}) = \mu(\{s \in I^{L_s(M)} \mid \forall k : s(k) \in x_{i(M \to N)^{-1}(k)}\}) = \prod_{k \in K} \mu_I(x_{i(M \to N)^{-1}(k)}) = \prod_{j \in J} \mu_I(x_j) = \mu(S)$.

In the case that $M|\alpha =$ sample, we can similarly consider only those sets $S$ of the form $x_{\mathsf{sample}} \times \{s \in I^{L_s(N)} \mid \forall j : s(j) \in x_i\}$. The position $\alpha$ in $M$ is not related to any position in $L_s(N)$, therefore $(M, \alpha) \notin K$. Again, we can calculate the measure $\mu(\{s \in I^{L_s(M)} \mid (s(M, \alpha), s \circ i(M \to N)) \in S\}) = \mu(\{s \in I^{L_s(M)} \mid s(M, \alpha) \in x_{\mathsf{sample}}, \forall k : s(k) \in x_{i(M \to N)^{-1}(k)}\}) = \mu_I(x_{\mathsf{sample}}) \prod_{k \in K} \mu_I(x_{i(M \to N)^{-1}(k)}) = \mu_I(x_{\mathsf{sample}}) \prod_{j \in J} \mu_I(x_j) = \mu(S)$. □

**Lemma D.6.** *For any initial term $M$, reduction strategy $f$ on $M$, natural number $n$, skeleton $N$ with $k$ holes, measurable set $T \subset \mathbb{R}^k$ and measurable set $S$ of samples in $I^{L_s(N)}$, $\mu(\{s \in I^{L_s(M)} \mid \exists r \in T, s' \in S : (M, s) \Rightarrow_f^n (N[r], s')\}) = \mu(\{s \in I^{L_s(M)} \mid \exists r \in T, s' \in I^{L_s(N)} : (M, s) \Rightarrow_f^n (N[r], s')\})\mu(S)$*

*Proof.* Suppose, to begin with, that $n = 0$. Either $\exists r \in T : N[r] = M$, in which case both sides of the equation are $\mu(S)$, or there is no such $r$, in which case both sides of the equation are 0.

For $n > 0$, suppose for induction that the lemma is true for $n - 1$, for all $N, T$ and $S$. If $\exists r \in T, s' \in S : (M, s) \Rightarrow_f^n (N[r], s')$, the $r$ and $s'$ are necessarily unique, therefore we may assume that $T$ is the product of $k$ measurable setsets of $\mathbb{R}$, $(T_j)_{0 \leq j < k}$, as all of the other cases follow by taking unions and limits of these rectangles. Let

$$R_s = \{Sk(O) \mid O \in Rch(M), r \in \mathbb{R}^k, O|f(O[r]) = \mathsf{sample}, O \to N[r]\}$$
$$R_d = \{Sk(O) \mid O \in Rch(M), r \in \mathbb{R}^k, O|f(O[r]) \neq \mathsf{sample}, O \to N[r]\}.$$

Together, $R_s$ and $R_d$ contain all of the skeletons of terms that can reduce to $N$, and they're disjoint. For each of these skeletons, there is a map $g_O$ from the reals in $O$ to the reals in $N$, i.e. $g_O(r) = r'$ if $O[r] \to N[r']$ for $O \in R_d$, and $g_O(s, r) = r'$ if $O[r] \to N[r']$ with $s$ as the value the sample takes in the reduction. All of these functions are measurable, as they are just rearrangements of vector components, or in the case that $O|f(O[r]) = \underline{x}(\mathsf{X}, \ldots, \mathsf{X})$, a combination of the measurable function $x$ with a rearrangement of vector components. In the case that $O \in R_s$, the function $g_O$ simply inserts the sample into the vector of reals at a certain index, call it $h_O$.

$$\mu(\{s \in I^{L_s(M)} \mid \exists r \in T, s' \in S : (M,s) \Rightarrow_f^n (N[r], s')\})$$
$$=\mu(\{s \in I^{L_s(M)} \mid \exists O \in R_s, r \in \mathbb{R}^{k-1}, s' \in I^{L_s(O)}, r' \in T, s'' \in S : (M,s) \Rightarrow_f^{n-1} (O[r], s') \Rightarrow_f (N[r'], s'')\})$$
$$+\mu(\{s \in I^{L_s(M)} \mid \exists O \in R_d, r \in \bigcup_l \mathbb{R}^l, s' \in I^{L_s(O)}, r' \in T, s'' \in S : (M,s) \Rightarrow_f^{n-1} (O[r], s') \Rightarrow_f (N[r'], s'')\})$$
$$= \sum_{O \in R_s} \mu(\{s \in I^{L_s(M)} \mid \exists r \in \mathbb{R}^{k-1}, s' \in I^{L_s(O)} :$$
$$g_O(r, s'(O, f(O[r]))) \in T, s' \circ i(O[r] \to N[g_O(r, s'(O, f(O[r])))]) \in S, (M,s) \Rightarrow_f^{n-1} (O[r], s')\})$$
$$+ \sum_{O \in R_d} \mu(\{s \in I^{L_s(M)} \mid \exists r \in \bigcup_l \mathbb{R}^l, s' \in I^{L_s(O)} :$$
$$g_O(r) \in T, s' \circ i(O[r] \to N[g_O(r)]) \in S, (M,s) \Rightarrow_f^{n-1} (O[r], s')\})$$
$$= \sum_{O \in R_s} \mu(\{s \in I^{L_s(M)} \mid \exists r \in \prod_{j \neq h(O)} T_j, s' \in I^{L_s(O)} :$$
$$s'(O, f(O[r])) \in T_{h(O)}, s' \circ i(O[r] \to N[g_O(r, s'(O, f(O[r])))]) \in S, (M,s) \Rightarrow_f^{n-1} (O[r], s')\})$$
$$+ \sum_{O \in R_d} \mu(\{s \in I^{L_s(M)} \mid \exists r \in \bigcup_l \mathbb{R}^l, s' \in I^{L_s(O)} :$$
$$g_O(r) \in T, s' \circ i(O[r] \to N[g_O(r)]) \in S, (M,s) \Rightarrow_f^{n-1} (O[r], s')\})$$
$$= \sum_{O \in R_s} \mu(\{s \in I^{L_s(M)} \mid \exists r \in \prod_{j \neq h(O)} T_j, s' \in I^{L_s(O)} : (M,s) \Rightarrow_f^{n-1} (O[r], s')\})$$
$$\mu(\{s' \in I^{L_s(O)} \mid s'(O, f(O[r])) \in T_{h(O)}, s' \circ i(O[r] \to N[g_O(r, s'(O, f(O[r])))]) \in S\})$$
$$+ \sum_{O \in R_d} \mu(\{s \in I^{L_s(M)} \mid \exists r \in \bigcup_l \mathbb{R}^l, s' \in I^{L_s(O)} : g_O(r) \in T, (M,s) \Rightarrow_f^{n-1} (O[r], s')\})$$
$$\mu(s' \in I^{L_s(O)} \mid s' \circ i(O[r] \to N[g_O(r)]) \in S)$$
$$= \sum_{O \in R_s} \mu(\{s \in I^{L_s(M)} \mid \exists r \in \prod_{j \neq h(O)} T_j, s' \in I^{L_s(O)} : (M,s) \Rightarrow_f^{n-1} (O[r], s')\}) \mu(T_{h(O)}) \mu(S)$$
$$+ \sum_{O \in R_d} \mu(\{s \in I^{L_s(M)} \mid \exists r \in \bigcup_l \mathbb{R}^l, s' \in I^{L_s(O)} : g_O(r) \in T, (M,s) \Rightarrow_f^{n-1} (O[r], s')\}) \mu(S)$$
$$= \sum_{O \in R_s} \mu(\{s \in I^{L_s(M)} \mid \exists r \in \prod_{j \neq h(O)} T_j, s' \in I^{L_s(O)} : s'(O, f(O[r])) \in T_{h(O)}, (M,s) \Rightarrow_f^{n-1} (O[r], s')\}) \mu(S)$$
$$+ \sum_{O \in R_d} \mu(\{s \in I^{L_s(M)} \mid \exists r \in \bigcup_l \mathbb{R}^l, s' \in I^{L_s(O)} : g_O(r) \in T, (M,s) \Rightarrow_f^{n-1} (O[r], s')\}) \mu(S)$$
$$= \sum_{O \in R_s} \mu(\{s \in I^{L_s(M)} \mid \exists r \in \mathbb{R}^{k-1}, s' \in I^{L_s(O)} : g_O(r, s'(O, f(O[r]))) \in T, (M,s) \Rightarrow_f^{n-1} (O[r], s')\}) \mu(S)$$
$$+ \sum_{O \in R_d} \mu(\{s \in I^{L_s(M)} \mid \exists r \in \bigcup_l \mathbb{R}^l, s' \in I^{L_s(O)} : g_O(r) \in T, (M,s) \Rightarrow_f^{n-1} (O[r], s')\}) \mu(S)$$
$$= \mu(\{s \in I^{L_s(M)} \mid \exists O \in R_s, r \in \mathbb{R}^{k-1}, s' \in I^{L_s(O)}, r' \in T, s'' \in I^{L_s(N)} : (M,s) \Rightarrow_f^{n-1} (O[r], s') \Rightarrow_f (N[r'], s'')\}) \mu(S)$$
$$+ \mu(\{s \in I^{L_s(M)} \mid \exists O \in R_d, r \in \bigcup_l \mathbb{R}^l, s' \in I^{L_s(O)}, r' \in T, s'' \in I^{L_s(N)} : (M,s) \Rightarrow_f^{n-1} (O[r], s') \Rightarrow_f (N[r'], s'')\}) \mu(S)$$
$$= \mu(\{s \in I^{L_s(M)} \mid \exists r \in T, s' \in I^{L_s(N)} : (M,s) \Rightarrow_f^n (N[r], s')\}) \mu(S)$$

$\square$

**Theorem VI.10.** *A closed term $M$ is AST with respect to* cbv *iff it is AST.*

*Proof.* For both red and $\Rightarrow_{\text{cbv}}$, the probability measure on the sample space can be used to define a measure on the reduction sequences of each finite length. In the case that the reduction sequences terminate after a finite number of steps by reaching a value, the reduction sequence defined using $\Rightarrow_{\text{cbv}}$ should be taken to continue by repeating the last term indefinitely, as red does, even though $\Rightarrow_{\text{cbv}}$ specifies no next step in this case. The distributions are equal for each length, by induction, as follows.

The base case, length 0, is trivial, because in both cases the distribution has probability 1 on the sequence containing only $M$.

For the inductive case, suppose that the distributions of reduction sequences of length $n$ are equal.

For any term $N$ which is not a value, there is a unique environment $E$ and redex $R$ such that $N = E[R]$. The position of the hole in $E$ is equal to $\text{cbv}(N)$, so that $N|\text{cbv}(N) = R$, as the cases in the definition of cbv match the cases in the definition of environments. If $R \neq$ sample, let $R'$ be the result of reducing $R$ (which is equal in both versions of the semantics), then $E[R'] = N[R'/\text{cbv}(N)]$ therefore $\text{red}(N, s) = (E[R'], s)$ and $(N, s) \Rightarrow_{\text{cbv}} (E[R'], s)$. The distribution of next terms, conditional on the previous term, in the case that that previous term reduces deterministically, is therefore equal for red and $\Rightarrow_{\text{cbv}}$.

In the other case that $R =$ sample, the first (term) part of $\text{red}(N, s)$ is $E[\pi_h(s)]$. If $k$ is the number of samples already taken in this reduction sequence, this is equivalent to $E[s_0(k)]$, where $s_0$ is the initial sample. For any sufficiently small neighbourhood of $N$ (i.e. containing only reduction sequences with the same skeletons), the probability of reaching this neighbourhood is independent of $s(k)$, because it depends only on the samples taken so far, and the samples are independent. The distribution of $s(k)$ conditional on reaching $N$ is therefore the uniform distribution on $I$. Similarly for the other semantics, if the distribution of remaining samples after $n$ steps is independent of the term by Lem. D.6, therefore the distribution of $s(Sk(N), \text{cbv}(N))$ conditional on $N$ is the uniform distribution on $I$. In either case, the distribution of the next term conditional on the previous term, in the case that it reduces randomly, is equal to the image under $r \mapsto E[\underline{r}]$ of the uniform distribution on $I$.

In any case, the distribution of reduction sequences after $n + 1$ steps, conditional on the reduction sequence after $n$ steps, as defined by $\Rightarrow_{\text{cbv}}$ and by red, is equal, any by the inductive hypothesis the distributions after $n$ steps are equal, therefore by integrating over the reduction sequence up to step $n$, the distributions on reduction sequences up to step $n + 1$ are equal.

The distributions on reduction sequences of any finite length as defined by $\Rightarrow_{\text{cbv}}$ and red are therefore equal, therefore so is the probability of having reached a value after $n$ steps, therefore so is its limit, the probability of termination, therefore the probability of termination is 1 iff the probability of termination with respect to cbv is 1. $\square$

It is also true that the distributions of terms produced by the standard semantics and by the reduction strategy cbv are the same, for much the same reason, although that is not relevant to our main results.

**Theorem VI.11.** *If $M$ terminates with some reduction strategy $f$ and trace $s$, it terminates with* cbv *and $s$.*

*Proof.* Suppose $M$ terminates with the reduction strategy $f$ and samples $s$. Let $(M, s) = (M_0, s_0) \Rightarrow_f \cdots \Rightarrow_f (M_n, s_n)$, where $M_n$ is a value. By Lem. D.2, the reduction sequence $(Sk(M_i))_i$ is related by $\sim_c^*$ to some skeletal reduction sequence $X$ which is in call-by-value order. For each skeletal reduction sequence $X_j$ in the sequence $(Sk(M_i))_i \sim_c \cdots \sim_c X$, it is possible to define a reduction sequence of terms such that $(M, s) = (N_{j,0}, s_{j,0}) \Rightarrow \cdots \Rightarrow (N_{j,n_j}, s_{j,n_j})$ and $Sk(N_{j,k}) = X_{j,k}$, by at each step applying the reduction at the same place as in $X_j$, essentially just filling in the reals in the skeletal reduction sequence to make it a reduction sequence of terms (it will be shown shortly that the correct branch is taken in each if-reduction). Assume for induction on $j$ that $(N_{j,n_j}, s_{j,n_j}) = (M_n, s_n)$. To prove that this holds true for $j + 1$ as well, it suffices to show that the corresponding term and samples immediately after the reductions involved in this $\sim_c$ step match, as from that point on, the $\Rightarrow$-sequences will match, as they start from the same point and have all the same reduction positions. Letting the term and remaining samples immediately before the reductions involved in this $\sim_c$ step be $(N, t)$ (which is the same in both $N_j$ and $N_{j+1}$ for the same reason again, they start at the same point and the sequences of reduction positions up to this point are the same), and the terms and samples after these reductions be $(O_1, t_1)$ and $(O_2, t_2)$, so that these match the $N$, $O_1$ and $O_2$ in the definition of $\sim_c$, $t_1 = t_2$ because both of them are, by Lem. VI.7, equal to the restriction of $t$ to the subset of $Rch(N)$ that starts with the reduction sequences to $O_1$ and $O_2$ respectively, and the potential positions in those subsets of $Rch(N)$ are identified with each other by $\sim$ because of the same case of $\sim_c$ that means $N_j$ and $N_{j+1}$ are related. If none of the relevant reductions are sample-reductions, $O_1 = O_2$ trivially. If there are samples taken (which must be the arbitrary reduction at $\beta$), it is also required that the same samples are taken in both branches. Similarly to Lem. VI.9, the positions of these sample-reductions in each branch are related by $\sim_p^*$, therefore they correspond to the same sample in $t$, therefore $O_1 = O_2$. The fact that the terms are equal also implies that the same branch of an if statements will be taken.

*e)* : There is therefore a reduction sequence $(M, s) = (N_0, s_0') \Rightarrow \cdots \Rightarrow (N_m, s_m')$ which ends in a value ($N_m = M_n$) and is in CBV order. It doesn't immediately follow that this is the $\Rightarrow_{\text{cbv}}$ reduction sequence starting from $(M, s)$ because the fact that $(N_i)_i$ is in CBV order only means that the reductions that do take place in this reduction sequence are in the correct order, not that the reduction sequence finishes once it reaches a value, or that there are no other reductions that would come earlier in CBV order that don't occur at all in $(N_i)_i$. Suppose that at some point $j$ in the reduction sequence, the reduction that occurs is not the next reduction in CBV order, but something else, and that the reduction that should be next is at position $\alpha = \text{cbv}(N_i)$, and the position of the actual reduction $N_j \to N_{j+1}$ is $\beta$. Suppose further, for a contradiction, that $N_j$ is not a value. Either $\alpha = \cdot$, the root position, or the reduction at the root position is dependent in some way on the reduction at $\alpha$ (e.g. $\alpha = \text{if}_1$, or $\alpha = \pm; @_1$). In any case, all the reductions in the rest of the reduction sequence are after $\alpha$ in CBV order, therefore none of them are $\leq \alpha$, and there continues to be an unreduced redex at $\alpha$, therefore the reduction at the root position never occurs, therefore $N_m$ is not a value, which is a contradiction, therefore at the point (if any) where $(N_i)_i$ departs from the $\Rightarrow_{\text{cbv}}$ reduction sequence starting at $(M, s)$, it has already reached a value, therefore $(M, s) \Rightarrow_{\text{cbv}} \ldots$ does reach a value eventually, i.e. $M$ terminates with the reduction strategy cbv and samples $s$. □

It is not quite true, however, that all reduction strategies produce the same values (barring some reduction strategies terminating where others don't), because it is possible for a reduction strategy to overshoot the value that the standard semantics produces by, for example, performing reductions inside of a lambda. Even for the same value of the samples, the final values may be different. A more restricted result is true, however, that if a term terminates with some reduction strategy $f$ and samples $s$, the value that $\Rightarrow_{\text{cbv}}$ terminates with can reduce via $\Rightarrow$ to the value that $\Rightarrow_f$ terminates with (as a corollary to this proof).

**Theorem VI.16.** *For any closed term $M$ and reduction strategy $r$ on $M$, if every term in $Rch_r(M)$ is AST, then every term in $Rch(M)$ is AST.*

*Proof.* Suppose that $(M, s) \Rightarrow_{\text{cbv}}^* (F, s')$ for some trace $s$ and term $F$ which is not AST. Let $P$ be the finite set of potential positions in $M$ such that the corresponding sample is used in the reductions $(M, s) \Rightarrow_{\text{cbv}}^* (F, s')$. As $F$ is not AST, the set $T_0 = \{t \in I^{L_s(M) \setminus P} \mid M \text{ terminates with } (t, s|_P) \text{ and cbv}\}$ has non-zero measure.

Take some order $p_1, \ldots$ on the elements of $P$, and define $T_n \subset I^{L_s(M)}$ recursively as follows. If the sample at $p_n$ is eventually used in the sequence $(M, (t, s|_P)) \Rightarrow_r^* \ldots$ for some positive measure subset of the traces $t$ in $T_{n-1}$, then let $T_n$ be the subset of $T_{n-1}$ where this sample is used within $m$ steps, for the minimal $m$ such that this is of positive measure. Otherwise, let $T_n$ be the subset of $T_{n-1}$ where the sample at $p_n$ is never used. Let $T$ be the final such $T_n$. Also, let $P_i$ be the set of $p_n$ such that the second case was taken in the definition of $T_n$ i.e. the sample at $p_n$ is never used in the reduction sequences starting at $(M, (t, s|_P))$ with strategy $r$, and let $k$ be the maximum value of $m$.

Let $T' = T \times \prod_{p \in P_i} I \times \prod_{p \in P \setminus P_i} \{s(p)\}$. Although $T'$ itself may have a measure of 0 as a subset of $I^{L_s(M)}$, it still has a natural measure-space structure of its own. It also has the properties that $M$ does not terminate with cbv and any trace in $T'$ (therefore $M$ also does not terminate with $r$ and any trace in $T'$, by Thm. VI.11), and for any trace $t \in T'$, if $(M, t) \Rightarrow_r^k (M', t')$, none of the potential positions in $M'$ correspond to any of the elements of $P \setminus P_i$. There are finitely many skeletal reduction sequences $k$ steps long starting at $M$ and following $r$, each one corresponding to some measurable subset of $I^{L_s(M)}$, so pick one such that its intersection with $T'$ is a non-null subset of $T'$, let this subset be $Q$.

Pick some $p \in Q$, and let $(M, p) \Rightarrow_r^k (N, p')$. The injection $i(M \to^k N)$ is the same as it would be for any other element of $Q$, and its image is disjoint from $P \setminus P_i$. The inverse image under composition with $i(M \to^k N)$ of $Q$ is therefore a non-null subset of $I^{L_s(N)}$, and $N$ fails to terminate with $r$ and every trace in this set, therefore $N$ is not AST with respect to $r$, therefore $N$ is not AST, but $N \in Rch_r(M)$, therefore not every element of $Rch_r(M)$ is AST.

Taking the contrapositive of this, we have the desired result. □

Ex. VI.17 Recall that

$$M = (\lambda n.P[n])\lfloor \text{sample}^{-1/2} \rfloor$$
$$P[n] = (\lambda p.\Xi[p]n)(\lambda x.\Theta[x]n))$$
$$\Xi[p] = Y\lambda fn.\text{if}(n = 0, \underline{0}, pn + f(n-1))$$
$$\Theta[x] = Y\lambda fn.\text{if}(n = 0, \underline{1}, x \times f(n-1)).$$

Let $r$ be the reduction strategy which reduces $M$ in the order

$$\begin{aligned}
&M \\
&\to^* P[\underline{n}] \\
&\to^* (\lambda p.\ \Xi[p]\ \underline{n})\ (\lambda x.\underbrace{x \times \cdots \times x \times}_{n} \underline{1}) \\
&\to \Xi[\lambda x.\underbrace{x \times \cdots \times x \times}_{n} \underline{1}]\ \underline{n} \\
&\to^* \underline{n^n} + \underline{(n-1)^n} + \cdots + \underline{0} \\
&\to^* \underline{\sum_{k=0}^{n} k^n}
\end{aligned}$$

(this is a little underspecified, but the details are not important).

A suitable sparse ranking function for it is

$$M \mapsto \frac{\pi^2}{3}$$
$$P[\underline{n}] \mapsto 2n+2$$
$$(\lambda p.\ \Xi[p]\ \underline{n})\ (\lambda x.\underbrace{x \times \cdots \times x}_{n-k} \times (\Theta[x]\ \underline{k})) \mapsto n+k+2$$
$$\Xi[\lambda x.\underbrace{x \times \cdots \times x}_{n} \times \underline{1}]\ \underline{n} \mapsto n+1$$
$$\underline{n^n} + \cdots + \underline{(n-k+1)^n} + \Xi[\lambda x.x \times \ldots \underline{1}]\ \underline{k} \mapsto k+1$$
$$\underline{\sum_{k=0}^{n} k^n} \mapsto 0.$$

The first step is justified because the probability of reaching $P[\underline{n}]$ for any specific $n \geq 0$ is $(n+1)^{-2} - (n+2)^{-2} = \frac{2n+3}{(n+1)^2(n+2)^2}$, therefore the expected next value of the ranking function is $\sum_{n=0}^{\infty} \frac{(2n+2)(2n+3)}{(n+1)^2(n+2)^2} = \frac{\pi^2}{3}$. The other steps are deterministic, and involve 1 or 0 Y-reductions each. Given that all of the reduction steps after the first are deterministic, and the number of Y reduction steps is not that hard to count, defining the sparse ranking function's values only at $M$ and $P[\underline{n}]$ would also have been reasonable, although in a similar term which had more random samples throughout, that would not have been so simple.

Providing a ranking function of some sort for this term in a similar level of detail would have been rather more complicated using only the standard reduction strategy. The recursion in $\Theta$ would have to be evaluated separately for every time it was used, so there would have been more (and more complex) terms to consider in the reduction sequence. Also, the number of Y-reductions from $P[\underline{n}]$ would have been $n^2+2n+1$ instead of $2n+2$, therefore the expected number of Y reductions starting from $M$ would be infinite, i.e. $M$ is not Y-PAST, therefore it is not even rankable. It would be possible to define an antitone sparse ranking function for it, but it is much more difficult to construct:

$$M \mapsto 1.47$$
$$P[\underline{n}] \mapsto 1 + \ln((n+1)n+1)$$
$$\Xi[\lambda x.\Theta[x]\ \underline{n}]\ \underline{n} \mapsto 1 + \ln((n+1)n+1)$$
$$\left.\begin{array}{r}\underline{n^n} + \cdots + \underline{k_1^n} + \\ \Xi[\lambda x.\Theta[x]\underline{n}](\underline{k_1 - 1})\end{array}\right\} \mapsto 1 + \ln((n+1)(k_1-1)+1)$$
$$\left.\begin{array}{r}\underline{n^n} + \cdots + \underline{(k_1+1)^n} + \\ \underline{\Theta[k_1]\ \underline{n}} + \\ \Xi[\lambda x.\Theta[x]\ \underline{n}](\underline{k_1 - 1})\end{array}\right\} \mapsto 1 + \ln((n+1)k_1+1)$$
$$\left.\begin{array}{r}\underline{n^n} + \cdots + \underline{(k_1+1)^n} + \\ \underbrace{\underline{k_1} \times \cdots \times \underline{k_1}}_{n-k_2+1} \times \\ \Theta[\underline{k_1}](\underline{k_2-1}) + \\ \Xi[\lambda x.\Theta[x]\ \underline{n}](\underline{k_1 - 1})\end{array}\right\} \mapsto 1 + \ln((n+1)k_1+k_2+1)$$
$$\underline{\sum_{k=0}^{n} k^n} \mapsto 0.$$

It wouldn't even be possible in this case to give a sparser version of the ranking function defined only at $M$ and $P[\underline{n}]$, because the amount that an antitone sparse ranking function must decrease at each Y-reduction step depends on the value of the ranking function at the next term where it is defined, therefore it is necessary to have these intermediate steps in order for the ranking function to be able to change sufficiently slowly.

## Contents